\documentclass[a4paper,fleqn,usenatbib]{mnras}

\usepackage{newtxtext,newtxmath}
\usepackage{siunitx}

\usepackage[T1]{fontenc}
\usepackage{ae,aecompl}

\usepackage{graphicx}	
\usepackage{amsmath}	
\usepackage{amssymb}	
\usepackage{subcaption}
\title[Energetic Particles in Protogalactic Outflows]{Hadronic Interactions of Energetic Charged Particles in Protogalactic Outflow Environments and Implications for the Early Evolution of Galaxies}

\author[Owen et al.]{
Ellis R. Owen,$^{1}$\thanks{E-mail: ellis.owen.12@ucl.ac.uk (ERO), xyjin@smail.nju.edu.cn (XJ), kinwah.wu@ucl.ac.uk (KW), lsuetyi@yahoo.com (SC)}
Xiangyu Jin,$^{1, 2}$
Kinwah Wu,$^{1}$
Suetyi Chan$^{1, 3}$
\\
$^{1}$Mullard Space Science Laboratory, University College London, Holmbury St. Mary, Dorking, Surrey, RH5 6NT, United Kingdom\\
$^{2}$School of Physics, Nanjing University, Nanjing 210023, China\\
$^{3}$School of Astronomy and Space Science, Nanjing University, Nanjing 210093, China
}

\date{Accepted XXX. Received YYY; in original form ZZZ} 

\pubyear{2018}

\begin{document}
\label{firstpage}
\pagerange{\pageref{firstpage}--\pageref{lastpage}}
\maketitle

\begin{abstract}
We investigate the interactions of energetic hadronic particles with the media
in outflows from star-forming protogalaxies. 
These particles undergo pion-producing interactions
  which can drive a heating effect in the outflow, while those advected by the outflow also transport energy beyond the galaxy, 
 heating the circumgalactic medium. 
We investigate how this process evolves over the length of the outflow 
  and calculate the corresponding heating rates in 
  advection-dominated and diffusion-dominated cosmic ray transport regimes. 
In a purely diffusive transport scenario,  
 we find the peak heating rate reaches $10^{-26}\;\! {\rm erg~cm}^{-3}\;\! {\rm s}^{-1}$ 
 at the base of the outflow where the wind is driven by core-collapse supernovae at an event rate of 0.1 $\text{yr}^{-1}$, but does not extend beyond 2 kpc. 
In the advection limit, 
 the peak heating rate is reduced to $10^{-28}\;\! {\rm erg~cm}^{-3}\;\! {\rm s}^{-1}$, but its extent can reach to tens of kpc. 
Around 10\% of the cosmic rays injected into the system can escape by advection with the outflow wind, 
  while the remaining cosmic rays deliver an important interstellar heating effect. 
We apply our cosmic ray heating model to the recent observation of the high-redshift galaxy MACS1149-JD1  
  and show that it could account for the quenching of a previous starburst inferred from spectroscopic observations.   
Re-ignition of later star-formation may be caused by the presence of filamentary circumgalactic inflows 
  which are reinstated after cosmic ray heating has subsided. 
\end{abstract}




\section{Introduction}
\label{sec:introduction}

Advection and diffusion are the two main mechanisms 
   for the transportation of high-energy charged hadronic (cosmic ray) particles 
   in galactic environments.  
These are evident in the Galactic interstellar medium (ISM) 
   \citep[see][]{Schlickeiser2002book, Strong2007ARNPS, Gaggero2015PRD, Korsmeier2016PRD, Yuan2017PRD},  
  and within the solar system 
  \citep[see][]{Jokipii1966ApJ, Orlando2008A&A, Abdo2011ApJ, 
    Potgieter2013LvRSP, Chhiber2017ApJS}.  
The interplay between the two processes   
   is determined by the extent of the bulk flows in the carrying medium 
    \citep[e.g.][]{Dorfi2012A&A, Uhlig2012MNRAS, Heesen2016MNRAS, Taylor2017PRD, Farber2018ApJ},   
    the structures and strengths of the local magnetic field   
     \citep[e.g.][]{Parker1964JGR, Jokipii1966ApJ, Berezinskii1990book,  
     Schlickeiser2002book, Alvarez-Muniz2002ApJ, 
     Aharonian2012SSR, Gaggero2012thesis, Snodin2016MNRAS} 
    and the amount of turbulence present in the system 
      \citep[e.g.][]{Berezinskii1990book, Schlickeiser2002book, Candia2004JCAP, 
      Gaggero2012thesis, Snodin2016MNRAS}.  
  
The general consensus is that cosmic rays (CRs) are accelerated to high energies in violent environments, 
   e.g. SN explosions, gamma-ray bursts,  
   large-scale shocks in the ISM or galactic outflows, AGN jets, galaxy clusters and 
  compact objects such as fast spinning neutron stars and accreting black holes 
  \citep[see][]{Berezinsky2006PRD, Brunetti2007ApJ, Pfrommer2007MNRAS, 
     Dar2008PhR, Reynoso2011A&A, Kotera2011ARAA}.  
Fermi processes \citep{Fermi1949PhRv} have been suggested as viable mechanisms   
   by which low-energy charged particles can be accelerated to attain relativistic energies. 
In astrophysical systems this might arise in shocks, 
   such as those resulting from SN explosions.  
Systems such as starburst galaxies, which have frequent SN events,  
  are therefore expected be abundant in energetic CRs
  \citep{Karlsson2008AIPC, Lacki2011ApJ, Lacki2012AIPC, Wang2014AIPC, Farber2018ApJ}.   
Likewise protogalaxies, which have vibrant star forming activity and hence high SN event rates, 
   should also be abundant in CRs.   
In a similar way to the shocks in the ISM generated by SN explosions,        
  large-scale shocks in the intergalactic medium (IGM) and intracluster medium (ICM) 
  can also be accelerators of CRs. 
There is evidence that 
  energetic CRs are an important ingredient in galaxy clusters   
  \citep{Takami2008ICRC, Brunetti2014IJMPD}, 
  providing pressure support to clusters' structure. 
These CRs may play an important role in regulating the energy budget of the ICM 
  through radiative losses, energy transportation and hadronic interactions.  

While we have some understanding of the effects of CRs on the thermal and dynamical properties 
  of the ISM and IGM in nearby astrophysical systems, 
  our knowledge of the impacts of CR particles  
  on the formation and evolution of structures on scales of galaxies or larger is very limited. 
The importance of CRs in protogalactic environments has gradually drawn more attention 
   \citep[e.g.][]{Giammanco2005A&A, Stecker2006ApJ, Valdes2010MNRAS, 
   Sazonov2015MNRAS, Bartos2015arXiv, Leite2017MNRAS, Owen2018MNRAS}. 
In particular, there are studies showing that 
  CR heating of the ISM could lead to the distortion of the stellar initial mass function (IMF) 
  and even quench the star formation process entirely
  \citep[see][]{Pfrommer2007, Chen2016MNRAS}. 
CRs can also drive the large-scale galactic outflows 
  \citep[see][]{Socrates2008ApJ, Weiner2009ApJ},  
  which transfer energy and chemically enriched material into intergalactic space. 
The resulting pre-heating of the IGM in turn alters cosmological structural formation processes, 
   and there are arguments that CRs might contribute to a certain degree of cosmological reionisation 
   \citep[see][]{Nath1993MNRAS, Sazonov2015MNRAS, Leite2017MNRAS}.

On sub-galactic scales, 
   the production of CRs is often attributed to 
   supernovae (SNe), compact objects (such as spinning neutron stars) and accretion-powered sources, 
   which are consequential of stellar evolution and hence star formation processes.    
However, the delivery of CR energy across a galaxy    
   depends on strength and structure of the galactic magnetic field 
   which, in turn, depends on the field evolution and hence the star-forming processes. 
SN explosions are energetic events. 
On the one hand, SN explosions would drive a large-scale galactic wind 
   \citep{Chevalier1985Nat, Socrates2008ApJ, Weiner2009ApJ},  
   but on the other hand, they inject enormous amount of mechanical energy into the ISM within the galaxy  
   which fuels the development of ISM turbulence
    \citep{Dib2006ApJ, Joung2009ApJ, Gent2013MNRAS, Martizzi2015MNRAS, Martizzi2016MNRAS}.  
SN explosions also help the magnetisation of the entire galaxy 
   \citep{Zweibel1997Nat, Zweibel2003ApJ, Beck2012MNRAS, Schober2013A&A, Lacki2013MNRAS}.  

In a magnetised medium with strong turbulence but no large-scale bulk flows,
   CR transport would be dominated by diffusion and this can lead to their effective containment 
   within the host galaxy~\citep[see][]{Owen2018MNRAS} 
   with their subsequent energy deposition into the media being regulated by hadronic, pion-producing interactions
  occurring above a threshold energy of 0.28 GeV~\citep{Kafexhiu2014PRD}.       
However, in the presence of a flow, 
  CRs entangled into a magnetised medium can be advected along. 
  This advection process would happen in large-scale galactic winds, 
  where diffusion still takes place, but on time scales far longer than the flow time scale 
   \citep[see][]{Berezinskii1990book, Schlickeiser2002book, Aharonian2012SSR, Heesen2016MNRAS}. 
As such, CRs can be advected into intergalactic space causing heating of the circumgalactic medium.  
Imaging observations of nearby starburst galaxies 
   have shown complex structural morphologies 
   in which winds and outflows are ``collimated" in a cone-like structure 
   while the gases and stars beneath retain a planar galactic disk-like structure.  
Such structural complexity 
  implies the coexistence of CR diffusion and advection 
  --- while in some regions the two processes would have comparable partitions in facilitating energy transport,  
  in other regions one of them would dominate.

Here we further investigate the contribution of CRs to ISM and IGM heating via hadronic processes 
  with a focus on the effects of CRs by galactic wind outflows.     
We organise the paper as follows. 
In \S~\ref{sec:outflows} 
  we discuss the properties of galactic outflows driven by SNe and CRs
   and the outflow model used for our investigation of CR heating. 
In \S~\ref{sec:particles_in_outflows}, 
   we present the formulation for CR propagation in the diffusion and advection dominated regimes. 
We also discuss the relevant mechanisms regulating the energy budget 
    of CRs advected by bulk flows
    and the hadronic processes by which the CR energy is deposited into the ISM and/or IGM. 
In \S~\ref{sec:results}, we show the results of our calculations of CR heating in protogalactic and outflow environments 
   and demonstrate how the heating effect depends on model parameters, and discuss     
the astrophysical implications. 
   An application to explain the inferred star-formation behaviour of the high-redshift galaxy MACS1149-JD1 \citep{Hashimoto2018Nat}
   is presented in 
   \S~\ref{sec:observation_comparison} 
   and conclusions are given in \S~\ref{sec:conclusions}.

Our calculations assume that CRs are energetic protons. This assumption is based on the idea that 
  CRs are produced and accelerated \citep[see][]{Berezinsky2006PRD, Kotera2011ARAA} 
  in a similar manner in the distant Universe as they are in the nearby Universe,
and that a substantial fraction of CRs detected on Earth (from the nearby Universe) are protons \citep{Abbasi2010PRL}.  
This allows us to ignore the composition evolution of CRs 
  as a first approximation.   
We also do not consider CR primary electrons explicitly 
  as these charged, low-mass leptonic particles have considerably higher radiative loss rates than charged hadrons. This means that
   they would not be a major contributor to the global energy transportation picture.  
Hereafter, unless it is necessary, we do not differentiate between CR particle species,  
  and CR protons are referred to as CRs. 

\section{Galactic Outflows}
\label{sec:outflows}

\subsection{Observational aspects and phenomenology} 
\label{subsec:phenomenology}

Galactic-scale outflows have been observed in star-forming galaxies nearby, 
  e.g. Arp 220 \citep{Lockhart2015ApJ},   
  and in the distant Universe 
  \citep{Frye2002ApJ, Ajiki2002ApJ, Benitez2002book, Rupke2005ApJS-a, 
        Rupke2005ApJS-b, Bordoloi2011ApJ, Arribas2014A&A}. 
In active star-forming regions,   
   the proximity of SNe allow 
   the confluence of gas flows induced by the SN explosions to develop into a larger-scale wind.   
The build-up of these confluent winds eventually 
  erupts as a large-scale galactic outflow, 
  usually with a bi-conical structure along the minor axis of the host~\citep{Veilleux2005ARAA}.        
The opening angles of the outflow cones are broad, of tens of degrees 
  \citep{Heckman1990ApJS, Veilleux2005ARAA}, 
 and their values vary between galaxies, 
 e.g. around \ang{26}$-$\ang{60} in NGC 253~\citep{Strickland2000AJ, Bolatto2013Nat} and 
 approximately \ang{60} in M82~\citep{Heckman1990ApJS, Walter2002ApJ}.  
Observations have shown  
   that the hot X-ray emitting gas in an outflow can reach up to 3~kpc 
     \citep{Strickland2000AJ, Cecil2002ApJ, Cecil2002RMxAACS}, 
   and the entire outflow structure could extend up to tens of~kpc 
     \citep[see][]{Veilleux2005ARAA, Bland-Hawthorn2007APSS, 
     Bordoloi2011ApJ, Martin2013ApJ, Rubin2014ApJ, Bordoloi2016MNRAS}.  
Galactic outflows are inhomogeneous, multicomponent, multiphase media, 
   with warm, partially ionised gases intermingled 
   with hot ionised bubbles and cooler, denser less ionised gas or neutral clumps.   
The outflow velocity has been measured from a few hundred km s$^{-1}$ in most cases, rising to a few thousand km s$^{-1}$ in a few extreme systems
  \citep{Rupke2005ApJS-b, Cecil2002RMxAACS, Rubin2014ApJ}, 
  and the total mechanical power in the flow is estimated 
  to be up to levels as high as $10^{43}$ erg s$^{-1}$ \citep[see][]{Cecil2002RMxAACS}. 
Galactic outflows play an active role in injecting mechanical energy into intergalactic space  
  and in distributing chemically enriched matter into the environment 
   \citep{Aguirre2001ApJ-a, Aguirre2001ApJ-b, Martin2002ApJ, Rupke2003AIPC, Adelberger2003ApJ, 
     Aguirre2005ApJ, Bertone2005MNRAS}. 
There are also arguments that 
   galactic outflows carry nG-strength magnetic fields 
   from the within the galaxy to the surrounding IGM and ICM, 
   where the seed fields would then be amplified 
   by turbulence, dynamo mechanisms and/or cosmic-scale shears to the $\mu$G-levels  
   inferred from observations    
   \citep{DeYoung1992ApJ, Goldshmidt1993ApJ, Dolag1999A&A, Dolag2002A&A, Bertone2006MNRAS, 
     Vazza2018MNRAS}. 
Of most relevance to this work, 
  galactic outflows are efficient vehicles to transport CRs and the energy they carry  
  across their source galaxy and to significant distances away from it 
  \citep[see][]{Heesen2016MNRAS}. 

\subsection{Outflow Wind Structure}
\label{sub:working_model}

In our calculations, 
  we adopt a working model  
  that sufficiently captures the most essential microphysics 
   and the associated global physics and astrophysics of the system. 
We focus on the redistribution of energy  
   through the advective and diffusive transport of CRs  
   and investigate the relative efficiency of CR heating in galactic outflows and the surrounding IGM
   between these two modes of CR transportation.     
Complexities such as the fine substructure of outflows, 
   the multi-phase nature of the flow material  
    and the re-acceleration of CR particles within the flow
   are worth separate further investigations in their own right, and so are not considered in detail in the present study, instead being left to future follow-up work.  

Galactic outflows can be powered by different mechanisms.
Most early models invoke thermally and/or SN-driven mechanisms~\citep{Larson1974MNRAS, Chevalier1985Nat, Dekel1986ApJ, Nath1997MNRAS, Efstathiou2000MNRAS, Madau2001ApJ, Furlanetto2003ApJ, Scannapieco2005ApJ, Samui2008MNRAS}
while more recent studies have also considered radiatively-driven outflows~\citep{Dijkstra2008MNRAS, Nath2009MNRAS, Thompson2015MNRAS}.
CRs have also been regarded as a means by which galactic outflows may be driven: indeed, a CR driving mechanism would offer a good explanation of the observed soft X-ray diffuse emission from the Milky Way~\citep{Everett2008ApJ}. At high redshift, actively star-forming galaxies would be abundant in CRs and thus are a clear candidate driver of outflows~\citep[see also][]{Samui2010MNRAS, Uhlig2012MNRAS}, while the 
high altitude winds which have the ability to inject CRs far into the circumgalactic medium are thought to be powered by CRs~\citep{Jacob2018MNRAS}.
CRs influence the density and structure of the flow compared to other driving mechanisms~\citep{Girichidis2018MNRAS} and also lose some of their energy in driving the outflow~\citep[e.g.][]{Samui2010MNRAS, Uhlig2012MNRAS}. These factors modify the heating effect that they are able to deliver when interacting with the wind fluid via hadronic interactions when compared to their role in winds driven by other mechanisms and, as such, mean that CRs must be self-consistently included in the modelling of the wind structure and dynamics.
At high-redshift, the impact of smaller galaxies, of mass around $10^9~\text{M}_{\odot}$, on their environment is argued to be more important than more massive galaxies~\citep[see, e.g.][]{Samui2008MNRAS, Samui2009NewA}.
As such we focus this study on these smaller systems, for which the impact will be greatest.

\begin{figure}
	\includegraphics[width=\columnwidth]{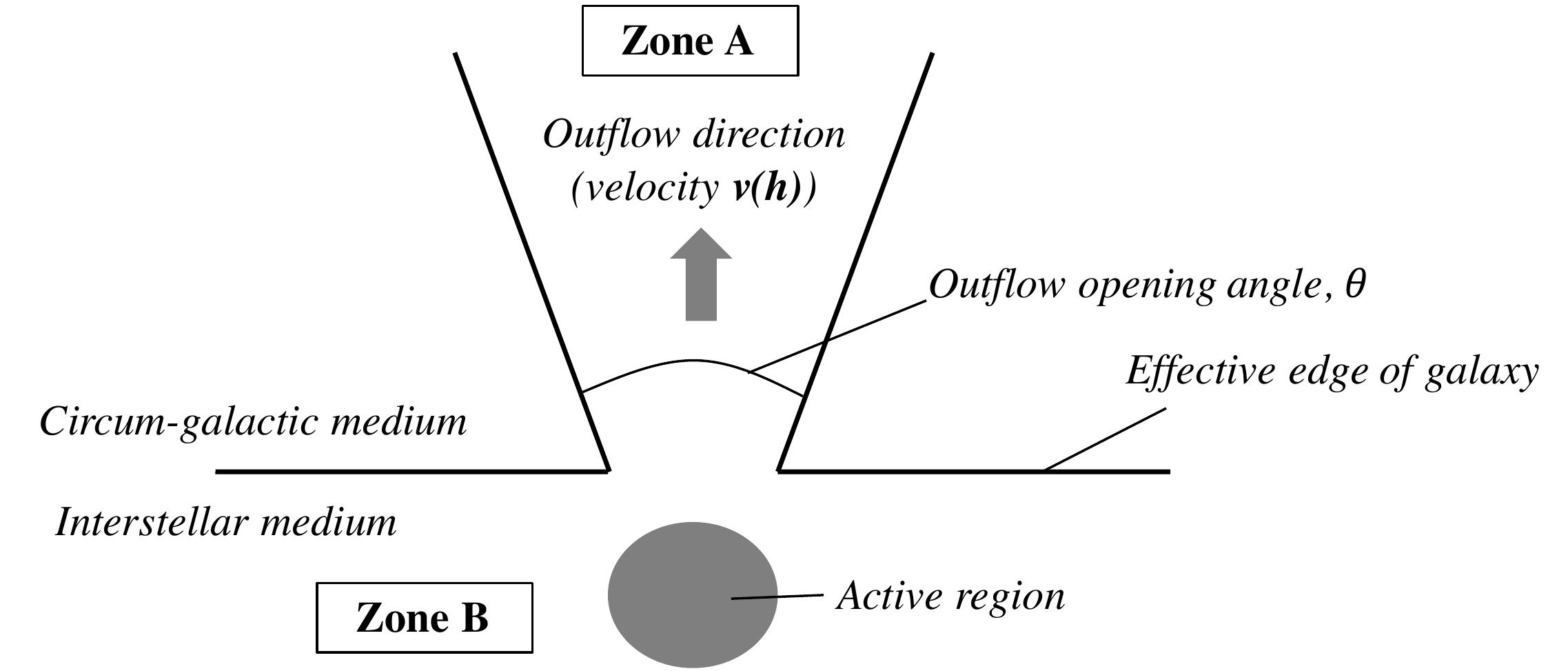}
    \caption{Schematic illustration of the `two-zone' outflow model geometry. Zone A represents the region in which CR transport would be predominantly advective in the presence of a galactic outflow while, in Zone B, CR transport would be mainly diffusive~\citep[see also outflow wind morphology in e.g.][]{Strickland2000AJ, Ohyama2002PASJ, Veilleux2005ARAA, Cooper2008ApJ}}
\label{fig:outflow_model}
\end{figure}

      The direction of emergence of a galactic outflow 
   is governed by the `path of least resistance'. 
In disk galaxies, 
  a bi-conical flow pattern above and below the galactic plane is generally observed \citep[see][]{Veilleux2005ARAA}, 
and this geometrical shape is expected 
  because a spherically expanding wind from an actively star forming region at the core of the galaxy 
  would be less obstructed by the the upper and lower edge of the galaxy that it encounters than the galactic plane. 
Emerging flows from spherical or near-spherical elliptical galaxies 
  would have a less well-defined morphological pattern.  
An isotropic spherical outflow could arise 
   if the outflowing wind from the galactic core region 
   encounters all edges of the galaxy at a similar time, 
   and if it is faced with similar inflowing pressures and resistances in all directions. 
   We consider disk galaxies with bi-conical outflows. 
A schematic of the outflow model is shown in Fig.~\ref{fig:outflow_model}~\citep[see also outflow wind morphology in e.g.][]{Strickland2000AJ, Ohyama2002PASJ, Veilleux2005ARAA, Cooper2008ApJ}, where two distinct `zones' are noted: Zone A is within the outflow cone where, we assume, CR transport is dominated by advection; Zone B is outside of the cone but within the galactic ISM and is the region in which CR transport is predominantly diffusive. 
We explore a range of opening angles, $\theta$, between \ang{45} and \ang{65} 
   covering a similar range of values to a large subset of those observed in nearby starburst galaxy outflows. 
A reference value of \ang{55} is chosen (if not otherwise specified) 
    as a working representation\footnote{Hydrodynamical simulations suggest that, 
    rather than remaining uniform throughout the extent of an outflow, 
    the opening angles start at a low value of \ang{10}$-$\ang{45} near their base 
    and then diverge to \ang{45}$-$\ang{100} 
    well above and below the galactic plane
    \citep{MacLow1999ApJ, Martel2001RMxAACS, Pieri2007ApJ, Bordoloi2016MNRAS},  
   but this finer substructure is not accounted for in our model - 
   instead we choose an opening angle which reflects that of the wider angle of the main part of the outflow.}. 

 \cite{Ipavich1975ApJ} considered a 1-D numerical magnetohydrodynamic model for a CR-powered wind emerging from a galaxy with a point-like mass distribution.
 The model is parametrised with energy and mass injection, 
  presumably provided by the SNe resulting from the starburst activity. 
A spherical geometry is assumed, 
  with the wind emerging radially from a small active star-forming region enveloping the galactic core.
Solving the associated magnetohydrodynamic equations yield several valid solutions depending on the boundary conditions adopted at a so-called critical point,
at which the flow becomes supersonic. One such solution is that of an outflow wind with an asymptotic velocity 
  for the outflow at distances sufficiently far from the star-forming galactic core.  
This idea was further developed by several authors since then, including ~\cite{Breitschwerdt1987ICRC, Breitschwerdt1991A&A, Breitschwerdt1993A&A, Everett2008ApJ, Bustard2016ApJ, Recchia2016MNRAS} and~\cite{Samui2010MNRAS}. The latter of these accounts for a CR-driven outflow in the presence of an NFW density profile, with the intention of application to high-redshift starburst galaxies. We largely follow the~\cite{Samui2010MNRAS} outflow model here and use it to compute
 the density profile and other relevant conditions 
  from which the advection of CRs and their subsequent hadronic interaction induced heating effect
  can be determined. 
  
\subsection{Outflow Model}
\label{subsec:wind_velocity} 


We denote $h$ as a coordinate variable along a flow streamline and, in a spherical symmetric geometry, 
  $h$ is the radial distance from the galactic core.   
  We model the CRs and wind fluid as two separate but interacting components of an outflow wind in which the CR component has negligible mass density but non-negligible energy density.
In the \cite{Samui2010MNRAS} model  
  the galactic outflow wind is a conic section of a spherical flow, 
  with an asymptotic velocity arising 
  at a sufficiently large distance from the galactic core region. This can be found by considering the steady-state spherically symmetric form of the fluid and CR equations~\citep{Ipavich1975ApJ, Breitschwerdt1991A&A}:
\begin{align}
\frac{1}{h^2}\frac{{\rm d}}{{\rm d} h}\left(\rho v h^2\right) &= 0 \label{eq:continuity_eq}\\
 v \rho \frac{{\rm d}v}{{\rm d}h} &= -\frac{{\rm d} P}{{\rm d}h} -\frac{{\rm d} P_{\rm C}}{{\rm d}h} - \rho \frac{{\rm d}\Phi}{{\rm d} h} \label{eq:momentum_eq}\\
\frac{1}{h^2}\frac{{\rm d}}{{\rm d} h}\left[\rho v h^2 \left(\frac{1}{2} v^2 + \frac{\gamma_{\rm g}}{\gamma_{\rm g}-1}\frac{P}{\rho}\right)\right] &= -\rho v \frac{{\rm d}\Phi}{{\rm d} h} + I \label{eq:energy_eq}\\
\frac{1}{h^2}\frac{{\rm d}}{{\rm d} h}\left[ \frac{\gamma_{\rm C}}{\gamma_{\rm C}-1} P_{\rm C} h^2 (v + v_{\rm A})\right] &= -I
\label{eq:cr_evolution}  \ ,
\end{align}
where~\eqref{eq:continuity_eq} is the mass continuity equation,~\eqref{eq:momentum_eq} is the momentum equation and~\eqref{eq:energy_eq} is the energy equation for the wind fluid, while~\eqref{eq:cr_evolution} is the evolution equation for the CR fluid component of the wind. $\rho$ is the density of the wind fluid, $v$ is the wind velocity, $P$ is the pressure of the wind fluid (gas), $P_{\rm C}$ is the CR pressure, $\Phi$ is the gravitational potential, $\gamma_{\rm g} = 5/3$ is the adiabatic index for the gas component, and $\gamma_{\rm C} = 4/3$ is the adiabatic index for the relativistic CR component. We specify the total mass injection rate into the wind as $\rho v h^2 = q$, from equation~\ref{eq:continuity_eq}, with $q$ as a mass injection rate due to SN mass-loading of the wind (see equation~\ref{eq:mass_loading_q_def}). $I$ is an energy exchange term between the CRs and baryonic wind fluid (see equation~\ref{eq:exchange_equation_pressure}).

We adopt a magnetic field strength and morphology along the outflow cone according to:
\begin{equation}
B(h, \mathcal{R}_{\rm SN}) = {B}_{\rm 0} \left(\frac{\mathcal{R}_{\rm SN}}{\mathcal{R}_{\rm SN, 0}}\right)^{1/2}\;\!{\left(1+\left[\frac{h}{h_{\rm B}}\right]^2\right)^{-1}} \ ,
\end{equation}
where $B_{\rm 0} = 5\mu\text{G}$, $\mathcal{R}_{\rm SN, 0} = 0.1~\text{yr}^{-1}$, and where $h_{\rm B}$ is introduced as a characteristic scale over which the magnetic field does not strongly vary within the host galaxy of the outflow.  
Physically, the variation of the magnetic field with $h$ would only be expected in regions of the model that are well within the outflow cone.
In regions which may better be regarded as interstellar environments, the magnetic field would vary less substantially with height. We find that a choice of $h_{\rm B} = 1.5~\text{kpc}$ yields a relatively uniform magnetic field within a 0.5 kpc starburst region, falling only by around 10\% from its peak value.
Beyond this, $B(h)$ reverts to an inverse-square law behaviour with $h$ thus ensuring the conservation of magnetic flux along the outflow.
The dependence of magnetic field strength at the base of the outflow (within the ISM of the host) on the square root of SN-rate follows from~\cite{Schober2013A&A}, which models the development of magnetic fields in young starburst galaxies via turbulent dynamo amplification.

Our choice of $B_{\rm 0}$ is reflective of interstellar environments, where energy densities of CRs at the peak of their spectrum are comparable to that of the magnetic field.
As CRs gyrate and stream along the magnetic field lines at speeds faster than the Alfv\'{e}n velocity $v_{\rm A} = B(h)/\sqrt{4\pi \rho}$, they amplify interstellar Alfv\'{e}n waves which have wavelengths comparable to the gyro-radii of the streaming CRs~\citep{Wentzel1968ApJ, Kulsrud1969ApJ, Kulsrud1971ApL}.
In this process, known as the streaming instability~\citep{Wentzel1968ApJ, Kulsrud1969ApJ}, this leads to a resonant scattering effect, which slows the CRs and transfers momentum and energy from the CRs to the ambient medium after dampening of the waves at a rate given by $|v_{\rm A} \cdot \nabla P_{\rm C}|$~\citep[e.g.][]{Wentzel1971ApJ, Ipavich1975ApJ, Breitschwerdt1991A&A, Uhlig2012MNRAS}.
Further losses by the CRs to the wind fluid result from the work done by the CR pressure gradient in a bulk wind velocity $v$, arising at a rate of $|v \cdot \nabla P_{\rm C}|$~\citep{Samui2010MNRAS}.
Together, this allows us to define $I$ as the total energy exchange term between the baryonic wind fluid and the energetic CR component, given by:
\begin{align}
I &= -(v+v_{\rm A}) \frac{{\rm d}P_{\rm C}}{{\rm d} h}
\label{eq:exchange_equation_pressure}
\end{align}
\citep{Samui2010MNRAS}, where the minus sign is due to the energy exchange resulting in a loss by the CR component and a gain by the wind fluid. 

\cite{Samui2010MNRAS} solve this system of equations when adopting an NFW~\citep[Navarro--Frenk--White,][]{Navarro1996ApJ} gravitational potential of the form
\begin{equation}
\Phi(h) = -\frac{3 {\rm G} M_{\rm tot}}{h} \ln\left\{h+ \frac{h}{R_{\rm s}}\right\} \ ,
\end{equation}
for $M_{\rm tot}$ as the total galaxy mass and where $R_{\rm S}$ is the scale height, being the ratio of the virial radius of the galaxy and the concentration parameter, $R_{\rm vir}/c_{\rm g}$. This potential is relevant to galaxies like that which we also wish to model here. 
In the system of equations, 
$q$ is the volumetric mass injection (which is non-zero only within the starburst region, i.e. $h<h_{\rm inj}$).
This may be quantified in terms of the SN event rate, $\mathcal{R}_{\rm SN}$, 
  and the mass ejecta $M_{\rm inj}$ resulting from a SN explosion 
  to estimate $\dot{M}$.  
  The level of mass injection per event varies with SN types. 
Type~II SNe offer a characteristic mass of around $10\;\! {\rm M}_{\odot}$,   
  and Type~I$\;\!$b/c of a few ${\rm M}_\odot$ 
  \citep[see, e.g.][]{Branch2010Nat, Perets2010Nat}.  
Note that lower mass stars take a longer time to complete their life cycles and so,
at high-redshifts, only the very high-mass stars 
  would have enough time to evolve to the SN stage within the host galaxy's evolutionary timescale. 
Moreover, a low metallicity environment  
   would yield a more top-heavy initial stellar mass distribution 
   \citep[e.g.][]{Abel2002Sci, Bromm2002ApJ, Bastian2010ARAA, Gargiulo2014MNRAS},     
 skewing the progenitor masses to favour the core-collapse SN channels more typical of massive stars. 
Thus, core-collapse SNe and hypernovae, 
    with progenitors of masses ${{M}}_{{\rm SN}} \sim 8.5 ~{\rm M}_{\odot}$ or higher 
    \citep[see, e.g.][]{Smartt2009MNRAS, Smartt2009ARAA, Smith2011MNRAS} 
    would occur frequently in star-bursting protogalaxies.  
These core-collapse SN and hypernovae are extremely energetic, 
   with $E_{\rm SN} \approx 10^{53}\;\!{\rm erg}$ per event \citep{Smartt2009ARAA}.     
       
The mass injection rate may be parametrised as  
\begin{equation}
      q = \mathcal{P}\left[{{\cal{R}}_{\rm SN} M_{\rm inj}}\right] 
        = \mathcal{P}\left[\frac{\alpha_{*} \;\! {\cal{R}}_{\rm SF} M_{\rm inj}}{{\overline{M}}_{\rm SN}}\right] \ , 
        \label{eq:mass_loading_q_def}
\end{equation}
    where ${\cal{R}}_{\rm SF}$ is the star formation rate 
    and $\overline{M}_{\rm SN}$ is mean SN progenitor mass.  
The parameter $\mathcal{P}$ is the mass-loading factor, which is a scaling factor specifying the mass loaded into the wind 
    for a given mass ejected from the progenitor star in a SN event.  
    Although ${\mathcal P}$ could have a value above 1 and our limited knowledge of the ISM environment and the properties of SNe in protogalaxies 
prevents us from deriving a strong constraint for appropriate values of this parameter
   ~\citep[see e.g.][which gives mass loading fractions of 10 and above in NGC 1569, 
       among others]{Martin2002ApJ}, 
   we conservatively adopt that ${\mathcal P} = 0.1$.    
   The parameter $\alpha_{*}$ is the fraction of stars that yield Type II SNe (and hypernovae), 
   which can be estimated as 
\begin{equation}
\label{eq:sn_alpha}
     \alpha_{*} = 
       \frac{\int_{M_{\rm SN^*}}^{M_{\rm max}} {\rm d}M M^{-\Upsilon}}
           {\int_{M_{\rm cut}}^{M_{\rm max}}{\rm d}M M^{-\Upsilon}}  \ . 
\end{equation}
As a conservative estimate, 
   we adopt a Salpeter IMF index of $\Upsilon = 2.35$\footnote{There is some evidence 
    of a redshift-dependent IMF 
    \citep{Lacey2007MNRAS, Dave2008MNRAS, vanDokkum2008ApJ, Hayward2013MNRAS}, 
    but this remains an open discussion 
    \citep[see e.g][for reviews]{Bastian2010ARAA, Cen2010ApJ}. 
Given the lower metallicities, higher cosmic microwave background temperature 
   which could influence molecular-cloud collapse, 
    and the tendency for star-formation at high redshifts to arise in the `burst' mode 
    rather than more gradually \citep[see][]{Lacey2007MNRAS}, 
    the IMF at high redshift may be more top-heavy -- 
    favouring the production of the very massive stars 
    compared to the IMF observed in the current epoch, or the Salpeter IMF. 
A Top-heavy IMF has been claimed for some nearby starburst systems  
   \citep{Weidner2011MNRAS, Bekki2013ApJ, Chabrier2014ApJ}, 
   e.g. in M82 \citep{Rieke1993ApJ, McCrady2003ApJ} and 
   NGC 3603 \citep{Harayama2008ApJ}, 
   and even for Galactic centre clusters \citep{Stolte2005ApJ, Maness2007ApJ}.   
In these cases, the Salpeter IMF index of 2.35 
    underestimates the number of high-mass stars and hence the SN events.}.     
We set the the maximum stellar mass\footnote{Note this is at the upper end of likely progenitor masses 
     to ensure that our calculation is conservative. 
     In reality a greater proportion of massive stars are more likely to arise in the protogalactic environments.} 
   which could reasonably yield a SN explosion to be
   $M_{\rm max} = 50\;\! {\rm M}_{\odot}$~\citep{Fryer1999ApJ, Heger2003ApJ},  
   the stellar mass cut-off $M_{\rm cut} \approx 1\;\!{\rm M}_{\odot}$ 
   and the minimum mass required for a core-collapse SN event $M_{\rm SN^*} = 8.5\;\!{\rm M}_{\odot}$ 
   \citep[][]{Smith2011MNRAS, Eldridge2013MNRAS}. 
This gives $\alpha_{*} \approx 0.05$,   
  implying a scaling relation $\mathcal{R}_{\rm SF} \approx 160\;\!{\rm M}_{\odot}~{\mathcal R}_{\rm SN}$ 
    between the star-formation rate $\mathcal{R}_{\rm SF}$ and the SN event rate $\mathcal{R}_{\rm SN}$~\citep[see also][]{Owen2018MNRAS} . 
In determining $\dot M$, we set $M_{\rm inj} = 2\;\!{\rm M}_{\odot}$.  
    We also use $\overline{M}_{\rm SN} = 10\;\!{\rm M}_{\odot}$
    as a characteristic progenitor mass 
    for core-collapse SNe (not strictly the mean, although we find our results are relatively insensitive to the exact choice of value here).
  
  The energy budget is specified by the energy injection rate by CRs from SN events.
  This relates to our system of equations by the conservation law arising from combining and integrating equation~\ref{eq:energy_eq} and~\ref{eq:cr_evolution} (see e.g.~\citealt{Breitschwerdt1991A&A, Samui2010MNRAS} for details).
  At large radii, it is clear that the constant of integration is $Q = q v_{\infty}^2/2$, the kinetic energy flow of the wind. This is the rate at which energy is taken out of the host system by the outflow. 
  The rate of energy injected per unit volume may be expressed as the sum of that injected thermally and that injected via CRs. The thermal injection rate is given by: 
  \begin{equation}
  \dot{\epsilon}_{\rm th} = \mathcal{Q}\left[\nu \;\! \xi\;\!{\cal{R}}_{\rm SN} {E_{\rm SN}}\right]  \ ,
  \label{eq:inj_therm_sne}
  \end{equation}
  where the fraction of available energy which goes into driving the outflow is encoded by $\nu = 0.1$~\citep{Murray2005ApJ, Samui2010MNRAS} and is also applicable to the CR component.
  The parameter $\mathcal{Q}$ is introduced as the thermalisation efficiency  
   which implicitly accounts for the fraction of SN energy loss  
   that is in radiative cooling and in transforming cool clumps into ionised gas.   
    Observations of nearby systems, e.g. M82 
    \citep{Watson1984ApJ, Chevalier1985Nat, Seaquist1985ApJ, Strickland2000ApJ, 
     Heckman2017arXiv} 
     suggest that both ${\mathcal Q} \sim 0.1- 1$ and ${\mathcal P} \sim 0.1 - 1$.     
However, these values are not well constrained and    
conflicting values are assigned for the same system in come cases  
      \citep[cf.][]{Bradamante1998A&A,  Strickland2000AJ, 
     Strickland2000ApJ, Veilleux2008ASSP, Zhang2014ApJ}.   
   We thus consider a conservative benchmark model of $\mathcal{Q} = 0.01$ for our calculations. 
   The other parameter, $\xi$, 
   is the fraction of the mechanical SN energy available
   in the presence of energy losses by neutrino emission.  
For core-collapse SNe, 
   around 99\% of the SN energy is carried away by streaming neutrinos 
    \citep[see][]{Iwamoto2006AIPC, Smartt2009ARAA, Janka2012ARNPS}, 
   and hence $\xi = 0.01$.  
 The energy injection rate via CRs is given as:
   \begin{equation}
  \dot{\epsilon}_{\rm CR} = \zeta \frac{\Omega_{\rm A}}{4\pi}\left[\nu\;\!\xi\;\!{\cal{R}}_{\rm SN} {E_{\rm SN}}\right]  
   \label{eq:inj_cr_sne}
  \end{equation}
  where $\zeta$ is introduced as the fraction of SN energy passed to CR power, which is then available for transfer to the outflow wind and/or hadronic interactions. We adopt a characteristic value of $\zeta=0.1$ for this, which is slightly conservative~\citep[see][]{Fields2001A&A, Strong2010ApJ, Lemoine2012A&A, Caprioli2012JCAP, Morlino2012A&A, Dermer2013A&A, Wang2018MNRAS}. We note that, as CRs are initially radiated isotropically away from the source region (the starburst core of the host system), we must use the solid angle fraction ${\Omega_{\rm A}}/{4\pi}$ between the interfacing outflow regions (i.e. `Zone A' in Fig.~\ref{fig:outflow_model}) and the core to properly account for the fraction of initially streaming CRs which are suitably directed to be able to drive the outflow wind
  (this stems from the distinct two zones of our model, between which CR transfer is taken to be negligible -- see the discussion about our `Two-Zone' approximation in section~\ref{sec:two_zone_heating_rates} for more details and our justification of this approach). For a single outflow cone, this is given by 
 \begin{equation}
\Omega_{\rm A} = 2 \pi \left[1-\cos\left(\frac{\theta}{2}\right) \right] \ .
\label{eq:zone_a_solidangle}
\end{equation}
 Combining equations~\ref{eq:inj_therm_sne} and~\ref{eq:inj_cr_sne} gives the total volumetric injection power by SNe as:
\begin{equation}
   \dot{\epsilon} =  \eta_{\rm SNe} \left[\nu\;\!\xi\;\!{\cal{R}}_{\rm SN} {E_{\rm SN}}\right]  
   = \eta_{\rm SNe} \left[\frac{\nu\;\!\xi\;\!\alpha_{*}\;\!{\cal{R}}_{\rm SF}E_{\rm SN}}{{\overline M}_{\rm SN}}\right] \ ,  
\end{equation}
in terms of SN event rate $\mathcal{R}_{\rm SN}$ or star-formation rate $\mathcal{R}_{\rm SF}$, where we introduce the combined SN efficiency term:
\begin{equation}
\eta_{\rm SNe} = \left(\mathcal{Q}+\zeta \frac{\Omega_{\rm A}}{4\pi}\right) \ .
\end{equation}
   In a CR-driven outflow, some amount of the injected energy from SNe is lost in driving the flow, leaving a fraction $f$ transferred into the wind kinetic energy.
   For the purposes of the volumetric energy injection term into the wind fluid, we may thus use
   \begin{equation}
   Q = f\dot{\epsilon} = f \eta_{\rm SNe}\left[\frac{\nu\;\!\xi\;\!\alpha_{*}\;\!{\cal{R}}_{\rm SF}E_{\rm SN}}{{\overline M}_{\rm SN}}\right] \ ,
   \label{eq:energy_inj_param}
   \end{equation}
  where $f$ typically takes values of a few percent, with the rest of the `driving' energy being lost as the wind climbs out of the gravitational potential of the host galaxy -- see \cite{Samui2010MNRAS} for an analytical expression for $f$ in an NFW profile which is adopted in the outflow model used here.
  
  \subsubsection{Velocity \& Density Profile}
  \label{subsubsec:velocity_density_profile}
  
We may manipulate equations~\eqref{eq:continuity_eq} to~\eqref{eq:cr_evolution} as follows. First, from equations~\eqref{eq:exchange_equation_pressure} and~\eqref{eq:energy_eq}, and specifying that for the flow model $h>h_{\rm inj}$, we may write
\begin{equation}
\rho v \left[v \frac{{\rm d}v}{{\rm d} h} + \frac{\gamma_{\rm g}}{\gamma_{\rm g} -1}\frac{{\rm d}}{{\rm d}h}\left(\frac{P}{\rho}\right)\right] = -\rho v \frac{{\rm d}\Phi}{{\rm d}h} - (v_{\rm A} + v)\frac{{\rm d}P_{\rm C}}{{\rm d}h} \ .
\label{eq:modified_energy_eq}
\end{equation}
Following~\cite{Samui2010MNRAS}, we now multiply equation~\eqref{eq:momentum_eq} by $v$ and subtract from equation~\eqref{eq:modified_energy_eq} above:
\begin{equation}
\frac{{\rm d}P}{{\rm d} h} = \gamma_{\rm g} \frac{P}{\rho} \frac{{\rm d}\rho}{{\rm d}h} - (\gamma_{\rm g} -1) \left(\frac{v_{\rm A}}{v}\right) \frac{{\rm d}P_{\rm C}}{{\rm d}h} \ .
\label{eq:gas_pressure_gradient}
\end{equation}
The Alfv\'{e}n velocity is given by $v_{\rm A} = B(h)/\sqrt{4\pi \rho}$ and $B(h) \;\! h^2$ is conserved throughout the most part of the outflow cone (scaled to give $B_0$ as the interstellar magnetic field at the base of the outflow). Thus, we may differentiate $v_{\rm A}$ to give:
\begin{equation}
\frac{1}{v_{\rm A}}\frac{{\rm d}v_{\rm A}}{{\rm d}h} = -\frac{2h}{h_{\rm B}^2 + h^2} - \frac{1}{2\rho}\frac{{\rm d}\rho}{{\rm d}h} \ .
\label{eq:alfven_diff_temp}
\end{equation}
In the limit where $h_{\rm B}<h$, this simplifies to
\begin{equation}
\frac{1}{v_{\rm A}}\frac{{\rm d}v_{\rm A}}{{\rm d}h} \approx -\frac{2}{h} - \frac{1}{2\rho}\frac{{\rm d}\rho}{{\rm d}h} \ ,
\label{eq:alfven_diff}
\end{equation}
which holds for all $h$ where the second term dominates, and so is a suitable approximation for our purposes.
Furthermore, by differentiating equation~\eqref{eq:continuity_eq},
\begin{equation}
\frac{1}{\rho}\frac{{\rm d}\rho}{{\rm d}h} + \frac{1}{v}\frac{{\rm d}v}{{\rm d}h} + \frac{2}{h} = 0 \ .
\label{eq:density_diff}
\end{equation}
Combining equations~\eqref{eq:cr_evolution} and~\eqref{eq:exchange_equation_pressure} with these results (equations~\ref{eq:alfven_diff} and~\ref{eq:density_diff})  gives
\begin{equation}
\frac{{\rm d}P_{\rm C}}{{\rm d}h} = \frac{\gamma_{\rm C} P_{\rm C}}{\rho}\left(\frac{v+v_{\rm A}/2}{v+v_{\rm A}}\right)\frac{{\rm d}\rho}{{\rm d}h} \ ,
\label{eq:cr_pressure_gradient}
\end{equation}
while the gas pressure $P$ can be determined from this and equation~\eqref{eq:gas_pressure_gradient} as:
\begin{equation}
\frac{{\rm d}P}{{\rm d}h} = \left\{\gamma_{\rm g}P - \gamma_{\rm C}(\gamma_{\rm g}-1)P_{\rm C}\left[\frac{v+v_{\rm A}/2}{v+v_{\rm A}}\right]\left(\frac{v_{\rm A}}{v}\right)\right\}\frac{1}{\rho}\frac{{\rm d}\rho}{{\rm d}h} \ .
\end{equation}
This, together with equation~\eqref{eq:cr_pressure_gradient} can be substituted back into equation~\eqref{eq:momentum_eq} to give
\begin{equation}
\rho v \frac{{\rm d}v}{{\rm d}h} + c_{*}^2\frac{{\rm d}\rho}{{\rm d}h} = -\rho \frac{{\rm d}\Phi}{{\rm d}h}
\label{eq:interim1}
\end{equation}
where $c_{*}$ is introduced as an effective sound speed, defined by
\begin{equation}
c_{*}^2 = \frac{\gamma_{\rm g} P}{\rho} - \frac{\gamma_{\rm C}P_{\rm C}}{\rho} \left\{\left(\frac{v+v_{\rm A}/2}{v+v_{\rm A}}\right)\left(\gamma_{\rm C}-\gamma_{\rm C}(\gamma_{\rm g} -1) \left[\frac{v_{\rm A}}{v}\right]\right) \right\} \ .
\end{equation}
Finally, using equation~\eqref{eq:density_diff} to substitute the density gradient in~\eqref{eq:interim1} allows the velocity gradient to be written as:
\begin{equation}
\frac{{\rm d}v}{{\rm d}h} = \frac{2v}{h}\frac{\left(c_{*}^2 - \frac{h}{2}\frac{{\rm d}\Phi}{{\rm d}h}\right)}{v^2-c_{*}^2} \ .
\label{eq:velocity_gradient}
\end{equation}

\begin{figure}
	\includegraphics[width=\columnwidth]{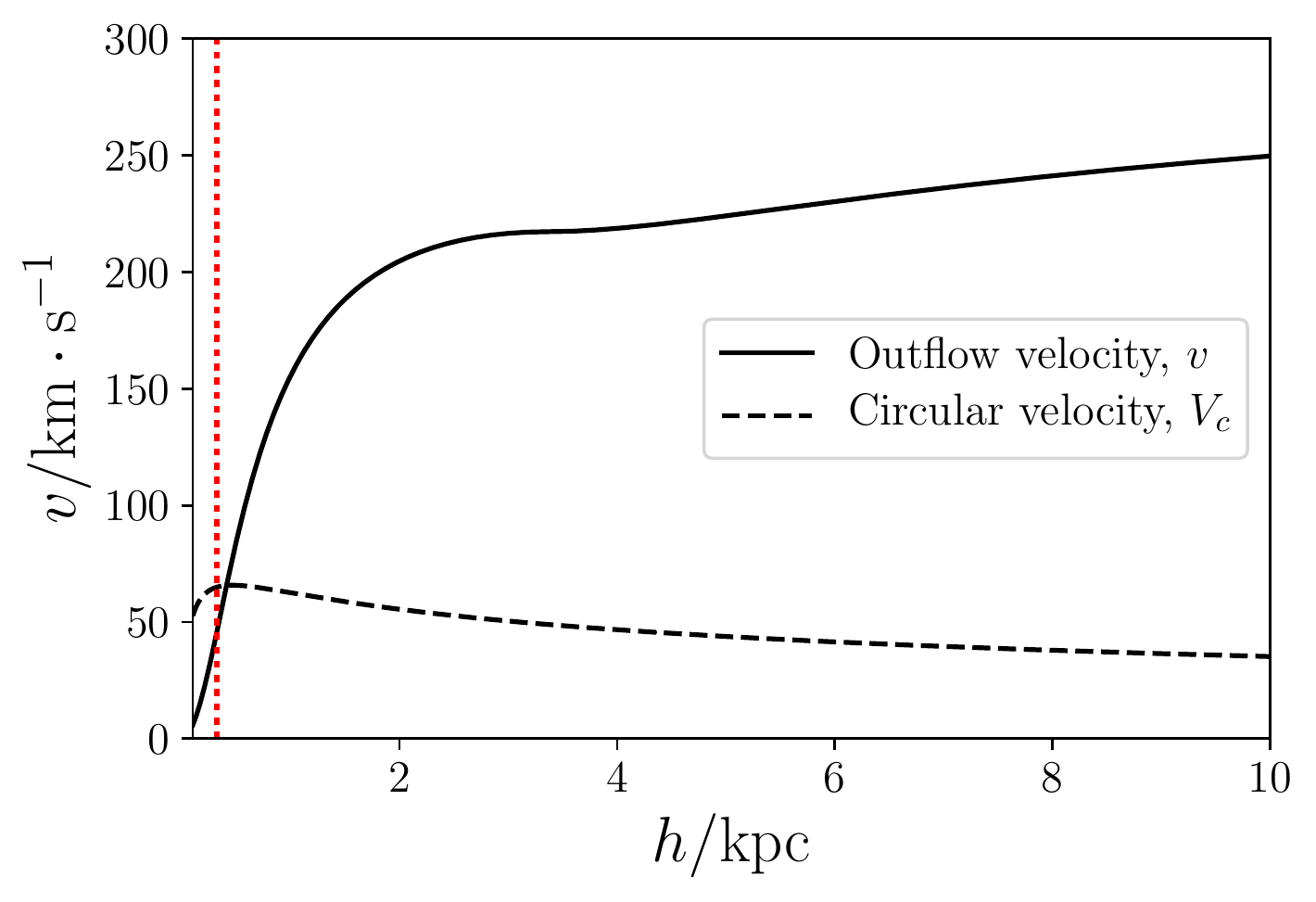}
    \caption{CR-driven outflow velocity profile (solid black line), which approaches a terminal velocity of around $v_{\infty} \approx 290~\text{km}~\text{s}^{-1}$ at large $h$ values at around $50~\text{kpc}$.
    This is calculated for our reference protogalaxy model with
    mass $10^9~\text{M}_{\odot}$ and SN rate of $0.1~\text{yr}^{-1}$. We find a mass outflow rate of 
    $q = 0.01~\text{M}_{\odot}~\text{yr}^{-1}$, 
    energy flux of 
    $\dot{\epsilon} = 3.0\times10^{38}~\text{erg}~\text{s}^{-1}$ per outflow cone
     and critical point location at $h_* = 0.32~\text{kpc}$ (indicated by the red vertical dashed line). The dashed black line shows the NFW profile circular velocity, which is a comparable to the flow velocity at the critical point.}
\label{fig:velocity_profile}
\end{figure}

The location at which the outflow velocity becomes supersonic is referred to as the critical point, $h_*$.
At the critical point, the flow velocity will be equal to the effective sound speed, i.e. $v = c_{*}$, thus the denominator of equation~\eqref{eq:velocity_gradient} will vanish. For a smooth velocity through the critical point (as would be expected physically), we require the numerator to vanish, and it must go to zero more quickly at this point than the denominator to ensure a regular function through this point, i.e.
\begin{equation}
c_{*}^2 -\frac{h}{2}\frac{{\rm d}\Phi}{{\rm d}h} = 0 \ .
\end{equation}
This allows for a useful, alternative estimate for the value of $c_{*}$ (and hence $v$) to be made at the critical point:  the gravitational potential gradient may be expressed in terms of the circular velocity of the system at the critical point $V_{\rm c,*} = \sqrt{{\rm G}M(h_{*})/h_{*}}$, i.e.
\begin{align}
 \label{eq:soundspeed_critical}
c_{*}^2 &= \frac{h}{2}\frac{{\rm d}\Phi}{{\rm d}h} \bigg|_{h_{*}}\\ \nonumber
 & = \frac{{\rm G}M(h_{*})}{2h_{*}} = \frac{F^2 \;\!  V_{\rm c, vir}^2}{2} \ ,
\end{align}
where $M(h_{*})$ is the enclosed mass of the system up to the critical point. As galaxy rotation curves are approximately flat at large radii, the circular velocity at the virial radius would be comparable to that at the critical point. This allows us to use the full mass of the system in place of $M(h_{*})$ to enable easier parametrisation, and means that $V_{\rm c, vir} \approx V_{\rm c,*}$. 
We introduce $F$ in equation~\ref{eq:soundspeed_critical} to account for the small difference between $V_{\rm c, vir}$ and $V_{\rm c,*}$, with $F/\sqrt{2}$ typically being of order 1 for all plausible model parameter choices~\citep[see also][]{Samui2010MNRAS}. 
For our reference model with mass $10^9~\text{M}_{\odot}$ and SN rate of $\mathcal{R}_{\rm SN} = 0.1~\text{yr}^{-1}$, we find a mass outflow rate of $q = 0.01~\text{M}_{\odot}~\text{yr}^{-1}$, energy flux of $\dot{\epsilon} = 3.0\times10^{38}~\text{erg}~\text{s}^{-1}$ per outflow cone, a critical point location at $h_* = 0.32~\text{kpc}$ and a corrective factor of $F = 1.05$.

The flow velocity at the critical point can then be used as a boundary condition from which equation~\ref{eq:velocity_gradient} can be integrated.
We adopt a numerical approach to do this, using a 4th order Runge-Kutta method~\citep[][]{Press2007book} to integrate both inwards and outwards from the critical point. 
To ensure a smooth solution over the critical point, we enforce a linear gradient across it locally using the method specified in~\cite{Ipavich1975ApJ}. 
Fig~\ref{fig:velocity_profile} shows the resulting solution for our reference protogalaxy model (solid black line). This shows that the wind tends towards a terminal velocity of around $v_{\infty} \approx 290~\text{km}~\text{s}^{-1}$, and demonstrates the comparability between the flow velocity and circular velocity around the critical point.

\begin{figure}
	\includegraphics[width=\columnwidth]{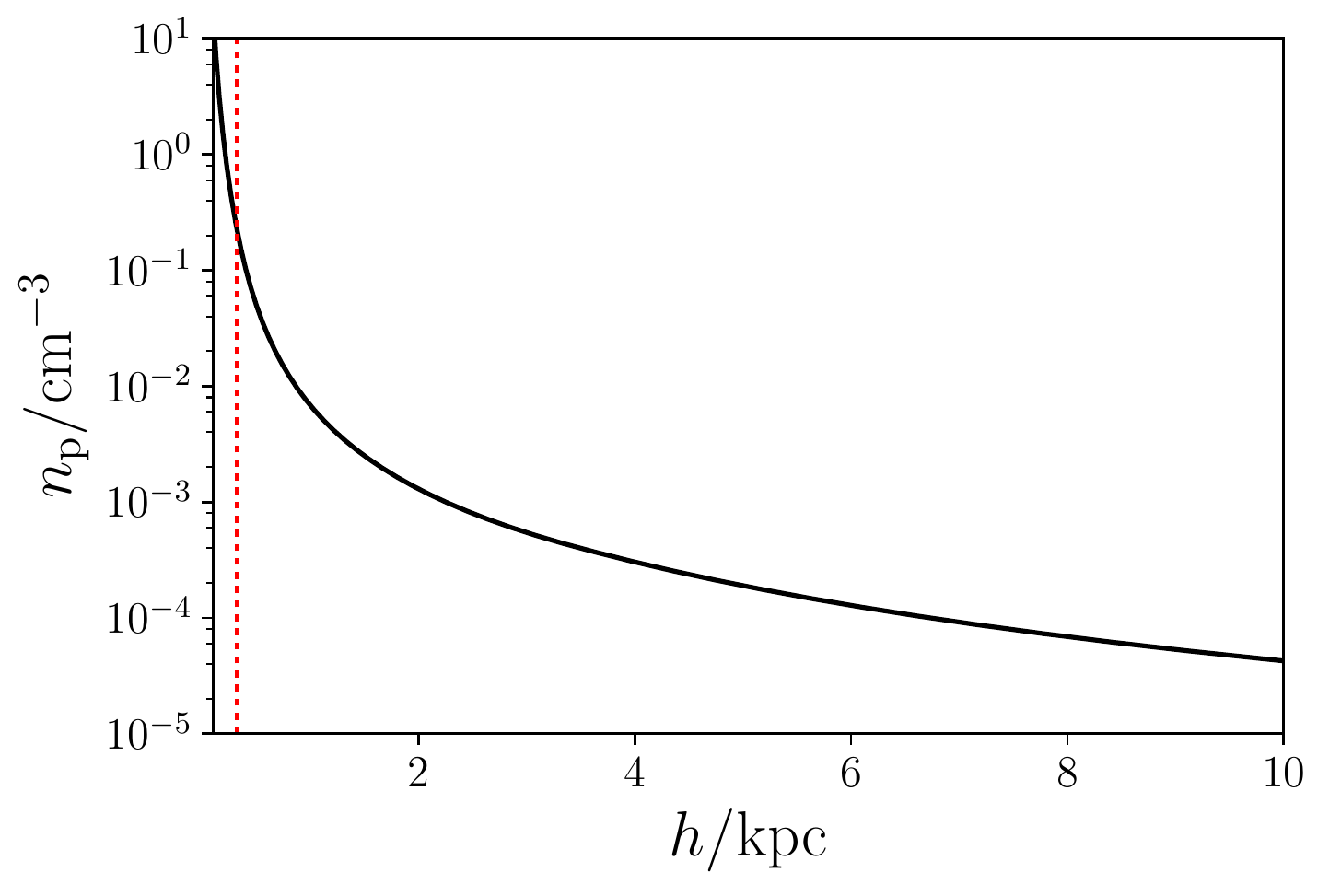}
    \caption{CR-driven outflow density profile (solid black line), calculated for our reference protogalaxy model with
    mass $10^9~\text{M}_{\odot}$ and SN rate of $\mathcal{R}_{\rm SN} = 0.1~\text{yr}^{-1}$. The critical point location is at $h_* = 0.32~\text{kpc}$ (indicated by the red vertical dashed line).
    This gives an ISM density (within the protogalaxy) of around $10~\text{cm}^{-3}$, and a temperature of around 10$^5$ K at the critical point.}
\label{fig:density_profile}
\end{figure}

An associated density profile can also be found numerically from equation~\ref{eq:density_diff}. 
This is an important component of the model because, in section~\ref{sec:absorption}, we will show that the local density of a medium determines the level of CR heating that can arise via hadronic interactions.
Fig.~\ref{fig:density_profile} shows the resulting density profile of the outflow when adopting the same reference model parameters used for the velocity profile.
This corresponds to an ISM density (within the protogalaxy) of around $10~\text{cm}^{-3}$, and a temperature of around 10$^5$ K at the critical point. 
CR and gas pressure profiles can be similarly calculated, but are not important for the analysis in this paper.

\section{Interactions of Energetic Cosmic Ray Particles in Protogalactic Outflows}
\label{sec:particles_in_outflows}  

\subsection{Cosmic Ray Interactions}

\subsubsection{Hadronic Processes}
\label{sec:absorption}

   Proton-proton (${\rm pp}$) interactions of CRs are expected to dominate over photo-hadronic interactions  
   at ${\rm GeV}$ energies and above in most galactic and protogalactic systems~\citep[e.g.][]{Mannheim1994A&A, Owen2018MNRAS}. 
These ${\rm pp}$-interactions produce a shower of secondary particles  
   which include hadrons, charged and neutral pions, leptons and neutrinos
   \citep[see][]{Pollack1963PhRv, Gould1965AnAp, Stecker1968ApJ, Almeida1968PR, Skorodko2008EPJA, Dermer2009book}. 
The energies carried by the energetic protons (CRs) will be distributed among their descendant particles 
   and through their subsequent interactions and decays. 
   In particular, through the transfer of energy to the secondary charged pion particles and then to leptons (mainly electrons and positrons), 
   the primary CR proton can deposit a fraction of its energy into the ambient medium. 

The major channels of the ${\rm pp}$-interaction are  
\begin{align}\label{eq:pp_interaction}%
\rm{p} + \rm{p} \rightarrow %
	\begin{cases}%
		&\rm{p}  \Delta^{+~\;} \rightarrow\begin{cases}%
				\rm{p} \rm{p} \pi^{0}  \xi_{0}(\pi^{0}) \xi_{\pm}(\pi^{+} \pi^{-}) \\[0.5ex]%
				\rm{p} \rm{p}  \pi^{+}  \pi^{-}  \xi_{0}(\pi^{0}) \xi_{\pm}(\pi^{+} \pi^{-}) \\[0.5ex]%
				\rm{p} \rm{n}  \pi^{+}  \xi_{0}(\pi^{0}) \xi_{\pm}(\pi^{+} \pi^{-})\\[0.5ex]%
			\end{cases} \\%
		&\rm{n} \Delta^{++} \rightarrow\begin{cases}%
				\rm{n} \rm{p} \pi^{+} \xi_{0}(\pi^{0}) \xi_{\pm}(\pi^{+} \pi^{-}) \\[0.5ex]%
				\rm{n} \rm{n} 2\pi^{+} \xi_{0}(\pi^{0}) \xi_{\pm}(\pi^{+} \pi^{-})\\[0.5ex]%
			\end{cases} \\%
	\end{cases} \ ,%
\end{align}%
     where $\xi_{0}$ and $\xi_{\pm}$ are the multiplicities of the neutral and charged pions respectively 
     while the $\Delta^{+}$ and $\Delta^{++}$ baryons are the resonances 
     \citep[see][]{Almeida1968PR, Skorodko2008EPJA}. 
The hadronic products will continue their interaction processes 
   until their energies fall below the interaction threshold $E_{\rm p}^{\rm th} = 0.28~{\rm GeV}$ 
   \citep{Kafexhiu2014PRD}\footnote{This threshold is determined from the energy required for the production of a pair of neutral pions, 
    being the lowest energy particle produced in the cascade, 
    where $E_{\rm p}^{\rm th} = 0.28 ~{\rm GeV} = 2m_{\pi^0} + {m_{\pi^0}}^2/2m_{\rm p}$, 
    for $m_{\pi^0}$ as the neutral pion rest mass and $m_{\rm p}$ as the proton rest mass.}, although 
    it is expected that this would arise after just a few interaction events \citep[see][]{Owen2018MNRAS}. 
   
The pionic products undergo decays, where most of the neutral pions ${\pi^0}$ will decay into two photons through an electromagnetic process,    
\begin{align}%
\label{eq:process1}
	\pi^0	&\rightarrow 2\gamma  \ ,   %
\end{align}%
  with a branching ratio of 98.8\%~\citep{Patrignani2016ChPh} 
  and on a timescale of $8.5 \times 10^{-17}\;\!{\rm s}$.  
Charged pions $\pi^\pm$ will produce leptons and neutrinos via a weak interaction,   
 \begin{align}%
 \label{eq:process2}
	\pi^+	&\rightarrow \upmu^+ \nu_{\rm \upmu} \rightarrow \rm{e}^+ \nu_{\rm e} \bar{\nu}_{\rm \upmu} \nu_{\rm \upmu}\     \nonumber \\%
	\pi^-	&\rightarrow \upmu^- \bar{\nu}_{\rm \upmu} \rightarrow \rm{e}^- \bar{\nu}_{\rm e} \nu_{\rm \upmu} \bar{\nu}_{\rm \upmu}       %
       \ , 
\end{align}%
 with a branching ratio of 99.9\%~\citep{Patrignani2016ChPh}  
 and on timescales of roughly $2.6\times 10^{-8}\;\!{\rm s}$.  
 
The $\gamma$-ray photons from the $\pi^0$ decay have minimal interactions with the surrounding medium  
   and they effectively stream away from their production site. 
The neutrinos from the $\pi^\pm$ decay also interact minimally with the medium.  
While the $\gamma$-ray photons and neutrinos allow for a net energy escape from the interaction region,  
   the leptons, which interact strongly with the ionised and magnetised ISM and/or outflow wind medium,  
   play a key role in mediating the energy transfer process.   
Although some of the energy carried by the charged leptons is lost through inverse Compton and synchrotron processes, 
  a non-negligible fraction can still be passed to the ISM 
  in lepton-hadron coulomb scattering and collisions.   

The energy transferred to the pions can be estimated from their production cross-sections. 
The parameterisation of the pion-production cross-sections proposed by \citet{BlattnigPRD2000} 
   gives a reasonable fit to the data~\citep[with only a minor discrepancy below 50-GeV, see][]{Owen2018MNRAS}, 
   when accounting for all pion-production branches.  
The ratios of the primary energy then passed to the different secondary species $\{\pi^{+}, \pi^{-}, \pi^{0}\}$ 
  follows as $\{0.6,\;\! 0.1,\;\! 0.3\}$ at 1-GeV 
  while this tends towards around $\{0.3,\;\! 0.4,\;\! 0.3\}$ at higher energies.   
Thus, the total fraction of CR primary energy passed to charged pion production is around 0.7, of which around 0.1 is lost to neutrinos~\citep{Dermer2009book}.  
On their decay to secondary electrons (and positrons - hereafter, we refer to all of the charged lepton secondaries as electrons for simplicity, without losing generality) and neutrinos, 
   around 75\% of the pion energy is passed to the neutrinos, while the electrons adopt around 25\%~\citep[see, e.g.][]{Aharonian2000A&A, Loeb2006JCAP, Dermer2009book, Lacki2012AIPC, Lacki2013MNRAS}. Overall gives the fraction of the CR primary energy ultimately passed to secondary electrons as around 0.15. This is equally split among each of the electrons produced. With a multiplicity of around 4 at GeV energies (see~\citealt{Albini1976NCimA, Fiete2010JPhG} for a fitted parametrisation of multiplicity data) which dominate the CR spectrum, this gives a typical secondary electron energy $E_{\rm e}$ of around 3.75\% of that of the CR primary proton energy, $E$. Thus, a GeV CR proton undergoing a ${\rm pp}$ interaction would be expected to inject approximately 4 secondary electrons, each with an energy of around 40 MeV. We introduce the parameters $f_{\rm e} = 0.0375$ as the fraction of primary energy passed to electrons, and $\xi_{\rm e}$ as their multiplicity. Both are technically functions of the primary proton energy, $E$ but we set them to be constant at their dominating $E = 1~\text{GeV}$ value for our calculations here with no discernible impact on our results.
 
 \subsubsection{Leptonic Processes \& Thermalisation}
 \label{sec:thermalisation}
 
In protogalactic environments, 
  the three main processes by which 
  the secondary electrons release their energies are radiative cooling 
  (via inverse Compton scattering with cosmological microwave background, CMB, photons and 
  starlight photons, and/or via synchrotron emission when interacting with the ambient magnetic field), 
  free-free cooling (mainly due to the electron-proton bremsstrahlung processes),  
  and Coulomb collisions in the ISM.   
In the high-redshift galactic environments considered here, 
  radiative losses are mainly caused by inverse-Compton scattering with the CMB and possibly starlight, if the host galaxy is able to sustain high star-formation rates (i.e. sufficient to yield a SN event rate of above $0.1~\text{yr}^{-1}$ -- see~\citealt{Owen2018MNRAS}). 
  Such loses arise at a rate of 
\begin{equation}
   \dot{E}_{\rm rad} = \frac{4}{3}\sigma_{\rm T} {\rm c} \left(\frac{E_{\rm e}}{m_{\rm e} {\rm c}^2}\right)^2 U_{\rm i}
\end{equation} 
   per particle \citep[e.g.][]{Rybicki1986book, Blumenthal1970PRD}, 
   where ${\rm c}$ is the speed of light, $\sigma_{\rm T}$ is the Thomson scattering cross-section, $E_{\rm e}$ is the electron energy 
    and $U_{\rm i}$ is the energy density in the radiation field (or magnetic field in the case of Synchrotron losses).
   The rate of free-free (bremsstrahlung) cooling per particle is 
\begin{equation}
   \dot{E}_{\rm ff} \approx \alpha_{\rm f} {\rm c} \sigma_{\rm T} n_{\rm p} E_{\rm e} \,
\end{equation}
   where $\alpha_{\rm f}$ is the fine structure constant,  
  and the energy loss of the electrons due to Coulomb interactions in the ionised ISM is 
\begin{equation}
  \dot{E}_{\rm C} \approx m_{\rm e} {\rm c}^2 n_{\rm p} {\rm c} \;\! \sigma_{\rm T} \ln\Lambda \ ,
\end{equation}
  where $\ln\Lambda\simeq 30$ is the Coulomb logarithm~\citep[see][]{Dermer2009book, Schleicher2013A&A}. 
  Further losses arises due to adiabatic expansion of the CR fluid as it propagates along an outflow. This is quantified by
  \begin{equation}
  \dot{E}_{\rm ad} = \frac{2}{3}\frac{1}{h^2}\frac{\partial}{\partial h}\left[ h^2 v(h) \right] E_{\rm e}
  \end{equation}
\citep{Longair2011book}, and applies equally to protons and electrons.
Hence, the fraction of energy carried by the CR electrons 
   that could be deposited into the ISM is simply 
\begin{equation}
\label{eq:efficiency_equation}
  f_{\rm C}(E_{\rm e}) = \frac{\;{\tau_{\rm C}}^{-1}}{{\tau_{\rm C}}^{-1} + {\tau_{\rm rad}}^{-1} + {\tau_{\rm ff}}^{-1} + {\tau_{\rm ad}}^{-1}}\;\!\Bigg\vert_{E_{\rm e}}\ ,
\end{equation}
  where $\tau_{\rm C}$, $\tau_{\rm rad}$, $\tau_{\rm ff}$ and $\tau_{\rm ad}$ 
  are timescales of Coulomb, radiative, free-free (bremsstrahlung) and adiabatic losses respectively.
Overall, we can account for the branching ratios and cooling processes by introducing the term 
  $f_{\rm therm}(E) = f_{\rm C}|_{E_{\rm e}}  \left( \xi_{\rm e} \;\!  f_{\rm e}\right)|_{E}$ into our calculations, 
  which allows us to properly estimate the fraction of CR primary energy deposited that is ultimately thermalised.
   
   This thermalisation of the CR electron secondaries does not occur immediately. \citet{Owen2018MNRAS} shows the time-scale over which this arises can be estimated as
   \begin{equation}
   \tau_{\rm th} \approx 0.39 ~\left(\frac{E_{\rm e}}{40~\text{MeV}}\right)~\left(\frac{n_{\rm p}}{10~\text{cm}^{-3}}\right)^{-1}~\text{Myr} \ ,
   \end{equation}
   which is shorter then the dynamical timescales estimated for the system in both the advective and diffusive CR transport regimes by at least an order of magnitude (as shown in Fig.~\ref{fig:timescales}) and so, for our purposes, we may assume that the thermalisation of the CR secondary electrons occurs rapidly and in the vicinity of the initial ${\rm pp}$ interaction. In line with this approximation, the rate at which energy is deposited into the ambient medium per unit volume is 
\begin{equation}
\label{eq:e_deposition}
     \dot{\epsilon}_{\rm p\pi} = {\rm c}\;\! E ~n(E) ~\hat{\sigma}_{\rm p\pi}(E)\;\!n_{\rm p}\;\!f_{\rm therm}(E) =  \frac{{\rm c} \;\! E ~n(E)\;\!f_{\rm therm}(E)}{l^*_{\rm p \pi}(E)}\ ,
\end{equation}
    where 
    $n_{\rm p}$ is the local number density of protons in the wind fluid, 
    $E$ and $n(E) = E \; {{\rm d} N(E)}/{{\rm d}E \; {\rm d}V}$  are the energy and differential number density of CR protons respectively,  
    and ${l^*_{\rm p \pi}} ( = 1/\hat{\sigma}_{\rm p\pi} n_{\rm p} ) $ is the mean-free-path of the interaction. 
The total inelastic cross-section of the ${\rm pp}$ interaction can be parametrised as 
\begin{equation}%
\label{eq:pp_cs}%
   \hat{\sigma}_{\rm p\pi} = \left( 30.7 - 0.96\ln(\chi) + 0.18(\ln\chi)^{2} \right)\left( 1 - \chi^{-1.9}   \right)^{3}~\rm{mb} 
\end{equation}%
   \citep{Kafexhiu2014PRD},
   where $\chi =  E/E^{\rm{th}}_{\rm p}$ and $E^{\rm{th}}_{\rm p}$ is the threshold energy, as introduced above. 
It follows that the rate of CR attenuation is 
\begin{equation}
\label{eq:cr_absorption}
   \frac{\rm d}{{\rm d}\;\! t} {n}(E) ~\bigg\vert_{\rm p\pi} = - \left[ {\rm c}  ~\hat{\sigma}_{\rm p\pi}(E)~n_{\rm p} \right] ~n(E) 
      =  - \left[ \frac{\rm c}{l^*_{\rm p \pi}(E)} \right] ~n(E)  \ ,
\end{equation}
  and the corresponding heating rate of the medium is 
    \begin{equation}
\label{eq:cr_heating_equation} 
   H({\mathbf x}) = \left\{ {\rm c} ~n_{\rm p}\int_{E_{0}}^{E_{\rm max}} {\rm d}E ~n(E) \;\!f_{\rm therm}(E) ~\hat{\sigma}_{\rm p\pi}(E) ~\right\}\Bigg\vert_{~\mathbf x} \ .
\end{equation}
The energy limits in the integral above will be discussed in \S \ref{sec:cr_spectrum}. 
    
\subsection{Cosmic Ray Energy Spectrum}
\label{sec:cr_spectrum} 

\subsubsection{Transportation \& Spectral Evolution}

The transport of CRs in a bulk flow is governed by 
\begin{align}
      \frac{\partial n}{\partial t}  = \  &  \nabla \cdot \left[D(E)\nabla n \right] - \nabla \cdot \left[{\mathbf{v}} n\right]  
        + \frac{\partial}{\partial E} \left[ \;\! b(E, {\mathbf{x}}) n \;\! \right]  \nonumber \\
     &  \hspace*{0.2cm} + Q(E, {\mathbf{x}}) - S(E, {\mathbf{x}}) \ , 
\label{eq:transport_equation}     
\end{align}
   \citep[e.g.][]{Schlickeiser2002book} where $n = n(E, {\bf{x}})$ is the differential number density of CR protons (i.e. the number density of CR particles per unit energy) 
   with an energy $E$ at a location ${\mathbf{x}}$.  
 The $\nabla \cdot \left[D(E)\nabla n \right]$ term describes the diffusion process,    
      specified by the diffusion coefficient $D(E)$ (see \S \ref{sec:diff_coeff}), while  
 the $\nabla \cdot \left[{\bf{v}} n\right]$ term describes the advection of CRs   
   in a bulk flow of velocity ${\mathbf{v}}$ (see \S \ref{subsec:wind_velocity}). 
The mechanical and radiative cooling of the CR particles is specified by the energy loss rate, $b(E, {\bf{x}})$,          
  the injection of CRs by the source term, $Q(E,{\mathbf x})$, 
  and the attenuation of CRs by the sink term, $S(E,{\mathbf x})$. 
The radiative loss timescale of protons is generally longer than 
  that of advection and diffusion 
  in typical galactic environments. 
In some galactic-scale outflows, adiabatic cooling could be important, so we retain this in our calculations
and simply set $b(E,{\mathbf x}) = \dot{E}_{\rm ad}$ in this first study. However a thorough investigation of the adiabatic cooling effect of CRs and the competition between advection and diffusion in different astrophysical settings
   deserves a separate investigation. 

We have adopted a scenario whereby CRs are produced in SN events, and  
   these are expected to be most frequent in active star-forming regions at the base of outflows.
As such, 
  we consider a situation where CR protons are injected at the base of the outflow wind cone 
  as an initial boundary condition for the solution of the CR transport equation. 
Since our current knowledge of the properties of CRs in protogalactic environments is very limited, 
  the initial energy spectrum of the CRs is uncertain. 
Given that the acceleration of CRs is a universal process, 
  e.g. via \cite{Fermi1949PhRv} processes in SN remnants, 
  we boldly assume 
  a differential energy spectrum of the freshly injected CRs similar to that observed in the Milky Way, 
  i.e. following a power-law 
\begin{equation}
\label{eq:ref_cr_eqn}
   \frac{{\rm d}\Phi(E)}{{\rm d}E \;\! {\rm d}\Omega} =  \mathcal{N}(E_0) \left(\frac{E}{E_0}\right)^{-\Gamma} \ .
\end{equation}
Here, $\Omega$ is introduced as the solid opening angle (of the outflow cone) and we adopt a power-law index of $\Gamma = 2.1$, in line with Milky Way observations of the galactic ridge -- 
   a region where abundant CR injection is likely to occur, and therefore is a reflection of the `fresh' CR spectrum as required here
   (see, e.g. \citealt[][]{Allard2007APh, Kotera2010JCAP, Kotera2011ARAA}, although slightly steeper indices of around 2.3--2.4 have been suggested in recent years for pure proton data in `fresh' acceleration regions, e.g.~\citealt{Adrian-Martinez2016PhLB, HESSA&A2018}).
$\mathcal{N}(...)$ is the normalisation, given by 
\begin{equation}
\label{eq:cr_norm}
     \mathcal{N}(E) = \frac{{\rm d}\Phi ({E})}{{\rm d}E \;\! {\rm d}\Omega} \;\! \bigg\vert_{\rm base} 
       = \frac{(1-\Gamma)E_0^{-\Gamma}}{E_{\rm max}^{1-\Gamma} - E_0^{1-\Gamma}}  
         ~\frac{v_{\rm CR} \; \epsilon_{\rm CR}}{E_0} \ ,
\end{equation}
with $E_{\rm max}$ as the maximum energy of interest, and 
   $E_0$ (the lowest energy under consideration) as the reference energy.
   We adopt a minimum energy bound of $E_0 = 1 \;\! {\rm GeV}$ 
   which corresponds to the approximate energy above which hadronic interactions may arise (via the ${\rm pp}$ mechanism -- see \citealt{Kafexhiu2014PRD}), 
   and a maximum energy of $E_{\rm max} = 10^{6} \;\! {\rm GeV}$ (i.e. 1 PeV) being the maximum realistic energy that could be reached by CRs 
   accelerated in SN remnants~\citep{Bell1978MNRAS, Kotera2011ARAA, Schure2013MNRAS, Bell2013}, while higher-energy particles would likely originate from outside the protogalaxy~\citep{Hillas1984ARAA, Becker2008PhR, Kotera2011ARAA, Blasi2014CRPhy}.
   With a power-law index of $\Gamma = 2.1$,
  the  $1 - 10^{6}~{\rm GeV}$ range harbours more than 99\% of the total energy content of the CRs 
   \citep[see][]{Benhanbiles-Mezhoud2013ApJ}. 
    $\epsilon_{\rm CR}$ is the CR energy density -- see section~\ref{sec:energy_densities} for details, and $v_{\rm CR}$ is the characteristic velocity with which the CRs propagate macroscopically. In the case of free-streaming CRs, this is the speed of light, ${\rm c}$. In a diffusion-dominated system, $v_{\rm CR}$ would be the diffusive speed $D(E)/\ell_{\rm diff}$ while, in an advection-dominated scenario in which CRs are trapped in a local magnetic field, but transferred along by the flow of the fluid in which they are entrained, this would be the bulk flow velocity\footnote{The microscopic CR propagation speed would remain as ${\rm c}$ in all cases, however in the diffusion and advection scenarios their macroscopic propagation appears to be much less due to the small-scale deflections and scatterings with the local magnetic field, such that their propagation can no longer be approximated as streaming.}.
    The differential CR flux can be used to write 
   the CR differential number density as
   \begin{equation}
\label{eq:differential_number_density}
 n(E) =  \frac{E \; {\rm d} N(E)}{{\rm d}E \; {\rm d}V} = \frac{\Omega \; E}{v_{\rm CR}} \frac{{\rm d} \Phi(E)}{{\rm d}E}  \ ,
\end{equation}
in line with the earlier definition in section~\ref{sec:absorption}.
  
   \subsubsection{Cosmic Ray Energy Densities}
   \label{sec:energy_densities}
   
      The CR energy density depends on whether the system is dominated by advection or diffusion, and is governed by the outflow velocity (in comparison to the diffusive speed).
   In an advection-dominated system, the CR energy density may be expressed as 
   \begin{equation}
\epsilon_{\rm CR, adv} = \frac{L_{\rm CR, eff}}{4\pi \ell_{\rm adv}^2~{v(h)}}\ ,
\label{eq:energy_density_adv}
\end{equation}
where we may approximate $v(h)$ with $v_{\infty}$, the terminal velocity of the outflow, for the purpose of modelling its large-scale redistribution of CR energy.
In a diffusion-dominated system, it is given by
\begin{equation}
\epsilon_{\rm CR, diff} = \frac{L_{\rm CR, eff}}{4\pi \ell_{\rm diff}~D(E)} \ .
\label{eq:energy_density_diff}
\end{equation}
  where $\ell_{\rm adv, diff}$ are the characteristic length-scales of the system when dominated by advection or diffusion.

Here, the power of the CRs, $L_{\rm CR, eff}$, is related to the power of the SN explosions injecting them into the system via  
\begin{equation}
\label{eq:cr_lum_orig}
    L_{\rm CR, eff} = (1-f) \;\! \zeta \left[{\cal{R}}_{\rm SN} { \xi  E_{\rm SN}}\right]  
     = (1-f) \;\! \zeta \left[\frac{ \xi \alpha_{*} E_{\rm SN} {\cal{R}}_{\rm SF}}{\rm{M}_{\rm SN}}\right] \ ,
\end{equation}
   where factor $f$ was first introduced in equation~\ref{eq:energy_inj_param} and accounts for the fraction of energy lost by the CRs in climbing out of the gravitational potential of the host galaxy (such that a fraction $1-f$ is retained by the CRs and so is available to undergo hadronic interactions). 
   The other symbols retain their earlier definitions (see \S \ref{subsec:wind_velocity}). When accounting for the flow solid angle, the factor $\Omega_{\rm A}/4\pi$ which appeared in equation~\ref{eq:inj_cr_sne}, is also required.

In a galaxy harbouring CRs with limited bulk flows and advection, particles diffuse throughout the volume of the host on kpc scales~\citep[see, e.g.][]{Owen2018MNRAS}. As such, we adopt $\ell_{\rm diff} = 1~\text{kpc}$ as the characteristic diffusion length-scale of the system when particle transport is well-within the diffusive regime. Conversely, if transport is dominated by advection, advective outflows extend for tens of kpc~\citep[see][]{Veilleux2005ARAA, Bland-Hawthorn2007APSS, 
     Bordoloi2011ApJ, Martin2013ApJ, Rubin2014ApJ, Bordoloi2016MNRAS}. As such, we adopt an advection length-scale for the propagating CRs of $\ell_{\rm adv} = 10~\text{kpc}$. In the case of an outflow system with a SN rate of ${\cal R}_{\rm SN} = 0.1~\text{yr}^{-1}$, outflow wind velocity of $v_{\infty} \approx 290~\text{km}~\text{s}^{-1}$ (see \S~\ref{subsec:wind_velocity}) and a diffusion coefficient of $3.0\times 10^{28}~\text{cm}^2~\text{s}^{-1}$ (appropriate for a 1-GeV CR in a 5-$\mu$G ambient mean magnetic field - see \S~\ref{sec:diff_coeff}), equation~\ref{eq:energy_density_diff} and~\ref{eq:energy_density_adv} would suggest associated energy densities of $\epsilon_{\rm diff} \approx 170.0~\text{eV}~\text{cm}^{-3}$ and $\epsilon_{\rm adv} \approx 0.59~\text{eV}~\text{cm}^{-3}$ in the diffusive and advective regimes respectively, i.e. the advection of CRs reduces their energy density by almost 2 orders of magnitude at the base of the outflow cone.
     
     These energy densities are largely consistent with the CR energy densities estimated from, e.g. M82, $\epsilon_{\rm CR} \approx 550~\text{eV}~\text{cm}^{-3}$, and NGC 253, $\epsilon_{\rm CR} \approx 260-350~\text{eV}~\text{cm}^{-3}$~\citep[][]{Yoast-Hull2015MNRAS}, starburst galaxies with similar SN rates to that adopted in the present model of $\mathcal{R}_{\rm SN} \approx 0.1~\text{yr}^{-1}$~\citep{Lenc2006AJ, Fenech2010MNRAS}. While M82 hosts a clear outflow, its CR energy density would suggest that CR propagation is still diffusion-dominated in the system overall. NGC 253 also appears to be predominantly diffusive throughout much of the galaxy, with only advective transport dominating above a height of around 2~kpc~\citep[][]{Heesen2007AN}, being consistent with the relatively high measured CR energy density. 
     
     Advection-dominated outflow systems would have considerably more rapid outflow velocities compared to their diffusion-dominated counterparts (to ensure that the advecting flow has a faster velocity than the diffusing CRs). A clear example of a starburst with a rapid outflow is NGC 3079, and CR propagation in this system would therefore be expected to be predominantly advective. This galaxy is known to harbour a remarkably fast outflow wind, of central velocity of around $1,100~\text{km}~\text{s}^{-1}$ but perhaps rising to nearly $3,000~\text{km}~\text{s}^{-1}$ in some regions~\citep{Filippenko1992AJ, Veilleux1994ApJ, Veilleux1994BCFHT, Veilleux1999AJ}\footnote{While NGC 3079 also hosts an active nucleus (AGN), analysis by~\citet{Cecil2001ApJ} has shown that the outflow wind is driven by the nuclear starburst rather than by the AGN, and so is a valid comparison here.}. Radio observations of NGC 3079 indicate average CR energy densities of around $8.0~\text{eV}~\text{cm}^{-3}$, with only a small variation throughout the host~\citep{Irwin2003MNRAS}. Given that the SN rate $\mathcal{R}_{\rm SN} \approx 0.3-0.5~\text{yr}^{-1}$~\citep{Irwin1988ApJ, Condon1992ARA&A, Irwin2003MNRAS}, if the system were fully diffusive, we would expect the CR energy density to be around $1,000 - 2, 500~\text{eV}~\text{cm}^{-3}$, i.e. 3-5 times that of M82 or NGC 253.  Instead, this estimated value is around 100 times less than the diffusive limit prediction and is therefore consistent with the energy density predicted by the advection-limit estimation.

\subsubsection{Cosmic Ray and $\gamma$-Ray Spectral Comparisons}
  \label{sec:cr_gamma_observations}

  We may compare our CR injection spectral model defined in equations~\ref{eq:ref_cr_eqn} and~\ref{eq:cr_norm} with $\gamma$-ray observations of the Galactic Ridge (GR) -- a region of abundant gas clouds and star-formation which is likely to be a useful tracer of CR interactions and their underlying spectrum from the resulting secondary $\pi^0$ decays. We model the expected CR spectral energy density in protogalaxies of characteristic SN rate $\mathcal{R}_{\rm SN} = 0.01, 0.1, 1.0$ and $10.0~\text{yr}^{-1}$, along with our model prediction for the Milky Way with $\mathcal{R}_{\rm SN} \approx 0.015~\text{yr}^{-1}$~\citep[e.g.][]{Dragicevich1999MNRAS, Diehl2006Nat, Hakobyan2011Ap, Adams2013ApJ} using equation~\ref{eq:ref_cr_eqn} and the diffusive CR energy density given by equation~\ref{eq:energy_density_diff}. We acknowledge that the Milky Way case differs from the protogalaxy models in that the SN types are more likely to be dominated by those resulting from lower mass stars with longer lifetimes than in a starburst protogalaxy. Such SN events have a lower characteristic energy of around $E_{\rm SN} \approx 10^{51}~\text{erg}$ (instead of the $E_{\rm SN} = 10^{53}~\text{erg}$ appropriate for core-collapse SNe with massive progenitors) with less energy loss to neutrinos -- $\xi$ is taken to be 0.9 for the Milky Way model~\citep[see, e.g. models and simulations in][which suggest neutrino losses are of around a few percent of the total Type 1a SN energy]{Wright2017PhRvD}, rather than the $\xi = 0.01$ value appropriate for Type II core-collapse SNe~\citep[e.g.][]{Iwamoto2006AIPC, Smartt2009ARAA, Janka2012ARNPS}. The size of the system is also different, with $\ell_{\rm diff} \approx 30~\text{kpc}$ for the Milky Way~\citep[see, e.g.][]{Xu02015ApJ}, compared to $\ell_{\rm diff} = 1~\text{kpc}$ adopted in our protogalaxy models.
  
\begin{figure}
	\includegraphics[width=\columnwidth]{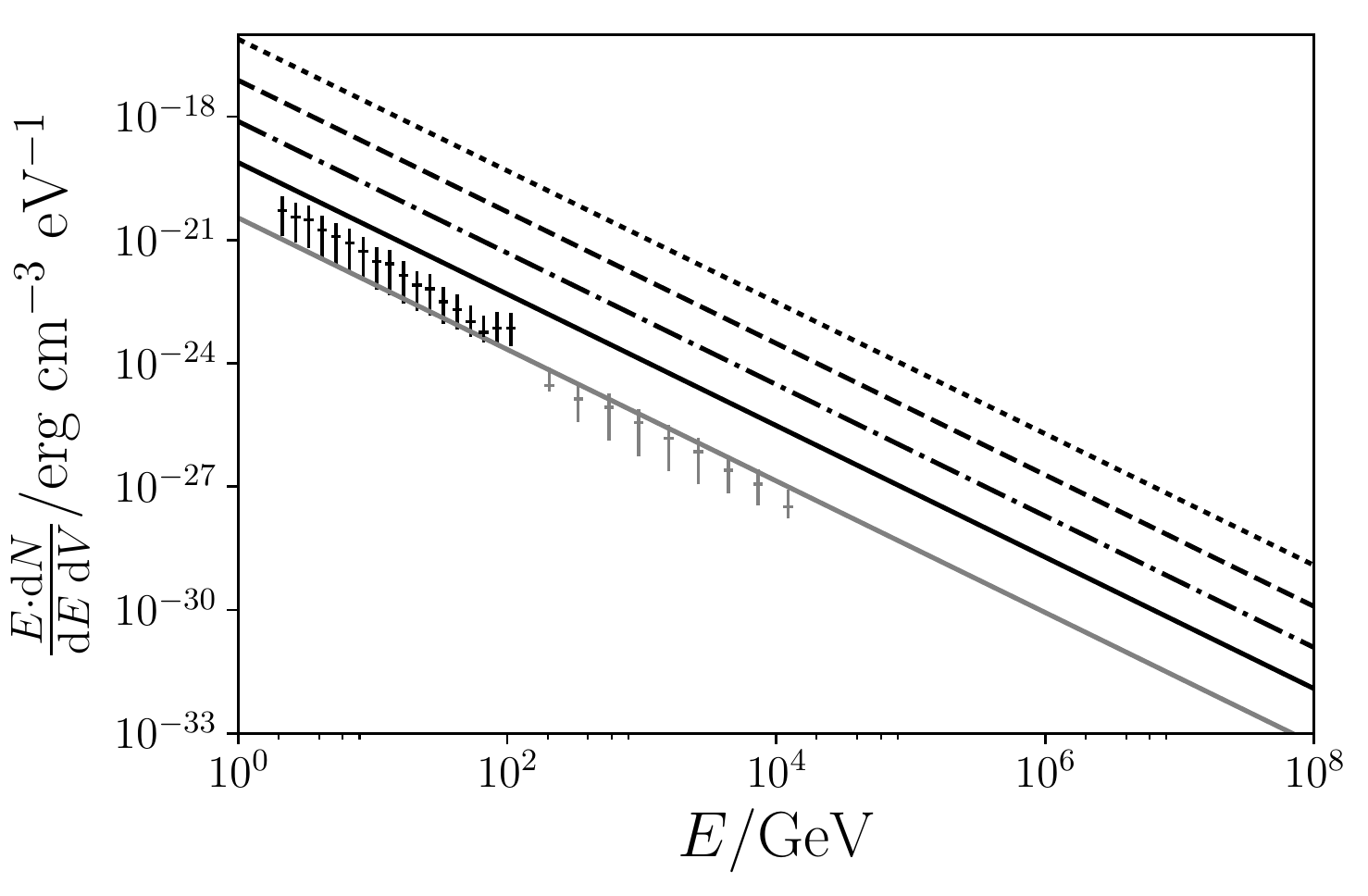}
    \caption{Initial injection spectral energy densities for CRs in the four protogalaxy models with $\mathcal{R}_{\rm SN} = 0.01, 0.1, 1.0$ and $10.0~\text{yr}^{-1}$ (the four black lines, solid, dashed-dotted, dashed and dotted respectively) and that scaled for the Milky Way (solid grey line). CR spectral energy density points inferred from $\gamma$-ray observations of the Galactic Ridge (GR) region between $l<|\ang{0.8}|$ in Galactic longitude, and $b<|\ang{0.3}|$ in Galactic latitude above 1-GeV with \textit{Fermi}-LAT and H.E.S.S. are shown in black and grey respectively, which suggest the model approach is largely consistent with observations -- points are derived from data published in \citet{Aharonian2006Natur} and \citet{Gaggero2017PhRvL}. We note that the uncertainties in the Milky Way (and protogalaxy) model parameters are likely to be much greater than the error bars indicated in the data points, so caution should be taken in drawing strong conclusions from this comparison. See main text for further details.}    
    \label{fig:CR_data_models}
\end{figure}

 $\gamma$ rays are produced by the decay of the $\pi^0$ secondaries produced in the CR $\text{pp}$ interactions according to process~\ref{eq:process1}. Since all of the $\pi^0$ energy is passed to the $\gamma$ rays, the relation between the CR spectral energy density and that of $\gamma$-rays is governed entirely by the inelastic cross-section for the production of $\pi^0$ secondaries -- see section~\ref{sec:absorption} for details. The CR energy flux is related to the $\gamma$-ray energy flux by
 \begin{equation}
 \label{eq:gamma_prod}
E_{\gamma} \frac{{\rm d} \Phi(E_{\gamma})}{{\rm d}E_{\gamma}} \approx \left(\frac{\sigma_{\pi^0}}{\hat{\sigma}_{\rm p\pi}}\right) \Biggr\vert_{E_0} \; E \; \frac{{\rm d} \Phi(E)}{{\rm d}E} \ ,
 \end{equation}
 where the cross-sections only show a weak energy-dependence (meaning that their values at $E_0$ are sufficient for our estimates)\footnote{This assumes local CR isotropy, and that the vast majority of CRs are attenuated by $\text{pp}$-interactions in the GR region. These assumptions should be assessed more carefully in future studies, and mean that the resulting estimates for CR number density from $\gamma$-ray emissions stated here are conservative and should be regarded as a lower limit.}. Equation~\ref{eq:gamma_prod} combined with equation~\ref{eq:differential_number_density} can be rearranged to allow the CR spectral energy density in a $\gamma$-ray emitting region to be estimated. Applying this to $\gamma$-ray measurements of the GR in the region between $l<|\ang{0.8}|$ in Galactic longitude, and $b<|\ang{0.3}|$ in Galactic latitude above 1-GeV allows the local injected CR energy density driving this $\gamma$-ray emission to be estimated, as shown in Fig.~\ref{fig:CR_data_models}. This indicates the injected spectral energy density for the four protogalaxy models considered in this study, with $\mathcal{R}_{\rm SN} = 0.01, 0.1, 1.0$ and $10.0~\text{yr}^{-1}$ (the four black lines, solid, dashed-dotted, dashed and dotted respectively), and also that for the Milky Way GR region, with the model scaled for the Galactic parameters discussed above for reference. This GR line is compared to values derived from $\gamma$-ray data from \textit{Fermi}-LAT (the black points, see~\citealt{Gaggero2017PhRvL})\footnote{We directly use the results from the {\it Fermi} analysis undertaken in~\citet{Gaggero2017PhRvL} here. These points used the Fermi Science tools V10R0P4 with 422 weeks of PASS 8 data, and event class CLEAN. See~\citet{Gaggero2017PhRvL} for further details of the $\gamma$-ray data analysis.} and H.E.S.S. (the higher energy grey points, from~\citealt{Aharonian2006Natur}, also shown in~\citealt{Gaggero2017PhRvL}), which are seen to be largely consistent with our scaled Galactic model. We note, however, that the uncertainties in our model parameters are likely to be much greater than the error bars indicated in the data points, so caution should be taken in drawing strong conclusions from this comparison.
  
From these data points, it is evident that there may be some motivation for a slightly steeper spectral index than that used in our protogalaxy models. However, it is not clear whether this  results from differences between a true protogalaxy environment and the conditions in the GR region (which may not be truly comparable to this level), or whether this could be due to systematics in the data, or shortcomings of the crude conversion between $\gamma$-ray flux and CR spectral energy density which we invoke here. The analysis in~\citet{Gaggero2017PhRvL} suggests a best-fit power-law index of around $-2.29\pm 0.27$ is appropriate for the H.E.S.S. data, while the combined \textit{Fermi}-LAT and H.E.S.S. analysis is consistent with a slightly steeper spectral index of $-2.49^{+0.09}_{-0.08}$ (with reduced $\chi^2$ of 3). These suggest that our adopted index of -2.1 is reasonable enough for our purposes, and we do not believe there is sufficient tension to adopt a steeper power-law that may not necessarily be any more or less physically motivated in a high-redshift protogalaxy -- particularly as the choice index does not strongly impact our results for any sensible range of values.

\subsection{Diffusion Coefficient}
\label{sec:diff_coeff} 

In a uniform magnetic field, the propagation of a charged particle describes a curve 
   characterised by the Larmor radius $r_{\rm L}$, which is given by
\begin{equation}%
  r_{\rm L} = \frac{3.3 \times 10^{12}}{|q|} \left(\frac{E}{10^{9}\;\!{\rm eV}}\right) \left( \frac{{\rm \mu G}}{B}\right) ~\rm{cm} \ ,%
\end{equation}%
   where $q$ is the magnitude of the charge of the particle.  
Propagation of CRs in a medium permeated by a turbulent, tangled magnetic field is more complicated. 
However, $r_{\rm L}$ can be used to derive a phenomenological prescription 
  for the CR diffusion process. 
The diffusive speed of the particles is expressed in terms of the diffusion coefficient 
   $D(E)$, which accounts for their scattering in the magnetic field and turbulence. 
   This can be quantified in the direction parallel or perpendicular to the magnetic field lines, 
   with that perpendicular to the lines typically being around two orders of magnitude smaller in 
   an ISM environment~\citep[see, e.g.][among others]{Shalchi2004ApJ, Shalchi2006ApJ, Hussein2014MNRAS}. 
   Radio observations suggest a principal magnetic field component is present in outflow cones directed perpendicularly to the disk of the host galaxy -- see, for instance M83~\citep{Sukumar1990IAUS}, NGC 4565~\citep{Sukumar1991ApJ}, NGC 4569~\citep{Chyzy2006A&A}, NGC 5775~\citep{Soida2011A&A} and NGC 4631~\citep{Hummel1988A&A, Hummel1991A&A, Brandenburg1993A&A, Mora2013A&A} among others.
   
   Thus the diffusion along the magnetic field lines directed along the outflow cone dominates the macroscopic propagation of CR particles, and
as a rough approximation we may parametrise the diffusion coefficient as a random walk process 
  with mean-free-path characterised by the local Larmor radius, i.e. as 
\begin{equation}
    D(E, h) = D_0 \left[ \frac{r_{\rm L}\left(E, \langle | B | \rangle|_h \right) }{r_{{\rm L},0}} \right]^{\delta} \ ,
\end{equation}
  where $\langle | B | \rangle|_h = |B(h)|$ is the characteristic magnetic field strength in the outflow at some position $h$,  
   and the normalisation $D_0 = 3.0\times 10^{28}$ cm$^2$ s$^{-1}$ is comparable to observations in the Milky Way ISM 
   \citep{Berezinskii1990book, Aharonian2012SSR, Gaggero2012thesis} 
   for a 1-GeV CR proton in a 5-$\mu$G magnetic field 
   (corresponding to a reference Larmor radius $r_{{\rm L},0}$).  
   The exponent $\delta$ encodes the effects of the interstellar turbulence, for which we adopt a value of $\delta=1/2$~\citep[see also][]{Berezinskii1990book, Strong2007ARNPS}, i.e. corresponding to a Kraichnan turbulence spectrum which is considered a suitable model for the ISM~\citep{Yan2004ApJ, Strong2007ARNPS} -- we argue it is reasonable to expect the processes driving turbulence in high redshift protogalaxies are not unlike those in the Milky Way.
   
   In diffusion-dominated systems, observations have not shown any strong evidence for large variations of the diffusion coefficient 
   in galactic outflows in nearby galaxies, e.g. in NGC 7462 \citep{Heesen2016MNRAS}. 
Despite the varying magnetic field in the presence of an outflow, diffusive propagation of CRs is not likely to extend far into the outflow cone. Thus, over the relevant length-scales, we argue that 
the expression for the coefficient above is effectively preserved along the flow such that $D(E, h) = D(E)$,    
   with its temporal evolution and the spatial variation 
   determined only by the temporal evolution and the spatial variation of the local characteristic magnetic field.
   In a system dominated by advection, magnetic field variations would presumably yield a more significant variation of the diffusion coefficient 
   along the outflow cone. However, in such systems, diffusion is not important over large distances with advective flows and streaming instabilities taking precedence -- so whether such variation of the diffusion coefficient is present is inconsequential to our analysis.

\subsection{Cosmic Ray Transport}
\label{subsec:transport}  

\begin{figure}
	\includegraphics[width=\columnwidth]{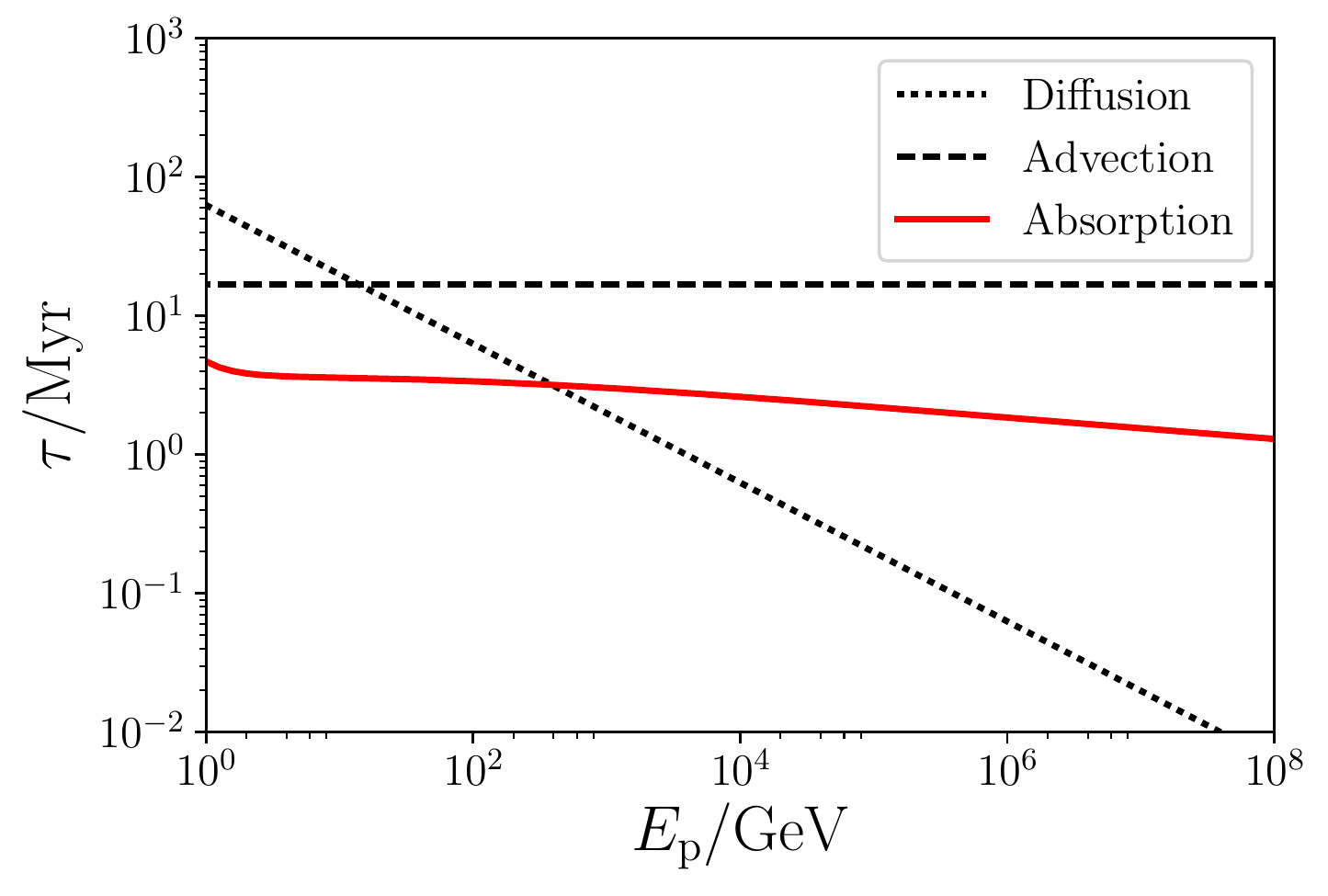}
    \caption{Timescales for attenuation (solid red line), 
       diffusion (dotted line) and advection of CRs in a bulk flow 
          of $290\;\!{\rm km~s}^{-1} \approx v_{\infty}$ (heavy dashed line). 
       This has assumed a uniform medium density of $n_{\rm p} = 10$ cm$^{-3}$ throughout, 
          and a magnetic field strength of 5 $\mu{\rm G}$, and 
       a characteristic propagation length-scale of $\ell = 5~\text{kpc}$.}
    \label{fig:timescales}
\end{figure}

The diffusion timescale is given by 
\begin{equation}
    \tau_{\rm diff}(E) = \frac{\ell^2}{4D(E)} \ ,  
\end{equation}
  and the advection timescale may be approximated as 
\begin{equation}
    \tau_{\rm adv} \simeq \frac{\ell}{v} \ .
\end{equation}
   where $\ell \approx 5~\text{kpc}$ is the characteristic CR propagation length-scale (note that this is not necessarily the same as $\ell_{\rm adv}$ and $\ell_{\rm diff}$ used previously, which were specific to the nature of the system under consideration -- here we instead chose a consistent length-scale over which processes can be compared, and which roughly corresponds to the distance over the bulk of CRs would be found in our outflow model -- see sections~\ref{subsubsec:advection} and~\ref{subsubsec:diffusion}).
The timescale over which the CR particles deposit their energy   
  is determined by the attenuation of the particles due to the {\rm pp}-interaction, i.e 
  it may be expressed in terms of {\rm pp}-interaction mean-free-path, as  
\begin{equation} 
  \tau_{\rm p\pi} (E)= \frac{l^*_{\rm p \pi}(E)}{\rm c} \ . 
\end{equation}  
Fig.~\ref{fig:timescales} shows the comparison of the three timescales  
  for CR protons with various energies,  
  with a 5-$\mu{\rm G}$ mean magnetic field, 
   a mean ISM number density $n_{\rm p} = 10$ cm$^{-3}$ and a flow velocity of $290~\text{km}~\text{s}^{-1}$ (being that of the terminal flow velocity established for the CR-driven outflow in section~\ref{subsubsec:velocity_density_profile}). We note that this is intended to illustrate the relative importance of the processes at work in this system, with the adopted conditions comparable to those at the base of the outflow (i.e. the host galaxy ISM) where densities are highest and much of the CR attenuation would arise.
   The true outflow model has substantially different densities, with the profile falling to several orders of magnitude lower by 5 kpc (see Fig.~\ref{fig:density_profile}), meaning that the CR attenuation timescale would be much greater at larger distances along the outflow cone. Nevertheless, the strong attenuation near the base of the outflow will dominate the timescales, meaning the estimate here remains a suitable approximation to illustrate the global picture of the system.
  
For all energies,
  $\tau_{\rm p \pi} < \tau_{\rm adv}$,  
 which would imply that CR protons are substantially attenuated near the base of the outflow, where conditions are most similar to those assumed in the Fig.~\ref{fig:timescales} approximation.
  However, over a length-scale comparable to the size of the host galaxy (of a radius of $\sim 1$~kpc), 
  the time over which advection would arise is an order of magnitude less than that shown in Fig.~\ref{fig:timescales}.
  This means that absorption and advection would operate over comparable timescales, and a non-negligible fraction of 
  CRs could be advected by the flow
  to reach distances beyond the galaxy from which they originate. Indeed, taking full account of the true density profile of the outflow means that attenuation would be substantially reduced compared to the situation indicated here -- meaning that a substantial fraction of the CR energy density could be deposited outside of the host galaxy instead of within it.
CR protons with $E_{\rm p} < 10^3 \;\! {\rm GeV}$ have a long diffusion time.  
With $\tau_{\rm diff} \gg \tau_{\rm p \pi}$, 
   in the absence of advection, 
   these CR protons will be contained within the galaxy 
   and eventually release all their energy through the {\rm pp}-interaction. 
CR protons with $E_{\rm p} \gg 10^3 \;\! {\rm GeV}$, 
  which have $\tau_{\rm diff} < \tau_{\rm p \pi}$ 
   and $\tau_{\rm diff} < \tau_{\rm adv}$ 
   (for outflows with $v \approx 290\;\!{\rm km~s^{-1}}$ or slower),  
   could, however, diffuse out of the galaxy 
   regardless of whether advection is present or not. 
   However, only a very small fraction of the total CR energy density are harboured in this part of the CR spectrum, 
   so their effects would not be of great astrophysical importance.
    
The timescale comparison gives a qualitative assessment of the relative importance 
  of the advective and diffusive processes in the context of CR heating. 
A more quantitative analysis requires us to solve the transport equation  
  (equation~\ref{eq:transport_equation}) explicitly, which we discuss in the remainder of this section. 
In our solution scheme, 
   we consider that 
    the system has settled into a steady state,
   which implies that we may set $\partial n/\partial t  = 0$. 
	We adopt a numerical scheme in which CR protons are only injected at the base of the outflow cone (i.e. from the actively star-forming region), 
   which practically transforms the source term $Q(E,{\rm x})$ into a boundary condition. 
   However, observationally, the SNe sources of CRs can be distributed some way into an outflow. In extragalactic studies, the majority of SNe are found in galaxies up to around half of their estimated scale radius
   ~(see, e.g. \citealt{Hakobyan2012A&A, Hakobyan2014MNRAS, Hakobyan2016MNRAS} which consider SNe in host galaxies up to 100 Mpc away).
    At higher redshifts, which would be most relevant to the starburst protogalaxies we model here, ISM conditions would presumably be more turbulent due to the higher SN activity and this may lead to a proportionally higher distribution of sources throughout the host system.
   Adopting a single boundary condition for the injection of CRs at the base of an outflow is thus insufficient to model the distribution of CR sources, particularly as we are intend to calculate the CR heating effect well within the ISM region of the outflow, down to 100 pc.
   We therefore calculate the outflow (both in the advection and diffusion cases) as the linear sum of scaled outflow solutions by a Monte-Carlo (MC) method, as outlined in section~\ref{subsubsec:extended_inj}.
   In the next section we show calculations of two regimes: firstly, when 
   the transport is dominated by advection 
   and, secondly, when the transport is dominated by diffusion. 
We solve the transport equation explicitly in these two regimes, 
   before accounting for the distribution of SN sources in the galactic core. 

\subsubsection{Advection Dominated Regime}
\label{subsubsec:advection}

In the advection dominated regime, we may drop the diffusion term.  
This reduces the transport equation (in the steady state) to 
\begin{equation}
\label{eq:modified_diff_eq_1}
      \nabla \cdot \big\{ {\bf{v}} n \big\} = \frac{\partial}{\partial E} \Big\{ b(E, {\bf{x}}) n \Big\}  - S(E,{\bf x}) \ .
\end{equation}
(Here and hereafter, unless otherwise stated, 
   we use the short-hand notation $n = n(E, {\bf x})$.) 
Suppose that the flow follows streamlines in the outflow cone.  
By symmetry, the flow is essentially 1-dimensional (specified by the co-ordinate $h$),  
  and the transport equation, when the outflow has settled into a velocity profile $v(h)$ (see section~\ref{subsec:wind_velocity}), may be expressed as 
\begin{equation}
   \frac{1}{h^2}\frac{\partial(h^2 v(h) n)}{\partial h} 
  = \left\{ \frac{\partial}{\partial E} \Big\{ b(E, h) n \Big\} -  {\rm c} ~n ~\hat{\sigma}_{\rm p\pi}(E)~n_{\rm p}(h)\right\} \ , 
\end{equation}
   where the sink term now takes the form $S(E,h)  = {\rm c} ~n ~\hat{\sigma}_{\rm p\pi}(E)~n_{\rm p}(h)$. 
With the substitution $Z(E, h) = n(E, h)\; v(h) \; h^2$, the transport equation becomes 
\begin{equation}
   \frac{\partial Z}{\partial h} 
     = \frac{1}{v(h)}\left\{ Z \frac{\partial b(E, h)}{\partial E} + b(E, h) \frac{\partial Z}{\partial E} -  {\rm c} \;\!
   \hat{\sigma}_{\rm p\pi}(E)\;\!n_{\rm p}(h) Z \right\} \ . 
\label{eq:advection_eq}
\end{equation}  
The variable $b(E, h) = -{{\rm d}E}/{{\rm d}t}$ is known when the cooling processes are specified, 
 and $\partial b(E, h) / \partial E$ can be found from this. 
When $b(E,h)$, the sink term and the boundary condition at the base of the outflow cone $h_0$ are set, 
   the transport equation can be solved numerically 
   using a finite difference method as described in Appendix~\ref{sec:appendixa}.  
In this work, we solve the equation for the case of the only non-negligible cooling process being that due to the adiabatic cooling of CRs propagating along the outflow cone, i.e. $b(E,h) =\dot{E}_{\rm ad}$, 
  subject to the boundary condition that $Z(E_0, h_0) = n(E_0, h_0)~v(h_0)~h_0^2$  
   with $h_0$ set to be $100~{\rm pc}$, the size of the starburst region \citep[see][]{Chevalier1985Nat, Tanner2016ApJ}
   and we use a reference energy $E_0$ at $1~{\rm GeV}$. 
Moreover,  $Z(E,h) \propto E^{-\Gamma(h)} \;\! h^2 \;\! v(h)$ 
  with a power-law index $\Gamma(h_0) = 2.1$ at $h_0$.  We invoke appropriate velocity profiles $v(h)$ modelled according to section~\ref{subsec:wind_velocity}.

\begin{figure}
 \includegraphics[width=\columnwidth]{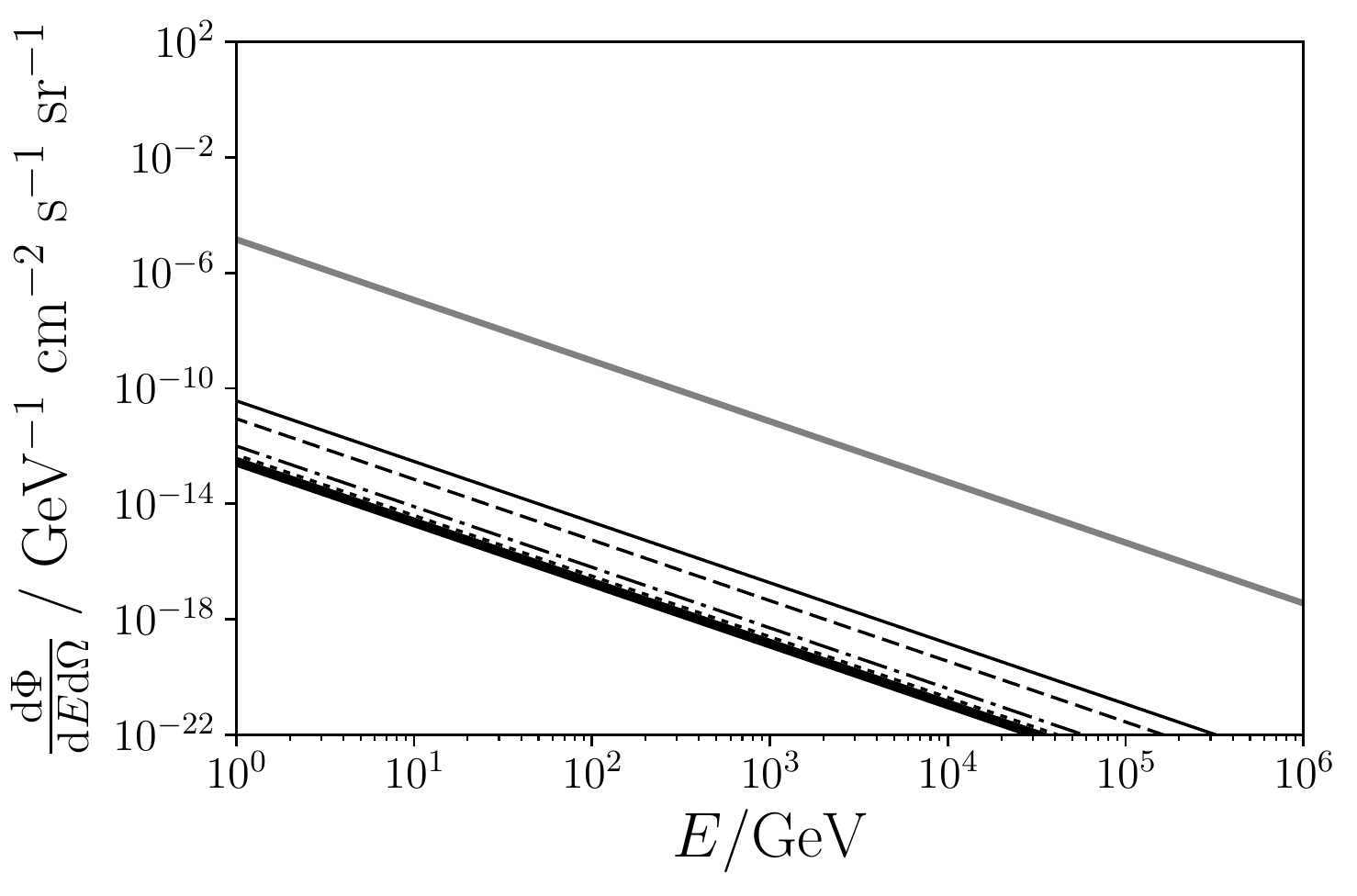} \\ 
 \includegraphics[width=\columnwidth]{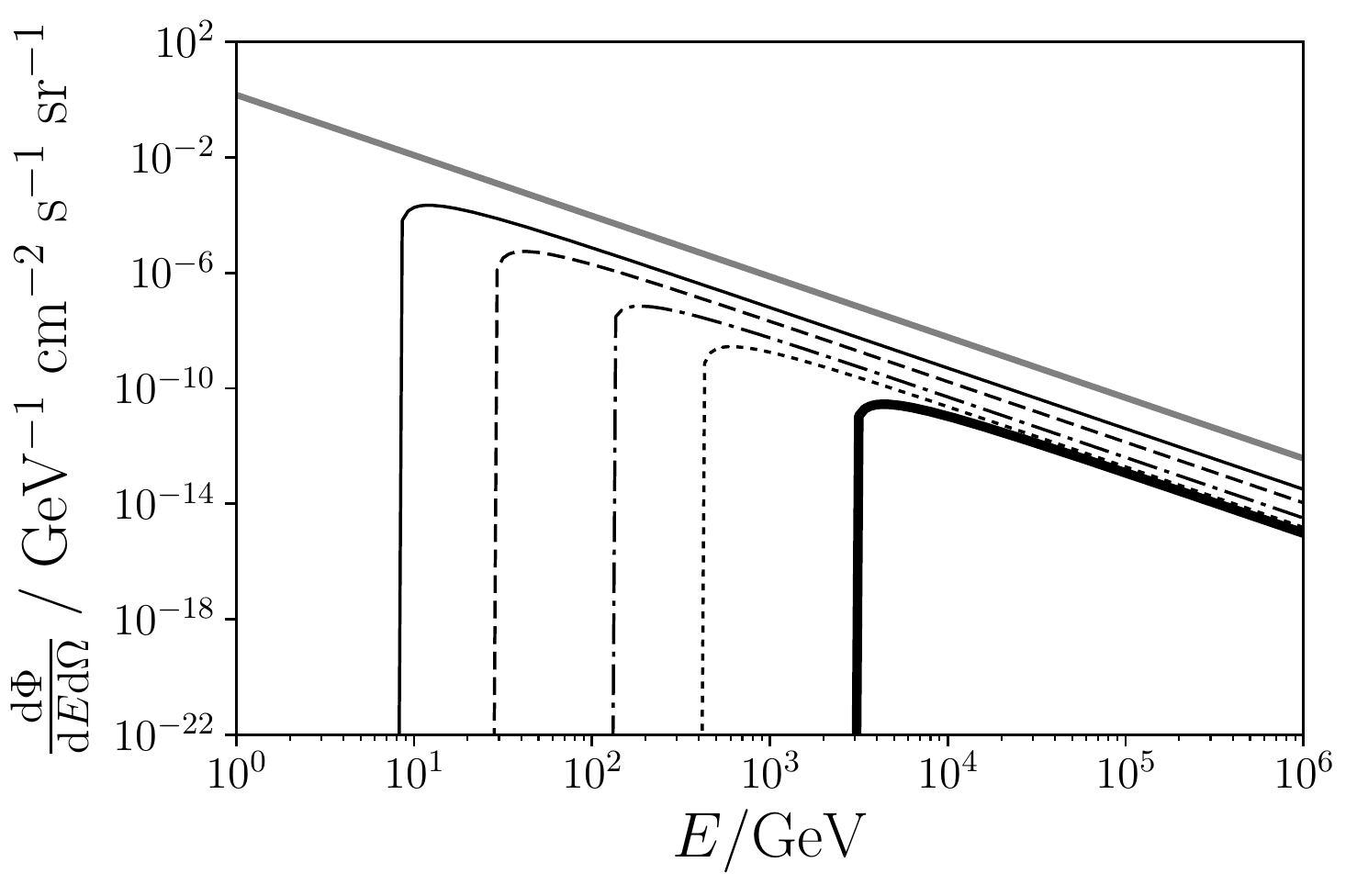}
    \caption{\textbf{Top}: 
    Normalised energy spectra of CRs subject to advection by the bulk galactic outflow. 
    Curves from top to bottom are respectively the energy spectra  
       at the base of the outflow (solid grey line), 
       at $h=10~{\rm kpc}$ (solid black line), 
       $h=20~{\rm kpc}$ (black dashed line), $h=30~{\rm kpc}$ (dashed-dotted line),
        $h=40~{\rm kpc}$ (dotted black line) and $h=50~{\rm kpc}$ (thick black line). 
     All spectra are practically power-laws with almost identical spectral indices, 
       implying insignificant evolution of the energy spectrum of CRs advected along with the flow. 
    \textbf{Bottom}:  Normalised energy spectra of CRs subject to predominantly diffusive propagation.  
    Curves from top to bottom are respectively the energy spectra 
    at the base of the outflow (solid grey line), at $h=0.25~{\rm kpc}$ (thin solid black line), 
    $h=0.5~{\rm kpc}$ (black dashed line), $h=1~{\rm kpc}$ (dashed-dotted line) 
    and $h=1.5~{\rm kpc}$ (dotted black line) and $h=2~{\rm kpc}$ (thick black line). 
 Strong suppression occurs at lower energies, 
    due to the attenuation timescale being shorter than the diffusion timescale at these energies. 
 On a length-scale of 1~kpc, 
   the diffusion timescale exceeds the attenuation timescale 
    when the energy of the CRs falls below about $5\times10^3~{\rm GeV}$. In both plots, the differential CR fluxes shown are the values calculated for a protogalaxy model with $\mathcal{R}_{\rm SN} = 0.1~\text{yr}^{-1}$.}
\label{fig:advection-diffusion_spectrum}
\end{figure}

By inspection of the transport equation, 
   we may see that in the absence of energy-dependent CR cooling,  
   softening (or hardening) of the CR energy spectrum over a large-scale galactic outflow 
   will not occur in the advection dominated regime. 
This conclusion can also be reached in a qualitative analysis 
  by comparing the attenuation and advection timescales and their energy dependencies. 
As shown in Fig~\ref{fig:timescales}, 
  $\tau_{\rm adv}$ is independent of the CR energy 
  and $\tau_{\rm p \pi} (E)$ has only very small variations across the energy range considered.  
Thus, without strong energy dependences in these two terms, 
     there should not be significant evolution of the energy spectrum of the CRs that are advected by the flow. 
Fig.~\ref{fig:advection-diffusion_spectrum} shows the CR energy spectra 
  (obtained by solving the transport equation numerically) 
  over distances up to $h=50~{\rm kpc}$  
  for an outflow with a full opening conic angle of \ang{55} 
  and the velocity profile determined in section~\ref{subsubsec:velocity_density_profile},
  which indicates negligible spectral evolution of the CRs along the flow.

\subsubsection{Diffusion Dominated Regime}
\label{subsubsec:diffusion} 
   
In the diffusion dominated regime, 
   the transport equation takes the form
\begin{equation}
 - \nabla \cdot \left[D(E)\nabla n \right] =  \frac{\partial}{\partial E} \left[b(E, {\bf{x}}) n\right] - S(E, {\bf{x}}) \ . 
\end{equation}
The outflow cone is axi-symmetric and so the transport equation is 1-dimensional, 
  specified by the coordinate $h$ (as in the advection dominated case). 
If the diffusion coefficient does not vary significantly along $h$, then we have 
\begin{equation}
 -  \frac{D(E)}{h^2}\frac{\partial}{\partial h} \left\{ h^2 \frac{\partial n}{\partial h}\right\} = 
   \frac{\partial}{\partial E}\left[b(E, h) n\right]  -  {\rm c} ~n ~\hat{\sigma}_{\rm p\pi}(E)~n_{\rm p}(h) \ .
\end{equation}
Substituting $Z(E, h) = h^2 n(E, h)$ into the equation yields  
\begin{align}
 -  D(E) & \left\{\frac{\partial^2 Z}{\partial h^2} -\frac{2}{h}\frac{\partial Z}{\partial h} + \frac{2 Z}{h^2} \right\} \nonumber \\
     & \hspace*{0.5cm}  = \frac{\partial}{\partial E}\left[b(E, h) Z\right]  -  {\rm c} ~Z ~\hat{\sigma}_{\rm p\pi}(E)~n_{\rm p}(h) \ .
\end{align}
After rearranging and expanding the energy derivative, we obtain   
\begin{align}
\label{eq:diffusion_equation2}
     \frac{\partial^2 Z}{\partial h^2} = -\frac{1}{D(E)} & 
     \Bigg\{ Z \frac{\partial b(E, h)}{\partial E} + b(E, h) \frac{\partial Z}{\partial E} \nonumber \\
         & -  {\rm c} ~Z ~\hat{\sigma}_{\rm p\pi}(E)~n_{\rm p}(h)\Bigg\} + \frac{2}{h} \frac{\partial Z}{\partial h} - \frac{2 Z}{h^2} \ . 
\end{align}

The transport equation is solved numerically, 
  with the scheme described in Appendix~\ref{sec:appendixa}. 
  This requires two Neumann boundary conditions  
  (i.e. step 1 in equation~\ref{eq:difference_diffusion_eq1}), 
  and these are obtained directly from considerations of SN event rates, 
  and the efficiency of CR production. 
Here we show how the two boundary conditions are constructed.  
  
For the first one, we begin with 
\begin{align}
   \frac{{\rm d}Z}{{\rm d}h}\bigg\rvert_{i,1} =& \left(\frac{E}{E_0}\right)^{-\Gamma_0} \left(\frac{{\rm d}Z}{{\rm d}h}\bigg\rvert_{\substack{E=E_0\\h=h_0}}-Z(E_0, h_0) \left(\frac{E}{E_0}\right)^{-1} \frac{{\rm d}\Gamma}{{\rm d}h}\bigg\rvert_{h=h_0}\right)
\end{align} 
  where ${{\rm d}Z}/{{\rm d}h}$ at the boundary needs to be specified. 
As we lack a prescription that accounts for the acceleration of CRs in the starburst region 
  with a transition to their transportation in the outflowing region, 
  we adopt the assumption that 
   ${{\rm d}Z}/{{\rm d}h}$ scales with that of CR electrons  
   initially at the lower boundary of the wind cone. 
Hence, we have   
\begin{equation}
  \frac{{\rm d}Z}{{\rm d}h}\bigg\rvert_{\substack{E=E_0\\h=h_0}} 
  = \left(\frac{{\rm d}Z_{\rm CRe}}{{\rm d}h}\bigg\rvert_{\substack{E=E_0\\h=h_0}} \right) ~R \ ,
\end{equation}
   where $R$ is the scaling factor. 
Radiative cooling processes are generally inversely proportional to the fourth-order of the mass of the charged particles.   
Thus, we set $R = (m_{\rm e}/m_{\rm p})^4$, 
  which implies that ${{\rm d}Z}/{{\rm d}h}$ is negligible for CR protons.   
This prescription is consistent with the CR proton flux being conserved at the boundary, 
  i.e. $v(h_0) Z(E) \vert_{h_0} = v(h_0) \{ n(E) h^2\}  \vert_{h_0} = {\rm constant}$. 
We may estimate the value of ${{\rm d}n_{\rm CRe}}/{{\rm d}h}$ (the gradient in CR electron number density) from radio observations.  
Observations of the nearby starburst outflows in NGC 7090 and NGC 7462 
   in the 6~cm and 22~cm bands \citep{Heesen2016MNRAS} indicate  
\begin{equation}
   \frac{{\rm d}Z_{\rm CR,e}}{{\rm d}h}\bigg\rvert_{\substack{E=E_0\\h=h_0}} = -0.2 \  
\end{equation}
   at the base of the wind.  
The second condition relates to the rate of change of the CR spectral index at the base of the outflow cone. 
Similarly, we assume a scaling with the CR electrons.  
Observations of the nearby starburst galaxies NGC 7090 and NGC 7462 
  \citep{Heesen2016MNRAS} 
  suggest that $  \Gamma_{\rm CR, e}' =  {\rm d}\Gamma_{\rm CR,e}/{\rm d}h = -1.4$ 
  at the base of the galactic outflow.  
Thus,  
\begin{equation}
   \frac{{\rm d}\Gamma}{{\rm d}h}\bigg\rvert_{h_0} = \Gamma_{\rm CR,e}' +\log R \ ,
\end{equation}
   where $R$ is as defined above.  
This requires that the cooling and spectral evolution at the lower boundary is insignificant, 
  which, in turn, ensures a negligible variation in the spectral index of the CR protons.  
   
\begin{figure}
	\includegraphics[width=\columnwidth]{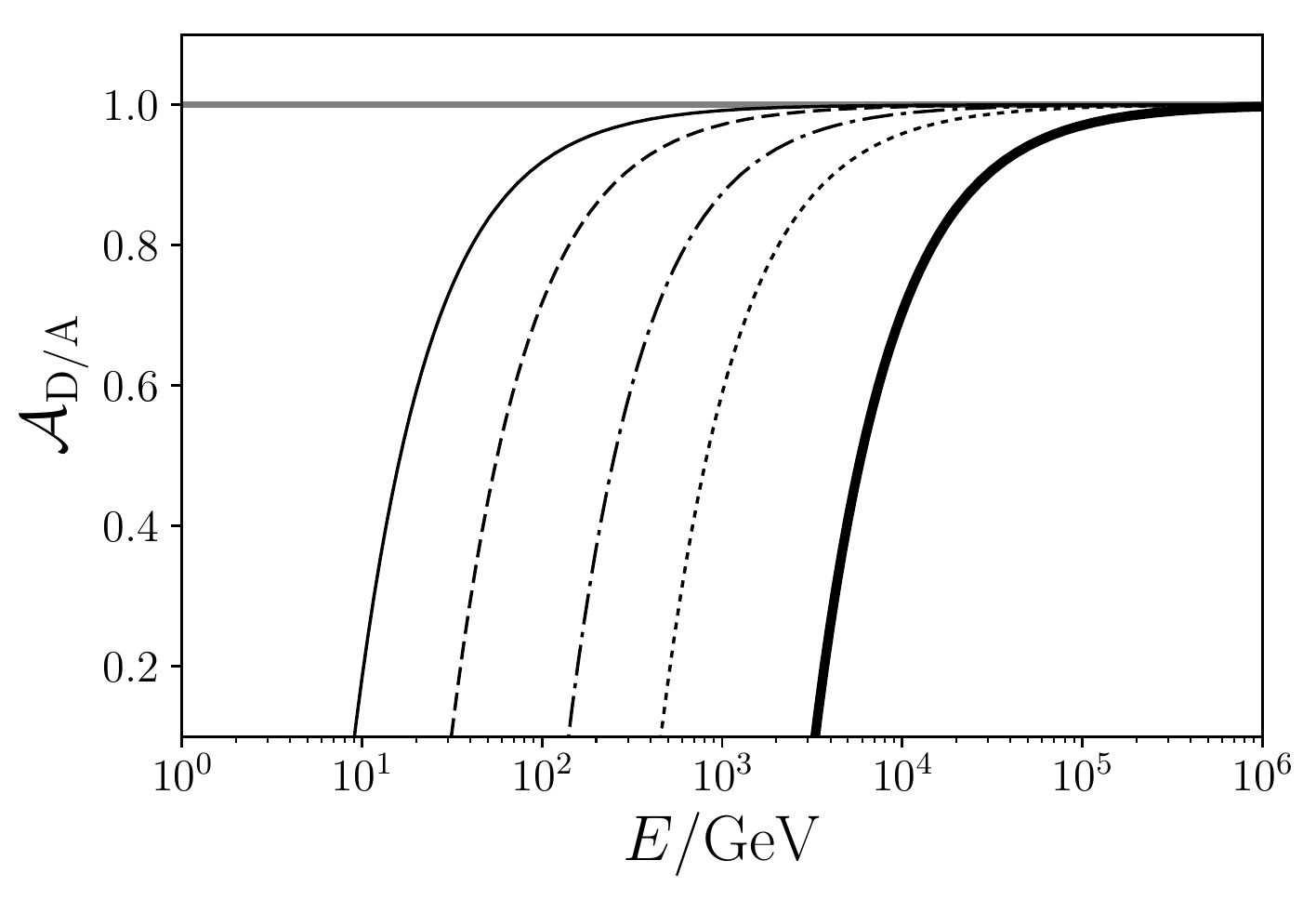}
    \caption{Plot to show the ratio of the normalised diffusion spectrum to the normalised advection spectrum over energy, to give an indication of the level of attenuation $\mathcal{A}_{\rm D/A}$ experienced by the CRs as they diffuse (this being the cause of the turn-over in the spectrum). Curves from top to bottom are for the base of the outflow (solid grey line), at $h=0.25~{\rm kpc}$ (thin solid black line), 
    $h=0.5~{\rm kpc}$ (black dashed line), $h=1~{\rm kpc}$ (dashed-dotted line),  
   $h=1.5~{\rm kpc}$ (dotted black line) and $h=2~{\rm kpc}$ (thick black line).}
    \label{fig:diff_adv_ratio}
\end{figure}

We adopt the same numerical solution scheme used in solving the transport equation in the advection dominated regime.  
The solution $Z$ is obtained 
  by integrating the transport equation with the two boundary conditions at the base of outflow cone  
  using a Runge-Kutta method  
  (as described in the Appendix~\ref{sec:appendixa}). 
The result is shown (for $h$ up to $2~{\rm kpc}$) in Fig.~\ref{fig:advection-diffusion_spectrum}\footnote{The values shown in Fig.~\ref{fig:advection-diffusion_spectrum} are calculated for a protogalaxy model with $\mathcal{R}_{\rm SN} = 0.1~\text{yr}^{-1}$. If we instead scale to a Milky Way-like model as described in section~\ref{sec:cr_gamma_observations} then the CR differential fluxes are a little lower. Assuming an inelasticity of approximately 0.3 for the production of $\gamma$-ray producing neutral pions (see section~\ref{sec:absorption}), the resulting $\gamma$-ray flux can be estimated. Given that the Milky Way is predominately diffusive in terms of CR propagation, the CR differential fluxes according to the above spectral model approach would be around $1.0 \times 10^{-5} \; {\rm ph} \; {\rm GeV}^{-1} \; {\rm cm}^{-1} \; {\rm s}^{-1} \; {\rm sr}^{-1}$ at 1-GeV, or $7.0 \times 10^{-12} \; {\rm ph} \; {\rm GeV}^{-1} \; {\rm cm}^{-1} \; {\rm s}^{-1} \; {\rm sr}^{-1}$ at 1-TeV. This is consistent with $\gamma$-ray flux measurements of the Galactic Ridge above $E_{\rm \gamma} = 10~\text{GeV}$ with \textit{Fermi}-LAT, H.E.S.S. and VERITAS in e.g.~\citet{Aharonian2006Natur, Gaggero2015arXiv, Archer2016ApJ, Gaggero2017PhRvL}.}. The ratio of the normalised diffusion spectrum to the normalised advection spectrum gives an indication of the level of attenuation $\mathcal{A}_{\rm D/A}$ experienced by the CRs as they diffuse, as this is the cause of the turn-over in the spectrum -- this is shown in Fig.~\ref{fig:diff_adv_ratio}, which illustrates the dominant effect of attenuation of the diffusing CRs up to around $5\times10^3$ GeV, at which point the diffusion process becomes more important.

\section{Results and Discussion}
\label{sec:results}

We consider starburst protogalaxies at high redshift, which studies have indicated could host substantial 
star-forming activity~\citep[e.g.][]{Hashimoto2018Nat, Watson2015Nat} and strong galactic outflows~\citep[e.g.][]{Steidel2010ApJ}.
Such star-forming activity makes these systems a likely host of abundant CRs -- but here we primarily address whether they would predominantly operate as CR calorimeters 
  \citep{Thompson2007ApJ, Lacki2011ApJ}, or
  whether strong outflow activity 
  could provide a means by which CRs can escape and interfere with the circumgalactic and/or intergalactic environment.
  To help our understanding of how CR containment in young star-forming galaxies progresses  
  and how the coexistence of diffusion and advection in different regions of the same host develops,
  we consider starburst protogalaxies that would be present at redshift $z \approx 7$, 
  being among most distant objects that may be observed with current and near future deep optical/UV surveys, 
  e.g. the Subaru HSC deep-field survey~\citep{Subaru_Proposal2012}. 
  
  \subsection{Hadronic Heating in Outflows}
  \label{subsec:adv_diff_competition}
  
 We have so far only considered outflow environments where the propagation of CRs is dominated by either advection or diffusion. 
 In reality, both advection and diffusion would presumably operate simultaneously and a proper treatment of CR transport in an outflow would require a complete solution of the full transport equation. 
 However, at any one location and particle energy, 
 the dynamics would usually be dominated by only one of these processes.
  For instance, in regions where bulk velocities are low, diffusion would likely be more important than advection. This would be the case in regions of near the base of the outflow cone, where the low outflow velocity would not be able to compete against CR diffusion -- indeed, such an effect is seen in numerical simulations~\citep[e.g.][]{Farber2018ApJ}. 
  The opposite would be true at high altitudes, where the flow velocity is greater and could advect CRs faster than they would typically be able to diffuse. 
  The relative importance of the contributions from each of these two process along an outflow would impact on the distribution of CRs and, by equation~\ref{eq:cr_heating_equation}, would govern the location at which they deposit energy and thermalise.
  
     \subsubsection{Concurrent Advection \& Diffusion}
   \label{subsubsec:combined_heating_rates}
  
 We may attain a reasonable approximation for the distribution of CRs in a system where both advection and diffusion operate
  by weighting the pure advection and pure diffusion limit solutions by their respective timescales at each position and energy, and summing these contributions together.
 Evaluating advection and diffusion timescales at each calculation increment accounts for both the variation of flow velocity over position as well as the variation of the diffusion coefficient over energy. 
 The associated effective hadronic heating rate $\bar{H}$ in the concurrent advection/diffusion picture along the outflow then follows as:
\begin{equation}
\bar{H}(h) = \left\{ {\rm c} ~n_{\rm p} \int_{E_{0}}^{E_{\rm max}} {\rm d}E ~\bar{n}(E)\;\!f_{\rm therm}(E) ~\hat{\sigma}_{\rm p\pi}(E) ~\right\}\Bigg\vert_{h}
\label{eq:combined_heating_calc}
\end{equation}
  with
  \begin{equation}
 \bar{n}(E)\vert_{h} = \left\{\omega_{\rm diff} \;\! n_{\rm diff}(E) + \omega_{\rm adv} \;\! n_{\rm adv}(E)\right\}\vert_{h} \ ,
  \end{equation}
  where
  \begin{equation}
  \omega_{\rm diff}(E, h) = \frac{\tau_{\rm diff}^{-1}}{\tau_{\rm diff}^{-1}+\tau_{\rm adv}^{-1}} \Bigg\vert_{\left\{E, h\right\}} \ ,
  \end{equation}
  and
  \begin{equation}
  \omega_{\rm adv}(E, h) = \frac{\tau_{\rm adv}^{-1}}{\tau_{\rm diff}^{-1}+\tau_{\rm adv}^{-1}} \Bigg\vert_{\left\{E, h\right\}} \ .
  \end{equation}
 Here, $\tau_{\rm adv} = h/v(h)$) is the position-dependent advection timescale, while $\tau_{\rm diff} = h^2/4 D(E)$ is the energy-dependent diffusion timescale.

\begin{figure}
	\includegraphics[width=\columnwidth]{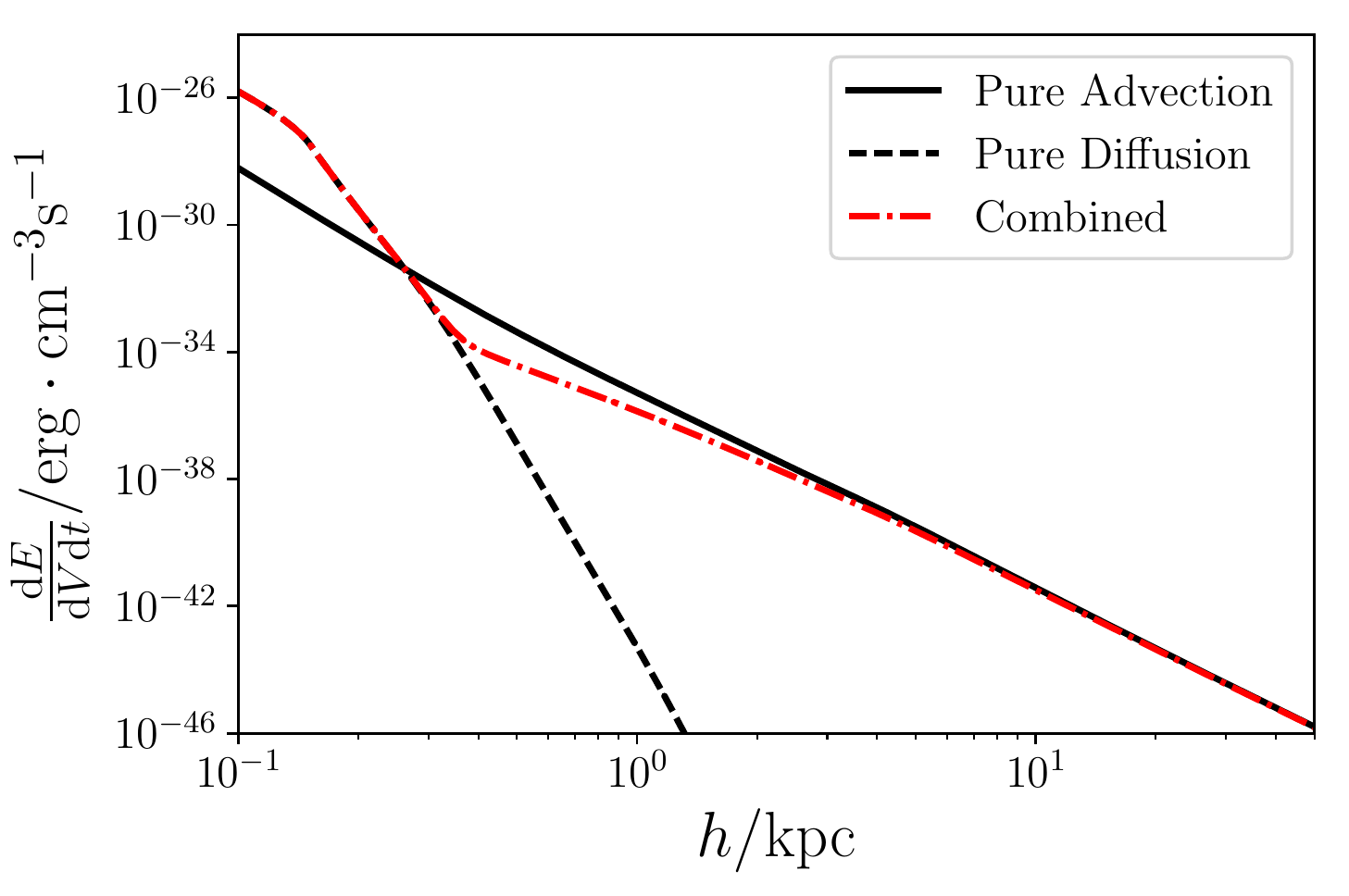}
    \caption{CR heating power in the outflow cone with altitude, $h$, in the advection limit (solid line) and diffusion limit (dashed line) for an outflow with opening angle \ang{55}, driven by a SN-rate of $\mathcal{R}_{\rm SN} = 0.1~\text{yr}^{-1}$,  with thermalisation efficiency $\mathcal{Q} = 0.01$, and mass loading $\mathcal{P}= 0.1$. The combined heating power where advection and diffusion operate concurrently (but with their contributions appropriately weighted by their importance) is indicated by the dashed red line.
   In the diffusion limit, the CRs deposit all their energy within the ISM of their host galaxy while, in the advection limit, the CRs can deposit their energy up to a few tens of kpc beyond the host. The combined model accounts for the weighted contribution of advective and diffusive CR transport and demonstrates the dominance of diffusion near $h=0$ where flow velocities are small, while advection is more important at higher altitudes.}
    \label{fig:advection_diffusion_heating}
\end{figure}

Individual advection-dominated and diffusion-dominated hadronic heating profiles are shown as the two black lines in Fig.~\ref{fig:advection_diffusion_heating}, while the concurrent advection/diffusion heating power is indicated by the dashed red line. 
This demonstrates how diffusive propagation is important in the inner regions of the outflow, while advection dominates at higher altitudes above 0.4 kpc. Above this point, the outflow velocity is sufficiently greater compared to the typical diffusive speed of the CRs (see also Fig.~\ref{fig:velocity_profile}). 
 This is calculated when adopting a 
 SN-event rate of $\mathcal{R}_{\rm SN} = 0.1~\text{yr}^{-1}$, 
 a conical galactic outflow of an opening angle of \ang{55}, 
 an outflow thermalisation efficiency of $\mathcal{Q} = 0.01$,  
 and mass-loading factor of $\mathcal{P} = 0.1$. 
 For reference, these choices yield an outflow terminal velocity of $v_{\infty} \approx 290~\text{km}~\text{s}^{-1}$ and are the same as those used to produce the profile in Fig.~\ref{fig:velocity_profile}.

  \subsubsection{Extended Starburst Region \& Computational Scheme}
   \label{subsubsec:extended_inj}
   
\begin{figure}
	\includegraphics[width=\columnwidth]{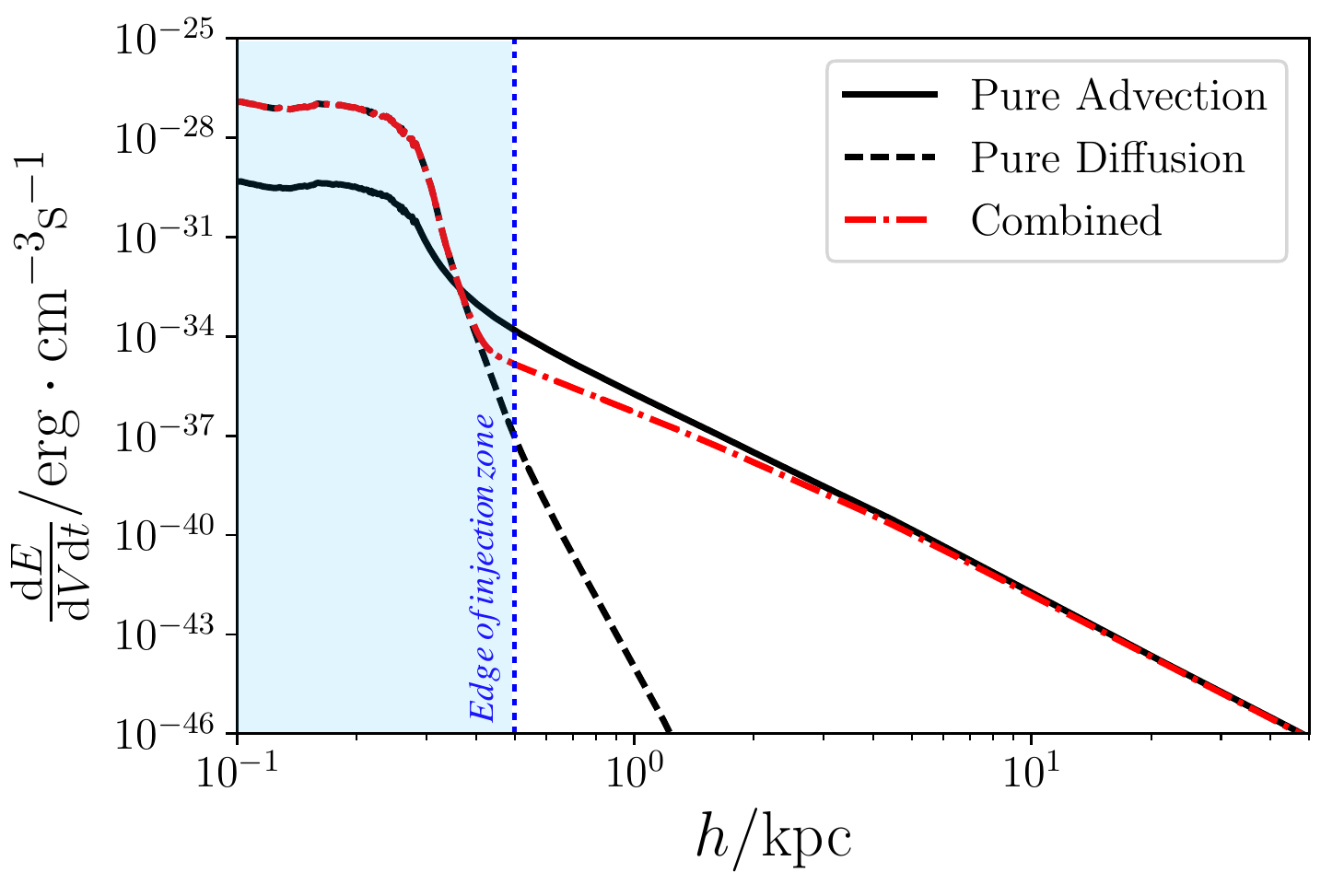}
    \caption{As per Fig.~\ref{fig:advection_diffusion_heating}, with an extended injection of CRs throughout an extended starburst core region, of radius 0.5 kpc. From the dashed red line accounting for the hadronic heating effect when considering both advective and diffusive CR transport, the transition between the two transport regimes is very stark with a reduction in heating power by around 7 orders of magnitude at an altitude of $h\approx 0.3~\text{kpc}$.}
    \label{fig:advection_diffusion_heating_dist}
\end{figure}

Observations have indicated that SN events can arise throughout the disk of their host, at least up to around half of their estimated scale radius 
   \citep{Hakobyan2012A&A, Hakobyan2014MNRAS, Hakobyan2016MNRAS}.
   While sections~\ref{subsubsec:advection} and~\ref{subsubsec:diffusion} and the previous heating profile in Fig.~\ref{fig:advection_diffusion_heating} are suitable descriptions for CRs in outflows emerging from a single point-like region in the centre of their host galaxy, 
   it is also necessary to consider the impact of a more physical extended core region.
     To do this, we adopt a Monte-Carlo (MC) scheme to simulate a spherical distribution of $N$ points up to 0.5 kpc from the protogalactic centre (being half the adopted scale-radius for the protogalaxy). We find that a choice of $N=1,000$ points yields a sufficient signal to noise ratio.
	The distribution of CRs calculated according to equation~\ref{eq:advection_eq} or~\ref{eq:diffusion_equation2} is scaled by $1/N$. 
	The scaled profile is then convolved with the MC spherical points distribution, with each point being taken to be a linearly independent $h_0$ boundary condition\footnote{For this, we adopt a uniform spherical injection region as a first model. More detailed injection distributions are beyond the scope of the current paper but, e.g. a singular isothermal self-gravitating spherical injection weighting is proposed by~\citet{Rodriguez2007MNRAS} -- or, see also~\citet{Silich2011ApJ, Palous2013ApJ} for other approaches. We found the choice of injection model, if reasonable, does not bear any strong influence on the results presented here.}.
	The ensemble of individual CR profiles are then superposed to give a resulting total CR distribution in the outflow, and this accounts for the extended distribution of driving CR sources. This approach is applied to both the pure diffusion and advection calculations as well as in the combined case where both processes operate.
	This resulting summed CR distribution can then be used to determine the hadronic heating profile using the same method as per equation~\ref{eq:combined_heating_calc}, and is shown in Fig.~\ref{fig:advection_diffusion_heating_dist} where lines retain their earlier definitions, and where the starburst extended CR injection region is indicated in blue. 
	This demonstrates the broadened profile for the extended injection, and shows how the transition between the advection and diffusion dominated transport zones is determined by the relative timescales over which they operate rather than the region in which the CRs are injected.
	A distinct picture of a lower `diffusion' region in an outflow emerges in Fig.~\ref{fig:advection_diffusion_heating_dist}, with 
 an `advection' region at higher altitudes where the flow velocity is faster. 
 This result follows the wind structure first introduced in~\citealt{Breitschwerdt1993A&A} (see also~\citealt{Recchia2016MNRAS}).

\subsubsection{Two-Zone Heating Rates}
\label{sec:two_zone_heating_rates}

\begin{figure*}
	\includegraphics[width=0.75\textwidth]{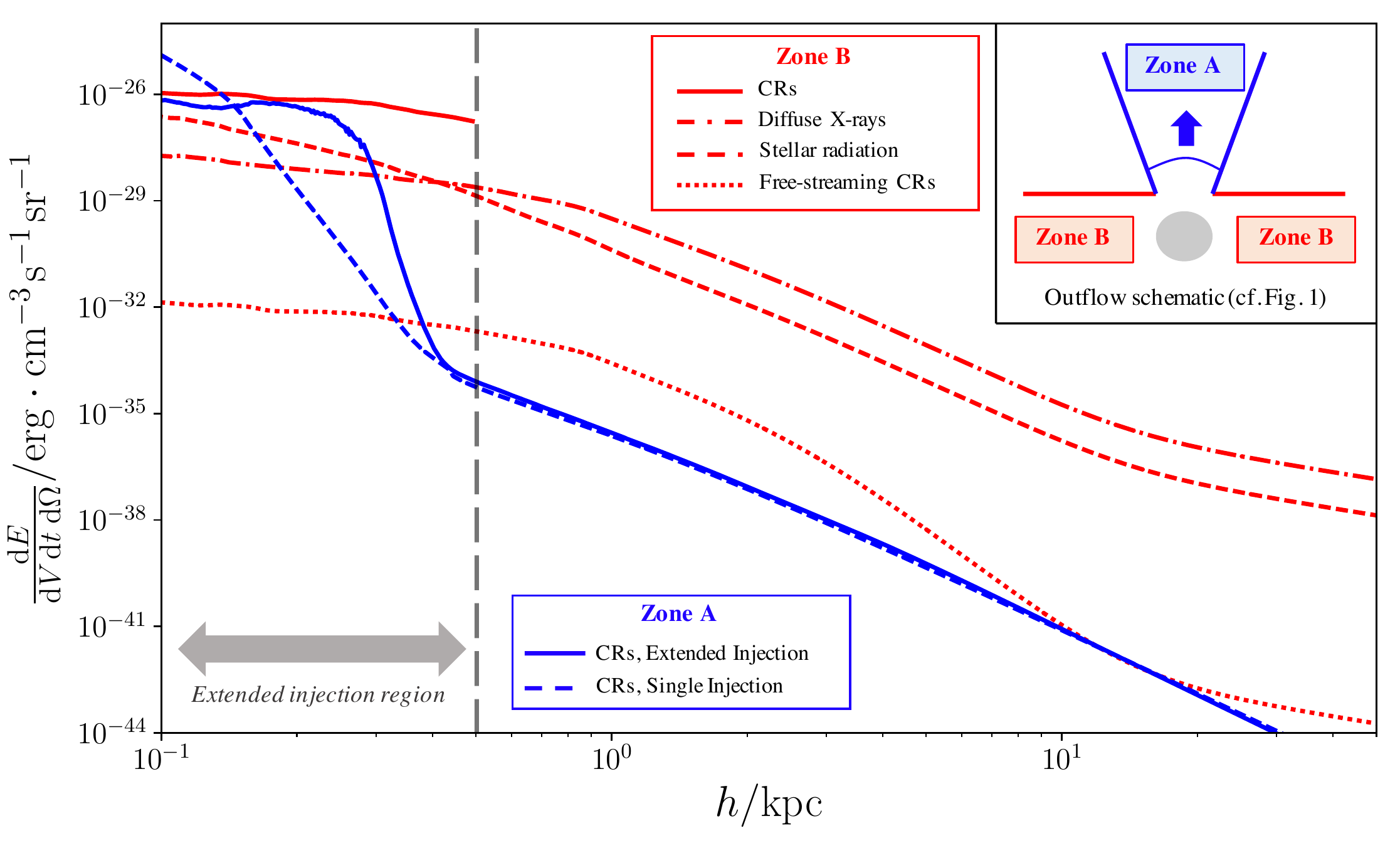}
    \caption{The CR heating power for the inner-galactic regions in Zone A (outflow, advection zone) and Zone B (diffusive ISM zone) shown, as labelled, by the blue and red lines respectively. The Zone B diffusive result is calculated according to the protogalaxy model specified in~\citet{Owen2018MNRAS}, with a SN event rate of $\mathcal{R}_{\rm SN} = 0.1~\text{yr}^{-1}$. The red dotted, dot-dashed and dashed lines show heating due to free-streaming CRs, and radiative heating by diffuse galactic X-rays and stellar radiation respectively are indicated for reference (see~\citealt{Owen2018MNRAS} for details on how these lines are estimated - although we note that the results shown in the earlier paper are for a higher ${\cal R}_{\rm SN}$ of 1.0 $~\text{yr}^{-1}$).
    To enable comparison, heating rates are normalised by the solid angle of their respective `Zone'. This accounts for the effective fraction of the total CR luminosity available for each Zone, which assumes the transport of CRs across Zone boundaries is negligible (see main text for details). The slight dip in the solid blue line results from elevated levels of CR attenuation in this part of the model, due to the higher densities associated with the inner regions of the outflow profile.}
    \label{fig:protogalaxy_heating_average}
\end{figure*} 

Our discussion up to this point has been predominantly concerned with the CR dynamics and heating distribution arising in the outflow cone, i.e. that labelled `Zone A' in the schematic in Fig.~\ref{fig:outflow_model}.
However, this only paints part of the picture:
 typically, a star-forming galaxy be unlikely to be enveloped entirely be a galactic outflow. 
 In disk galaxies in particular, the outflow morphology would normally be bi-conical in nature (cf. section~\ref{sec:introduction} and, e.g.~\citealt{Strickland2000AJ, Ohyama2002PASJ, Veilleux2005ARAA, Cooper2008ApJ}).
 Thus, as illustrated in Fig.~\ref{fig:outflow_model}, there would usually be a substantial region of the host galaxy that is not directly influenced by the outflow -- and in this region (`Zone B'), CR propagation would presumably operate predominantly by diffusion over kpc scales.
  It follows that CR heating in Zone B would therefore exhibit similar characteristics to the inner region of Fig.~\ref{fig:advection_diffusion_heating_dist}, where the advective flow is too slow to have any important effect on the redistribution of CRs.
  
 We may broadly use the results calculated in~\citet{Owen2018MNRAS} to model the expected CR heating effect in Zone B (and the lower regions of Zone A), where purely diffusive CR transport was invoked for a protogalaxy of otherwise similar specifications to that considered in the present work. 
 In the earlier study, the volumetric CR heating rate was calculated throughout a protogalactic ISM due to its irradiation by the entire CR emission from the galaxy. By contrast, in the current work we assume that CRs cannot readily propagate between the two zones (hereafter the `Two-Zone approximation'), which would mean that the CR irradiation experienced within each zone would be limited to the fraction of the total galactic CR emission which passes into that zone.
 We may evaluate this fraction by considering the solid angle subtended by each of the zones on the central starburst core, i.e. the CR power passing into Zone A would be 
 \begin{align}
 L_{\rm CR, A} &= L_{\rm CR, eff} \frac{2 \Omega_{\rm A}}{4\pi} \nonumber \\
 &=  L_{\rm CR, eff} \left[1-\cos\left(\frac{\theta}{2}\right) \right] 
 \label{eq:cr_zone_a_power}
 \end{align}
 (where the factor of 2 accounts for the bi-polar nature of the outflow) and that into Zone B follows as
  \begin{align}
 L_{\rm CR, B} &= L_{\rm CR, eff} \frac{\Omega_{\rm B}}{4\pi} \nonumber \\
 &= L_{\rm CR, eff} \cos\left(\frac{\theta}{2}\right) 
 \end{align}
 for $L_{\rm CR, eff}$ as the total CR power (see equation~\ref{eq:cr_lum_orig}).
 
 To properly attribute the results of~\citealt{Owen2018MNRAS} at $\mathcal{R}_{\rm SN} = 0.1~\text{yr}^{-1}$ to the CR heating levels expected in Zone B of the present study, a scaling by solid angle in both regions must be used. 
The resulting heating rate per steradian can then be used to compare heating power across the two zones, and is shown for Zone A in blue and Zone B in red in Fig.~\ref{fig:protogalaxy_heating_average}. In Zone B, the CR heating power is only calculated up to around 0.5 kpc, i.e. well within the ISM of the host -- this is because the structure of the magnetic fields connecting the ISM to the circumgalactic medium is unclear and falls beyond the scope of this discussion.
We further include lines for diffuse X-ray heating, heating from stellar radiation and heating from freely streaming CRs (which would arise if no magnetic field were present), as indicated by the relevant legend. These are intended for use as a comparison; for details regarding how these lines are calculated, we refer the reader to~\citealt{Owen2018MNRAS}.
In Fig.~\ref{fig:protogalaxy_heating_average}, it is evident that in Zone B, the CR heating rate is dominant and even exceeds that due to the more conventional radiative processes. 
In Zone A, the CR heating rate is much reduced compared to the radiative heating lines, but still has the potential to play an important role in influencing the thermal properties of the ISM  -- particularly in the diffusion-dominated inner region, where it is maintained at a comparable level to Zone B.
 
 We justify our use of the Two-Zone approximation as follows: we anticipate that the magnetic structure within the outflow would be perpendicular to the plane of the host galaxy and thus perpendicular to the magnetic field orientation within the Zone B ISM region.
  Indeed, such perpendicular magnetic structure in outflows is seen in simulation work where, e.g. the action of a CR-driven dynamo yields a perpendicular magnetic field configuration compared to the host galactic plane~\citep{Kulpa2011ApJ}, or by the advection of the magnetic fields by the flows themselves~\citep{Bertone2005MNRAS}, by magnetic amplification via the CR streaming instability~\citep{Uhlig2012MNRAS} along the outflow.
This magnetic structure would also be consistent with polarised radio synchrotron emission above and below the planes of galaxies known to host outflows in the nearby Universe, with the polarisation direction aligned with the orientation of the outflow cone~\citep[see, e.g.][]{Hummel1988A&A, Sukumar1990IAUS, Sukumar1991ApJ, Hummel1991A&A, Brandenburg1993A&A, Chyzy2006A&A, Soida2011A&A, Mora2013A&A}.
We argue that the principle mechanism for CRs to permeate the Zone A/Zone B interface would be via diffusion. 
With magnetic field lines aligned in a direction parallel to the inter-zone boundary, diffusion across the interface would be severely hampered -- the cross-boundary diffusion coefficient would effectively be perpendicular the the local magnetic field lines, and so would be around two orders of magnitude smaller than that along the field directions~\citep[e.g.][]{Shalchi2004ApJ, Shalchi2006ApJ, Hussein2014MNRAS}, and substantially less than the effective ISM diffusion coefficient.
The detailed substructure of the magnetic fields in these interfacing regions is not yet fully understood~\citep{Veilleux2005ARAA}, but we argue that our prescription is consistent with existing work on relevant scales and that adopting an alternative model for CR transport across this boundary at this point would not imply an interpretation that is any more physical than that adopted here. We acknowledge that, in future studies, it will be critical to assess the magnetic fields in these interfacing regions across a range of length-scales to properly determine the permeability of the Zone A/Zone B interface to diffusing CRs.
 
\subsubsection{Energy Deposition}
\label{subsubsec:energy_deposition}
  
\begin{figure}
	\includegraphics[width=\columnwidth]{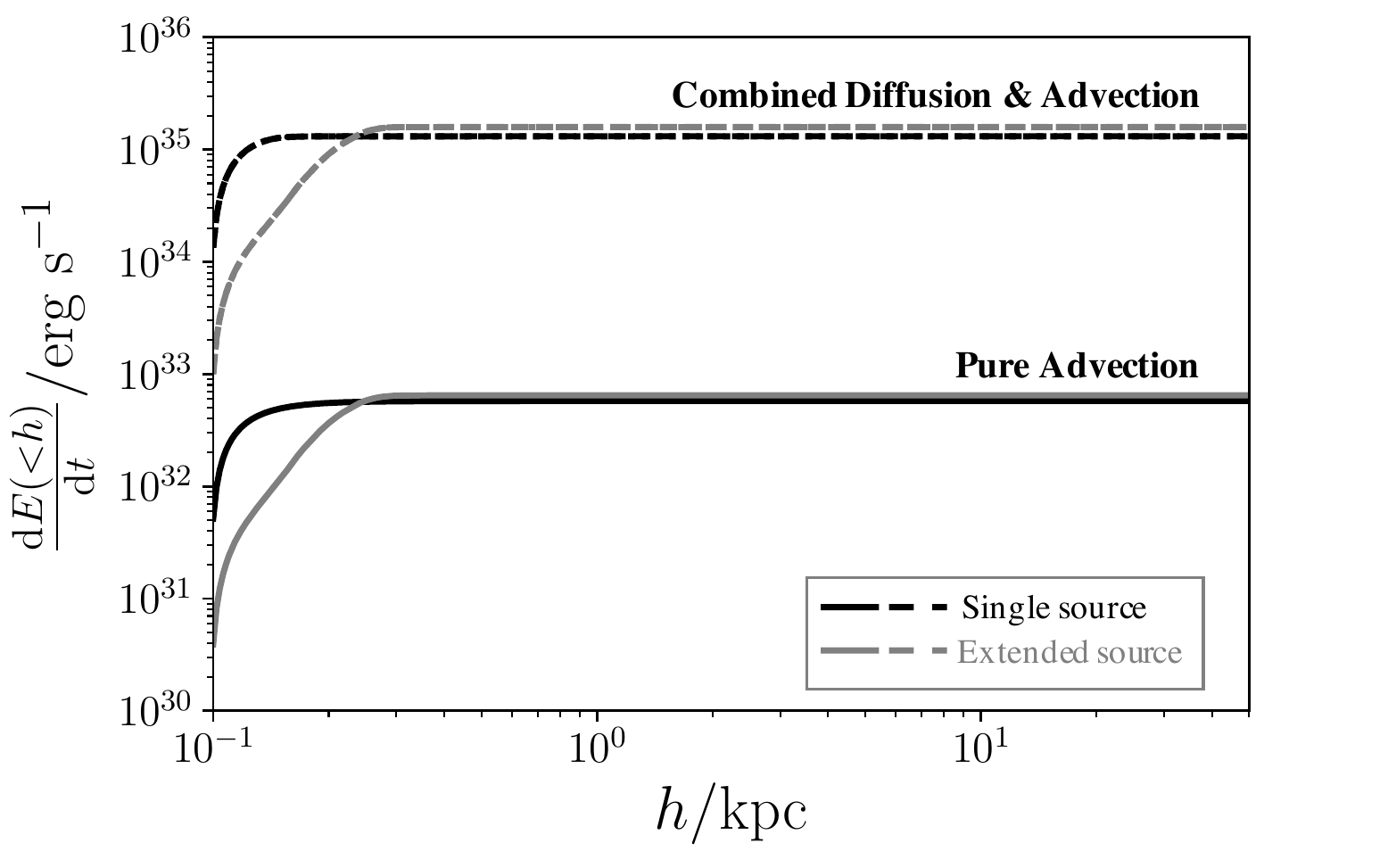}
    \caption{Relative volume-integrated heating profiles of the CRs along the outflow up to 50~kpc. The lower, solid lines show the case for pure advective transport of the CRs, while the upper dashed lines give the result when adopting a concurrent advection/diffusion CR propagation model as discussed in section~\ref{subsubsec:combined_heating_rates}. These combined lines are largely coincident with the result for pure diffusion, because both cases yield the majority of CR heating in the lower regions of the outflow -- advection is only important at greater altitudes, meaning that this integrated heating plot is insensitive to the difference between pure diffusion and combined advection/diffusion.
    The difference between a single boundary condition injection (black lines) and an extended CR source distribution (grey lines) is shown (see section~\ref{subsubsec:extended_inj} for details). Standard parameters have been used ($\mathcal{R}_{\rm SN} = 0.1~\text{yr}^{-1}$; opening angle \ang{55}).}
    \label{fig:total_heating_volume}
\end{figure} 
 
  For a SN event rate of $\mathcal{R}_{\rm SN} = 0.1~\text{yr}^{-1}$, 
  the total CR luminosity due to SN events in the model protogalaxy is $3.0\times10^{41}~\text{erg}~\text{s}^{-1}$ (see equation~\ref{eq:cr_lum_orig}), 
   of which $1.1\times10^{39}~\text{erg}~\text{s}^{-1}$ is passed into the outflow cone (when accounting for the geometry of the system and energy lost by the CRs in driving the outflow), being available to heat the ambient gases via hadronic interactions. 
 In Fig.~\ref{fig:total_heating_volume}), it can be seen that the total integrated heating effect up to 50 kpc -- the characteristic extent of a CR-driven galactic outflow~\citep[see, e.g.][for a discussion on the extent of outflows]{Veilleux2005ARAA, Bland-Hawthorn2007APSS} -- along the cone is around $2.0\times10^{35}~\text{erg}~\text{s}^{-1}$, with much of this energy being deposited within the inner 0.3 kpc (even if the injection of CR energy is extended throughout a starburst region). 
We note that the thermalisation process of CR primaries is relatively inefficient, so their total heating power is much less than the energy released by the CR protons as they undergo hadronic interactions (see section~\ref{sec:thermalisation} for a discussion on the respective loss channels compared to thermalisation). 
For instance, if we account for the power-law spectral energy distribution of CRs and adopt the mean energy of 8.2 GeV as a characteristic value, a CR would thermalise only a fraction of 
 $1.5\times10^{-3}$ of its initial energy on average\footnote{This fraction would vary along the outflow cone due to the density profile of the wind fluid. The value quoted here, and those hereafter, is an average weighted by the heating profile to give a characteristic value for the hadronic heating efficiency along the outflow.}. This means that the total power released by the CRs would be $1.4\times10^{38}~\text{erg}~\text{s}^{-1}$ as they interact, being around 12.0\% of the available CR power in the outflow cone. 
  This fraction is consistent with the level found in the numerical models of~\citet{Girichidis2018MNRAS} (between 5 and 25\%), which consider a combined transport scenario where outflow velocities close to the mid-plane are small, much like the diffusion limit and combined transport picture considered in the present work. 
  \citet{Farber2018ApJ} also considers a pure advection scenario. They find substantial CR energy is harboured within a few kpc of the galactic plane 
   because outflow velocities are low near $h=0$. The advected CRs therefore spend more time in the vicinity of the galactic plane and base of the outflow, and so are more likely to undergo hadronic interactions there. 
  Moreover, the CR number density distributions resulting from the pure advection models of the~\citet{Farber2018ApJ} study suggest a difference of around 4 orders of magnitude between the galactic disk and halo (when they invoke CR diffusion and coupling, the contrast falls to a little under 2 orders of magnitude). When further accounting for the density contrasts between the same locations (around 1 order of magnitude in the advection scenario), the CR hadronic heating rates in the plane and halo at around $h\approx1~\text{kpc}$ would presumably differ by around 5 orders of magnitude -- i.e. similar to the contrast in heating power found in comparable locations in this work.
  By contrast, \citet{Pakmor2016ApJ} find substantial CR energy (of order a few $\text{eV}~\text{cm}^{-3}$) up to heights of 5 kpc or more in similar systems when adopting even lower outflow velocities (around 100$~\text{km}~\text{s}^{-1}$) than those considered here -- however, their calculations focus predominantly on CR transport, and the authors indicate that the hadronic interactions of the CRs had not been included in their simulations.

\subsection{Model Parameters}
\label{sec:model_parameters}

\begin{figure}
	\includegraphics[width=\columnwidth]{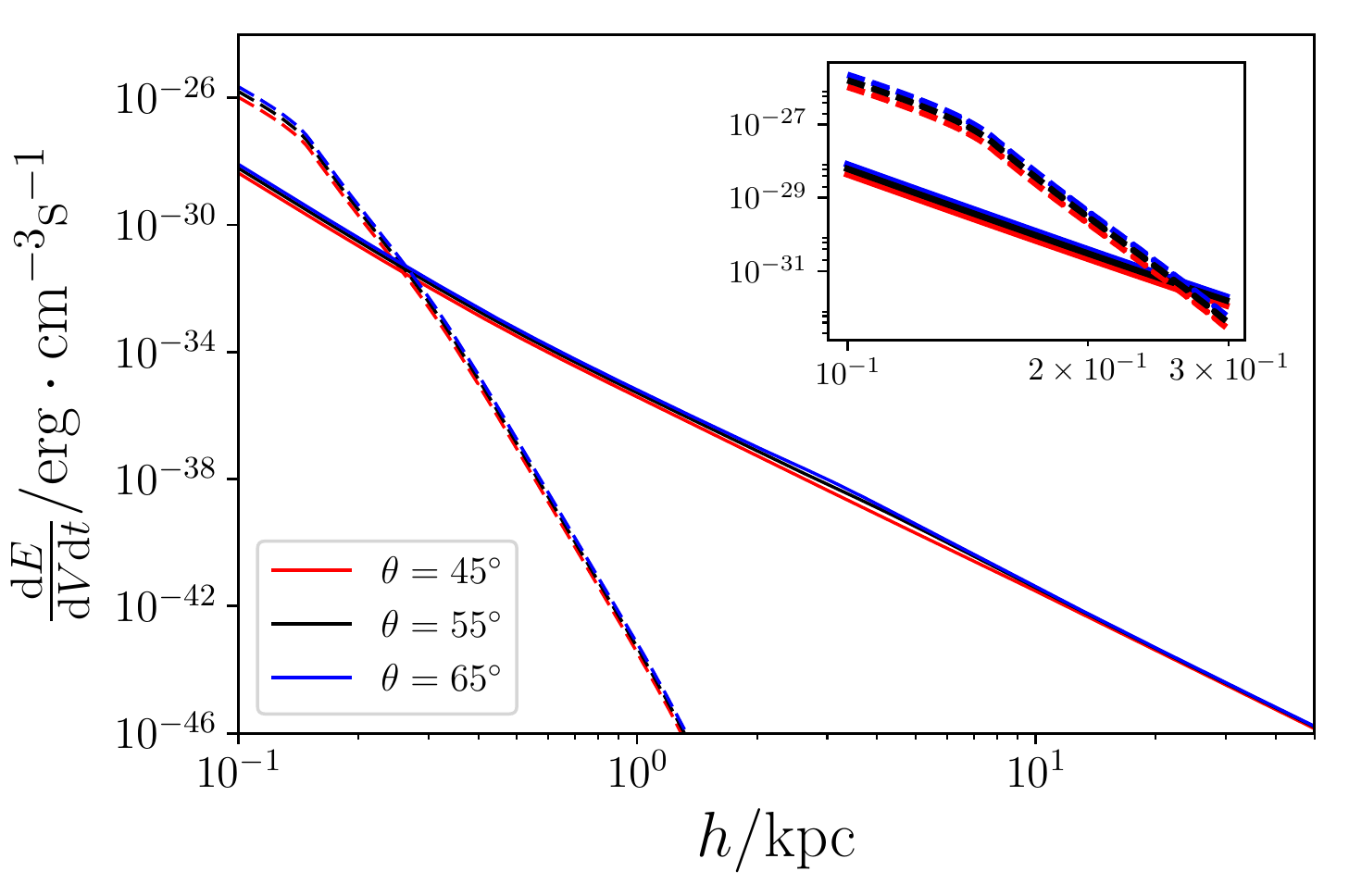}
    \caption{Advection (solid lines) and diffusion (dashed lines) heating rates with opening angles of \ang{45} (red line), \ang{55} (black) and \ang{65} (blue) where variation in the opening angle yields only a minimal change in the result. Up to around 10 kpc, a larger opening angle here increases the solid angle subtended by the cone on the active, star-forming region. This increases the surface over which CRs may be injected into the outflow at the base, and thus proportionally increases the number of CR particles present compared to a smaller opening angle. The inset sub-plot shows a detailed view of the line between 0.1 and 0.3 ${\rm kpc}$.}
    \label{fig:heating_opening_angle}
\end{figure} 

\begin{figure}
	\includegraphics[width=\columnwidth]{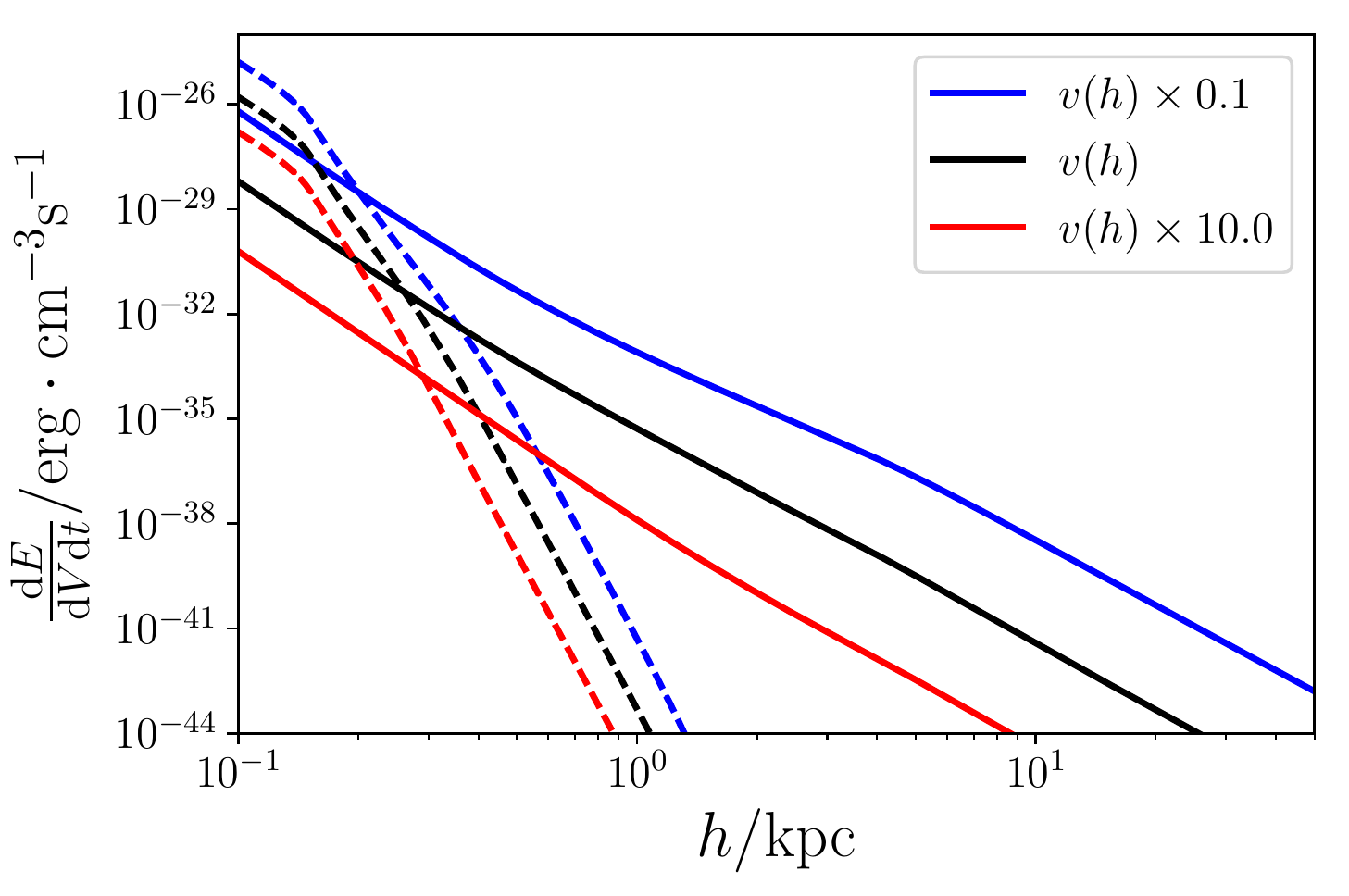}
    \caption{Heating profiles in the diffusion (dashed lines) and advection (solid lines) limits, with outflow profile according to standard parameter choice, yielding a terminal velocity of $v_{\infty} \approx 290~\text{km}~\text{s}^{-1}$ (black line), with velocity increased by a factor of 10, with corresponding decrease in the density profile (red line) and with the velocity decreased by a factor of 10 and corresponding change in density profile (blue line).}
    \label{fig:velocity_effect}
\end{figure}

\begin{figure}
	\includegraphics[width=\columnwidth]{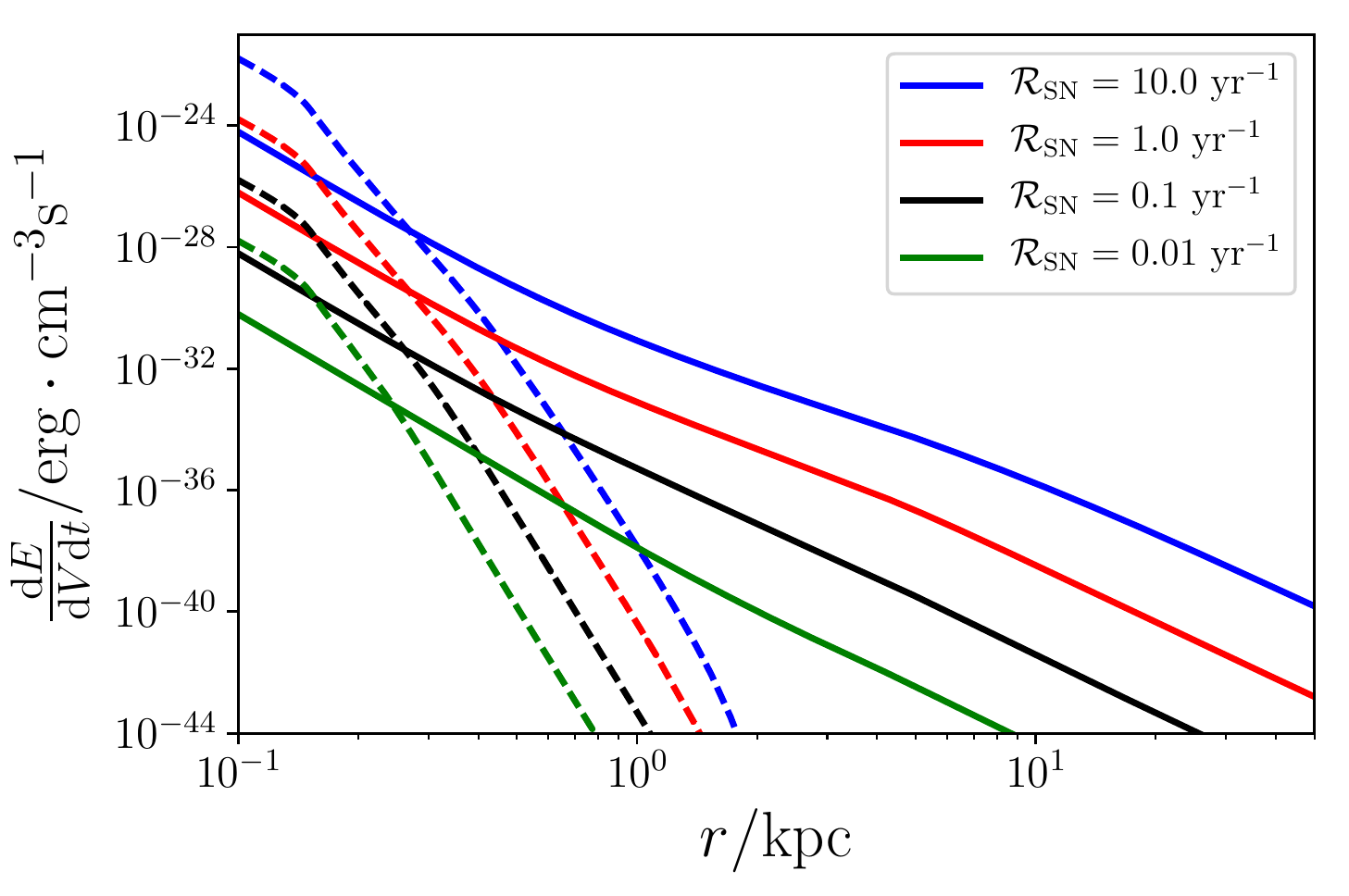}
    \caption{Impact of SN rate on CR heating. We show the profiles in the diffusion (dashed lines) and advection (solid lines) limits, with SN rates $\mathcal{R}_{\rm SN} = 0.01, 0.1, 1.0$ and 10.0 represented by the green, black, red and blue lines respectively. The CR heating rate scales directly with the square of SN rate (see text for further details).}
    \label{fig:sn_rate_heating}
\end{figure}

\begin{figure*}
	\includegraphics[width=\textwidth]{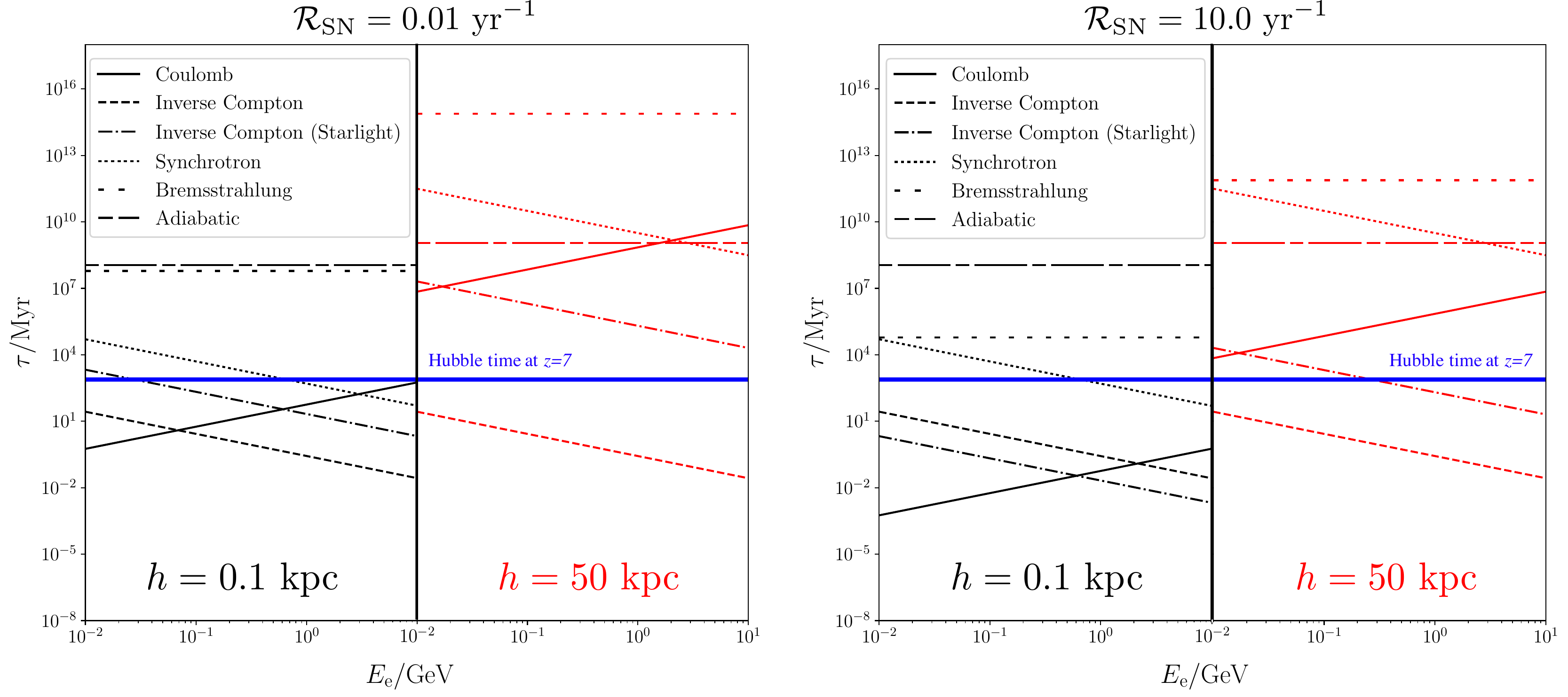}
    \caption{Impact of SN rate $\mathcal{R}_{\rm SN}$ and outflow altitude $h$ on CR secondary loss timescales. The thermalisation process is via Coulomb cooling, so efficiency is greater where the solid Coulomb line is lower. The left panel shows timescales for a relatively quiescent system with $\mathcal{R}_{\rm SN} = 0.01~\text{yr}^{-1}$, while the right panel is for an active system with $\mathcal{R}_{\rm SN} = 10.0~\text{yr}^{-1}$ at both low (black lines) and high (red lines) altitudes. For reference, the Hubble timescale at $z=7$ is shown by the solid blue horizontal line.}
    \label{fig:sn_rate_timescales}
\end{figure*}

We find that the critical quantities which govern the behaviour of the system 
are the SN event rate $\mathcal{R}_{\rm SN}$ and, to a lesser extent, the outflow velocity.
Other parameters bear less importance -- for example the influence of the outflow opening angle is demonstrated in Fig~\ref{fig:heating_opening_angle} where the black line is the result for the standard opening angle of \ang{55} while the blue and red lines show a variation of +\ang{10} and -\ang{10} respectively (i.e. \ang{65}, blue and \ang{45}, red). 
While the impact here is minimal, 
it can be seen that, for larger opening angles, the heating rate is slightly higher: this is due to the greater fraction of the starburst core being subtended by the outflow cone and the associated higher injection rate of CR particles.

It is also of interest to consider the impact of directly modifying the outflow velocity. This is because different wind-driving mechanisms can yield vastly different outflow rates~\citep{Veilleux2005ARAA}.
To ensure a reasonably self-consistent estimate, the density profile must be scaled in conjunction with the velocity:
from equation~\ref{eq:continuity_eq}, the product $\rho \;\! v(h) \;\! h^2$ is conserved. So, for an equivalent mass injection rate, a velocity scaled by a factor $\Psi$ would require a corresponding scaling of the density profile by factor $\Psi^{-1}$. 
We note that this also results in a change in the associated mechanical energy of the wind, which is also scaled by a factor of $\Psi$ as follows from $\rho\;\! v_{\infty}^2$. 
The effect of scaling the system in this way is shown in Fig.~\ref{fig:velocity_effect}, with the result essentially following directly from the initial condition for the CR energy density at the base of the outflow combined with the variation in the outflow density profile (which results from scaled velocity profile). 
In equation~\ref{eq:energy_density_adv} it can be seen that the CR number density, and hence any heating effect the CRs may drive, is inversely proportional to $v_{\infty}$. 
Thus, if the velocity is reduced by an order of magnitude, the CR number density increases by an order of magnitude. 
However, the density of the outflow would also increase by an order of magnitude in such a scaling. Overall, this would increase the CR heating power in the advection limit by two orders of magnitude as it is proportional to both the CR number density and the density of the outflow wind fluid (which provides the target hadrons for the CR heating effect).
In the diffusion limit, the picture is more straightforward: this time there is no dependence of the CR number density on the scaling, so the the CR heating power would only increase by one order of magnitude.
We find that there is little bearing on the fraction of CRs absorbed in the outflow cone when a scaling is applied: when reducing the flow velocity by a factor of 10, the absorbed fraction falls slightly to 11.9\% (compared to 12.0\% in the baseline case -- see section~\ref{subsubsec:energy_deposition}), while the fraction increases to 13.7\% if increasing the outflow velocity by the same factor.

The most directly influential parameter is the SN event rate, $\mathcal{R}_{\rm SN}$, which essentially specifies the energy budget of the system. 
The CR heating rates scale with $\mathcal{R}_{\rm SN}^2$, as seen in Fig.~\ref{fig:sn_rate_heating}, where a plausible range of values from $0.01~\text{yr}^{-1}$ (fairly quiescent, barely star-bursting system) to $10.0~\text{yr}^{-1}$ (extremely violent star-forming environment, perhaps possible in a system undergoing a major merger\footnote{For example, some studies have suggested that Arp 220 could have a SN event rate of 2-4$~\text{yr}^{-1}$~\citep{Lonsdale2006ApJ, Varenius2017arXiv}. In~\citet{Lonsdale2006ApJ}, 4 new radio sources were observed in 12 months, and the authors argue that this is consistent with 4 new SN events and a corresponding SN rate, although this is based on limited statistics. The more recent study by~\citet{Varenius2017arXiv} suggests a rate of $4\pm 2~\text{yr}^{-1}$ on the basis of the number of events observed, and by assuming only a small fraction of SN events are observable. So, while an event rate of 4$~\text{yr}^{-1}$ may be possible, the true level is likely to be lower than this~\citep{Page2018}. We adopt the maximum value of 10$~\text{yr}^{-1}$ as an extreme case in the present work for completeness, but acknowledge this should be treated very much as an upper-limit.}) are explored.
Both the density of the outflow wind fluid (being the target for the hadronic interactions) and CR emission 
are dependent on $\mathcal{R}_{\rm SN}$. The CR heating power depends on the product of these two quantities, and so the dependence on $\mathcal{R}_{\rm SN}^2$ follows from this.
The 12\% fraction of CR proton energy absorbed in the outflow by hadronic interactions (as calculated in section~\ref{subsubsec:energy_deposition}) is independent of SN event rate, as is the outflow velocity profile. In both cases, this is because the contributing quantities which are dependent on $\mathcal{R}_{\rm SN}$ scale antagonistically. 

Fig.~\ref{fig:sn_rate_timescales} shows the timescales for various loss processes of secondary CR electrons near the base of the outflow at $h=0.1~\text{kpc}$ compared to the upper regions $h=50~\text{kpc}$. 
Since the microphysics which affect these secondary electrons is instrumental in governing the energy fraction of the CR primaries which can thermalise, it is useful to understand why the efficiency of CR hadronic heating may vary depending on the environment: 
the left panels (where lines are plotted in black) show that, when changing from a SN event rate of $\mathcal{R}_{\rm SN} = 0.01~\text{yr}^{-1}$ to $\mathcal{R}_{\rm SN} = 10.0~\text{yr}^{-1}$, the intensity of starlight should increase so as to proportionally reduce the inverse Compton timescales. In the absence of an outflow, this would lead to greater inverse Compton losses (due to scattering off a more intense stellar radiation field), a lower CR thermalisation efficiency and a resulting smaller increase in CR heating rate than a direct proportionality with $\mathcal{R}_{\rm SN}$ would suggest~\citep[see][]{Owen2018MNRAS}.
However, in the presence of an outflow, the density profile is enhanced by greater mass-loading rates from the SN winds, thus yielding an environment which is better able to thermalise CR secondaries. Moreover, this denser environment promotes hadronic interactions to boost CR heating. 
This is sufficient to roughly maintain the proportionality seen in Fig.~\ref{fig:sn_rate_heating}, even though inverse Compton losses are more severe in more intense starlight (when $\mathcal{R}_{\rm SN}$ is greater).
The underlying complexity of this behaviour also accounts for the slight deviation from a direct and consistent scaling between the lines in the advection limit at high altitudes in  Fig.~\ref{fig:sn_rate_heating}.
While most of the processes in Fig.~\ref{fig:sn_rate_timescales} operate at an appreciable rate (i.e. with timescales much shorter than the Hubble timescale) at the base of the outflow, only inverse Compton losses off the uniformly distributed CMB photons (and some starlight inverse Compton for the highest energy particles when the SN event rate is sufficiently high) are important at high altitudes. This demonstrates that the higher in altitude the CRs reach within an outflow, the less likely they are to thermalise. Indeed, above a few kpc, it can safely be assumed that the CRs have been fully advected out of their host galaxy and would instead impart their effects on the external environments (e.g. in the circumgalactic and inter-galactic medium).

\subsection{Implications of Cosmic Ray Heating Power in and around Protogalaxies}

The CR heating power due to hadronic interactions could be as high as $10^{-25}~\text{erg}~\text{cm}^{-3}~\text{s}^{-1}$ in a protogalaxy with a SN-rate $\mathcal{R}_{\rm SN} = 0.1~\text{yr}^{-1}$, if outflows or galactic winds do not operate, such that strong CR containment within the host galaxy arises~\citep{Owen2018MNRAS}. 
We find that, if outflows are present, in the lower regions of the outflow cone (where CR diffusion still dominates over advection) a comparable heating power of $10^{-26}~\text{erg}~\text{cm}^{-3}~\text{s}^{-1}$ can also be attained.
In the pure advection limit (i.e. when neglecting diffusion), a comparatively reduced CR heating rate is redistributed along an outflow cone compared to the diffusion limit, while a substantial fraction of the CRs do not engage with hadronic interactions in the outflow wind at all, instead escaping from the system entirely. These CRs may deposit their energy over much greater regions in the vicinity of the source galaxy and beyond. This could affect pre-heating and ionisation processes in the wider Universe~\citep{Nath1993MNRAS, Sazonov2015MNRAS, Leite2017MNRAS}, or alter the dynamics of the circumgalactic medium.

\subsubsection{Cosmic Ray Containment}

In the diffusion limit, CRs can become contained within the ISM of their host galaxy where they deposit a substantial fraction of their energy. This is consistent with $\gamma$-ray observations of nearby starbursts - e.g. M82 - which are $\gamma$-ray bright, suggesting that some non-negligible fraction of the CRs interact to produce pion decay $\gamma$-ray emission within the galaxy, rather than being advected away~\citep[see, e.g.][]{Abdo2010A&A, Wang2014AIPC, Yoast-Hull2015MNRAS, Heckman2017arXiv, Wang2018MNRAS}. Since CR heating preferentially targets denser regions of the ISM (due to the particles of the interstellar gas being the principal targets in CR hadronic interactions), this could severely affect the ability of molecular clouds and cores to collapse into stars by raising their effective Jeans' mass. This would push the initial mass function of forming stars to a more top-heavy form -- or could even quench star-formation entirely, leading to a quiescent period until the ISM gas has sufficiently cooled to allow star-formation to resume. While the magnetic fields of star-forming regions would presumably also influence the local level of CR heating experienced, it is difficult to envisage a situation where magnetic shielding can act to such a level that there would be no impact by CR heating at all -- indeed, the opposite effect may be true if magnetic field vectors are arranged in a way so as to preferentially direct diffusing CRs into denser, star-forming regions. 
We leave such detailed modelling to future work. 

\subsubsection{Cosmic Ray Escape}
\label{sec:cr_escape}

The picture is very different in the advection limit: in section~\ref{subsubsec:energy_deposition}, we find that around 12\% of the total CR power within an outflow can be absorbed by hadronic processes, 
with the remaining 88\% effectively being transported into the circumgalactic medium (and beyond). Although the exact fraction would vary depending on the details of the model adopted and parameter values used, this result would suggest that a substantial fraction of CR energy in an outflow is actually able to be transported away by advection if caught up by the flow. Moreover, our discussion in section~\ref{sec:model_parameters} would indicate that this picture is not strongly sensitive to the choice of model parameters, with the majority of CRs in an outflow cone escaping for any reasonable parameter choices.
In terms of energetics, equation~\ref{eq:cr_zone_a_power} indicates that around 11.2\% of the total CR luminosity of a host galaxy would pass into an outflow (with opening angle \ang{55}). With the 12\% hadronic absorption fraction of this arising within the cone, it follows that around 10\% of the total CR luminosity of the source galaxy would escape into the circumgalactic and/or intergalactic medium -- a CR power that would correspond to around $3.1\times 10^{40}~\text{erg}~\text{s}^{-1}$ at a SN event rate of $\mathcal{R}_{\rm SN} = 0.1~\text{yr}^{-1}$.  This represents a substantial contribution to the exterior energy budget around a starburst galaxy, contributing a power greater than the expected diffuse X-ray emission (at around $10^{38}~\text{erg}~\text{s}^{-1}$ -- see, e.g.~\citealt{Watson1984ApJ}) and is an intriguing result: while an outflow can facilitate the advection of substantial levels of CR energy into the extragalactic environment, this would only correspond to around 10\% of the total CR energy being provided to the system by SN events. 
  Thus, even in the presence of outflows, the CR calorimetric ability of starburst galaxies is likely to still remain relatively effective with a large fraction of CR energy still being contained and deposited within the ISM. This contained fraction would perhaps be slightly enhanced shortly after the onset of star formation before sustained outflow activity 
 has had time to develop, or when star formation is bursty and distributed throughout the galactic disk 
 so as to prevent the formation of a concentrated star-forming region in the central galaxy core (this would be required to drive a galactic-scale outflow). In terms of CR feedback effects, this would represent a `best of both worlds' scenario with CRs being important both inside and outside their host galaxy when large-scale outflows begin to emerge.

In protogalaxies, CRs have a greater ability to couple with their ambient medium than radiation through mechanisms not open to photons, e.g. hadronic processes or magnetic scattering. We have seen that one of the implications of these advected CRs includes the elevated external heating effect calculated in this paper, up to a level of around $10^{-34}~\text{erg}~\text{cm}^{-3}~\text{s}^{-1}$, even when several of kpc away from their source galaxy. 
This opens up new questions about the impacts of such an effect on, e.g. pre-heating for cosmic reionisation~\citep{Sazonov2015MNRAS, Leite2017MNRAS} and the ability for such a process to be maintained. Moreover, advected CRs may have a role in amplifying intra-cluster and intergalactic magnetic fields, e.g. through resistive generation~\citep[see][for application of this process to escaping CRs from high-redshift protogalaxies and clusters at the cosmic dawn]{Miniati2011ApJ, Miniati2012ASPC, Beck2013MNRAS_B, Lacki2015MNRAS}, and/or by driving the growth of non-resonant magnetohydrodynamical instabilities in weak, pre-existing seed magnetic fields~\citep[see][]{Bell2004MNRAS, Miniati2011ApJ, DAngelo2015ICRC, Samui2018MNRAS}. There may, however, be even more important impacts than this.

Clusters, proto-clusters, and groups/pairs of protogalaxies are supported by outward pressure gradients against their gravitational potentials, essentially in hydrostatic equilibrium~\citep[see, e.g.][]{Suto2013ApJ, Biffi2016ApJ}. In low-redshift systems this is thought to be dominated by gas pressure from hot, thermal intergalactic baryons. However, at higher redshifts (above $z\gtrsim 1$, see~\citealt{Lacki2015MNRAS}) when star-formation rates (and hence SN rates) in the Universe were greater than in the current epoch, and substantial CRs may have been able to escape from their host environments, CR pressure could be begin to dominate~\citep[see, e.g.][]{Ginzburg1963SvA, Lacki2015MNRAS, Butsky2018arXiv}. In a pair of protogalaxies, this could have the effect of pushing neighbouring galaxies away, and this could alter the distribution of highly star-forming protogalaxies hosting outflows (e.g. perhaps those observed as Lyman-$\alpha$ emitters) compared to their more quiescent counterparts. A CR-advecting outflow may also have the ability to heat external gas and thus reduce (or even halt) gas inflow to the host galaxy. In turn, this may hamper star-formation by depriving the system of the cool, inflowing gases potentially responsible for driving the starburst phase~\citep[][]{Keres2005MNRAS, Dayal2013MNRAS, Lu2015MNRAS, Toyouchi2015ApJ, Yabe2015ApJ, Dayal2018PhR}, and this would lead to quenching or a reduction of the star-formation rate of the system by `strangulation'~\citep[e.g.][]{Peng2015Natur} -- see also the following section~\ref{sec:observation_comparison}, where we demonstrate the impact CR containment and subsequent outflow advection may have in the recently observed high-redshift galaxy MACS1149-JD1~\citep{Zheng2012Natur, Hashimoto2018Nat}.

\section{Application to the High-Redshift Starburst Galaxy MACS1149-JD1}
\label{sec:observation_comparison}

\begin{figure*}
	\includegraphics[width=0.9\textwidth]{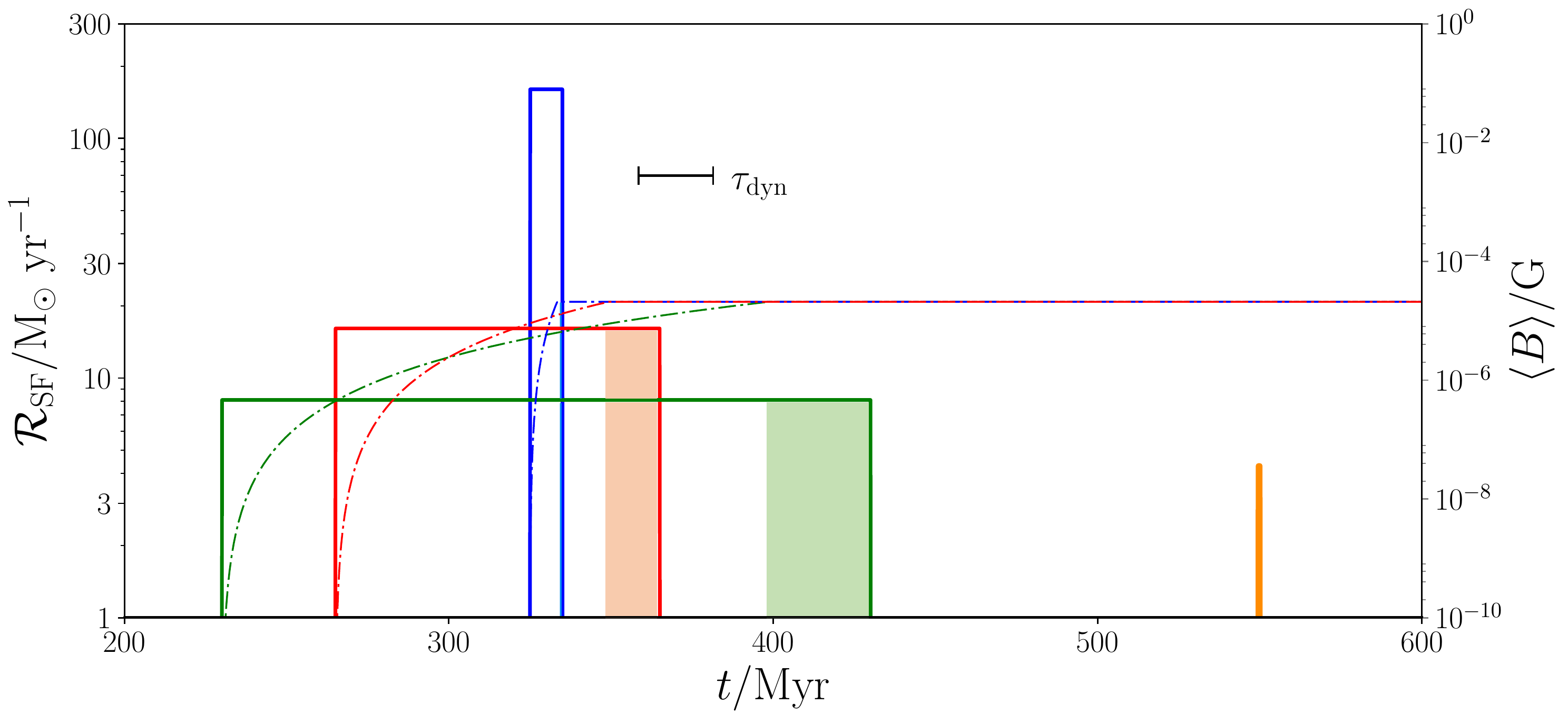}
    \caption{Plot to show the co-evolution of the star-formation history and magnetic field in the high-redshift star-forming galaxy MACS1149-JD1. CR containment and heating power would also co-evolve with the magnetic field strength. The three star-formation history models of~\citealt{Hashimoto2018Nat} are shown as 
 (1) a low-intensity star-formation phase, rate $\mathcal{R}_{\rm SF} \approx 8~\text{M}_{\odot}~\text{yr}^{-1}$ for a duration of 200 Myr (green solid line); 
 (2) a more active burst of star-formation, rate $\mathcal{R}_{\rm SF} \approx 16~\text{M}_{\odot}~\text{yr}^{-1}$ for a duration of 100 Myr (red solid line); and 
 (3) an extreme short and intense starburst at a rate of $\mathcal{R}_{\rm SF} \approx 160~\text{M}_{\odot}~\text{yr}^{-1}$ for a duration of 10 Myr (blue solid line). 
  The inferred star-forming activity of the galaxy of $4.2^{+0.8}_{-1.1}~\text{M}_{\odot}~\text{yr}^{-1}$ at $z=9.11$ is shown by the solid orange line. Star-formation rates $\mathcal{R}_{\rm SF}$ are indicated by the y-axis on the left. Corresponding magnetic field evolution models following~\citealt{Schober2013A&A} are shown by the dashed-dotted lines, with magnetic strength indicated by the y-axis to the right of the plot. The shaded region beneath the model lines indicate the period during which the magnetic field has fully saturated, before star-formation has been quenched. If CR heating is sufficiently prolonged, star-formation could be quenched within the system's dynamical time-scale of $\tau_{\rm dyn} \approx 23~\text{Myr}$, as indicated by the scale-line.}
    \label{fig:star_formation_histories}
\end{figure*} 

A spectroscopic analysis of the doubly-ionised oxygen forbidden O\;III line, at 88$\mu$m\footnote{This line is found to be particularly strong in chemically un-evolved environments, being a useful tracer of star-formation in metal-poor young star-forming galaxies~\citep[e.g.][]{DeLooze2014A&A, Cormier2015A&A} including those observed at high-redshift -- 
   see~\citet{Inoue2016Sci, Rigopoulou2018MNRAS}.}, of the high-redshift galaxy MACS1149-JD1 
   was conducted by \citet{Hashimoto2018Nat}, 
   obtaining a redshift of $z=9.11$.      
This corresponds to a time when the Universe was around 550~Myr old~\citep[][]{Wright2006PASP, Hashimoto2018Nat}. 
From the O\;III emission line,  
  \citet{Hashimoto2018Nat} also deduced  
  a star-formation rate of $\mathcal{R}_{\rm SF} = 4.2^{+0.8}_{-1.1}~\text{M}_{\odot}~\text{yr}^{-1}$ at the observational epoch.  
The wider spectrum and the Balmer break revealed an additional, older population of stars 
  reminiscent of strong starburst activity a few hundred Myr earlier, 
  implying MACS1149-JD1 had experienced two distinguishable star-formation episodes by the time it was observed. 
A two-phase star-formation history has also been found in other nearer galaxies~\citep{Dressler1983ApJ, Couch1987MNRAS, French2015ApJ}.   
It requires a mechanism to rapidly switch the galaxy from a strong starburst state into a quiescent state.  
The quenching of starbursts in galaxies is believed to be caused by feedback, 
  which halts the gas inflow and/or gravitational collapse of over-dense gases  
  (e.g. via turbulence or heating of the ambient gas, as suggested in~\citealt[][]{French2018ApJ}). 
  Given that abundant molecular gas reservoirs, which are able to fuel star-formation, are found to be present 
  in over half of quenched post-starburst systems studied, a lack of dense gas cannot account for the quenching. Instead, star-formation must be prevented by some other mechanism~\citep{French2015ApJ, Rowlands2015MNRAS, Alatalo2016ApJ}. 

\cite{Hashimoto2018Nat} presented three scenarios for the star-formation history of MACS1149-JD1 
  to account for the observed spectra. 
  These begin respectively at $z=17.0, 15.4$ and $13.3$ 
  with duration and rates corresponding to a total stellar mass of the galaxy of $1.1^{+0.5}_{-0.2} \times 10^9~\text{M}_{\odot}$ by $z=9.11$.  
All scenarios allow a quiescent period of at least 100~Myr between the end of the first starburst phase 
  and the beginning of the later star-formation phase observed at $z=9.11$. 
The scenarios are distinguished by their star-formation rate and the duration of their star-formation episode:  
  (1) a slow star-formation rate $\mathcal{R}_{\rm SF} \approx 8~\text{M}_{\odot}~\text{yr}^{-1}$ for a duration of 200~Myr; 
  (2) a moderate star-formation rate $\mathcal{R}_{\rm SF} \approx 16~\text{M}_{\odot}~\text{yr}^{-1}$ for a duration of 100 Myr; 
  and (3) a shorter, but more intense star-formation rate of $\mathcal{R}_{\rm SF} \approx 160~\text{M}_{\odot}~\text{yr}^{-1}$ for a duration of 10 Myr. 
The termination time of the star-formation episode was not firmly specified. 
 
 \subsection{Termination of Star Formation}
 
We propose that the first starburst episode revealed by the \cite{Hashimoto2018Nat} observation 
  was terminated by progressive CR heating, 
  causing a feedback process.    
The rapid evolution of the massive metal-poor stars gave rise to SNe~\citep[see][]{Abel2002Sci, Bromm2002ApJ, Clark2011Sci},  
  which produced shocks and injected turbulence into the ISM, and allowed for the acceleration of CRs.   
The shocks and turbulence also facilitated the growth of the galactic magnetic field via a turbulent dynamo mechanism \citep{Balsara2004ApJ, Balsara2005ApJ, Beck2012MNRAS, Schober2013A&A} in MACS1149-JD1.  
Initially, the magnetic field was not strong enough and energy carried by CRs was mainly transported to the IGM instead of deposited into the ISM. 
As the starburst progressed,
  magnetic field amplification continued, eventually reaching saturation.  
By that time, the magnetic field had not only attained suitable strength to trap sufficient amounts of CRs within the galaxy 
   (despite a fraction being advected out), 
   but could also sustain a prolonged period of CR heating of power above all other radiation sources of stellar origin \citep{Owen2018MNRAS}. 
The sequence of these processes should operate on a timescale comparable to the duration of the star-formation episode, 
  otherwise the star-formation would not have continued to proceed. 
This sets the requirement that the longest possible timescale would determine the progression of the heating process that eventually 
  led to quenching.   
A heuristic argument points out that the timescale on which the magnetic field evolves to saturation 
   and the dynamical response to the sustained CR heating would be comparable. 
Thus, the starburst episode would have two stages: 
  (i) an initial magnetic field growth stage and (ii) a subsequent delayed quenching stage due to sustained CR heating.  
In the magnetic field growth stage, 
  CRs are contained 
 as the magnetic field increases in strength (even when it has not fully saturated). As such, a prolonged sustained CR heating effect is established. 
At this point, the infall of cold gas has not been shut down, so star formation proceeds.
  In turn, this continues to drive the magnetic field amplification. 
  
The termination of star-formation can only arise when there is no remaining avenue by which it can be sustained at any appreciable rate.
This is achieved by the internalised CR heating of the system, as well as the strangulation of the cold inflows~\citep[e.g.][]{Peng2015Natur}.
The internal CR heating stage commences when the magnetic field amplification nears its saturation strength, and the ISM is heated for a prolonged period. 
Although the internal thermodynamics of the galaxy are modified accordingly, 
  the gas inflows fuelling star-formation will not cease instantaneously.  
  To halt the inflows, we argue that CRs are advected by outflows resulting from bust-like star-formation throughout the host galaxy.
    If MACS1149-JD1 had a starburst phase star-formation rate of $\mathcal{R}_{\rm SF} \approx 16~\text{M}_{\odot}~\text{yr}^{-1}$, this would correspond to a SN event rate of $\mathcal{R}_{\rm SN} = 0.1~\text{yr}^{-1}$ (see section~\ref{subsec:wind_velocity}), which would cause the magnetic field to saturate at $\mu$G levels within around 140 Myr~\citep{Schober2013A&A, Owen2018MNRAS}.
  In section~\ref{sec:cr_escape}, we showed that the advected CR power would be of order $10^{40}~\text{erg}~\text{s}^{-1}$ for such a system, being typically around 10\% of the available total CR luminosity from the SN events, and this would predominantly be injected into the circumgalactic environment where the CRs would preferentially interact with the high-density cold inflowing filaments to develop an appreciable CR heating power -- possibly comparable in strength to that experienced in the ISM of the host galaxy (possibly as high as $10^{-25}~\text{erg}~\text{cm}^{-3}~\text{s}^{-1}$). 
  
  This additional heating effect would raise the temperature of the inflowing gases to a level where they cannot drive star formation effectively and begin to evaporate.
  We may use the virial theorem to estimate the timescale over which filamentary inflows may halted in this manner: 
  the virial temperature is the temperature above which gravitational collapse of gas is halted and, presumably, any heating to temperatures above this level would lead to evaporation.
  If assuming inflows persist over filaments of lengths of up to 50 kpc~\citep{Dekel2009Natur, Stewart2013ApJ, Goerdt2015MNRAS, Dayal2018PhR} and that they are the sole driver of the star-formation activity arising at a rate of $\mathcal{R}_{\rm SF} \approx 16~\text{M}_{\odot}~\text{yr}^{-1}$ with a 30\% mass conversion efficiency~\citep{Turner2015Natur, Meier2002AJ, Behroozi2015ApJ, Sun2016MNRAS} with an inflow velocity of 400 km s$^{-1}$ (from the velocity offset of the Lyman-$\alpha$ line in MACS1149-JD1 -- see~\citealt{Hashimoto2018Nat}), the steady-state mass of these inflows would be around $6.7\times10^{7}~\text{M}_{\odot}$. If the typical diameter of these flows is similar to the galaxy which they feed, i.e. 1 kpc, this would give a virial temperature of 
  $T_{\rm vir} \approx 5,000~\text{K}$~\citep{Binney2008book}.
  If adopting a number density of $10~\text{cm}^{-3}$ for the filamentary inflows (i.e. comparable to the mean density of the ISM in the host galaxy), the estimated CR heating rate of $10^{-25}~\text{erg}~\text{cm}^{-3}~\text{s}^{-1}$ would suggest it would take of order only a few Myrs for the virial temperature of the inflows to be exceeded.
  At this point, the supply of gas to the galaxy would be strangulated, halting star-formation within a dynamical timescale $\tau_{\rm dyn}$ (being the time required for the system to respond to the strangulation and internal heating).
 Since the strangulation and ISM quenching timescales are comparatively short, the magnetic saturation and dynamical timescales alone specify the timescale over which star-formation would be quenched. 
     
In Fig.~\ref{fig:star_formation_histories} the three star-formation scenarios for the first starburst phase proposed by \cite{Hashimoto2018Nat} are shown. 
Along with these, the corresponding evolution of the galactic magnetic field as in \cite{Owen2018MNRAS},  
  using the parameters obtained by \cite{Hashimoto2018Nat} for MACS1149-JD1 and 
  following the prescription of \cite{Schober2013A&A} are also shown. 
The dynamical timescale for galaxies similar to MACS1149-JD1 is $\tau_{\rm dyn} \approx \sqrt{3\pi/(16\;\!G\rho)}\approx 23~\text{Myr}$\footnote{This is calculated by assuming a characteristic ISM density of $10~\text{cm}^{-3}$ rather than using the dynamical mass in \citet{Hashimoto2018Nat}, which has large uncertainties through their lensing parameter $\mu$}.   
Imposing the requirement of a delayed response to the quenching of star-formation after CR containment is attained (as described above)
  rules out the most intense starburst scenario among those proposed by \cite{Hashimoto2018Nat}.  
It also sets an upper limit for the star-formation rate of below $20~\text{M}_{\odot}~\text{yr}^{-1}$ 
   and a latest limit of before 260~Myr (i.e. at redshift $z \approx 15.4$) for the initial star-formation episode in this galaxy.
   
\subsection{Reinstatement of Star Forming Activity}
   
The cause of the resumption of star-formation after a quiescent period of 100~Myr remains to be explained.   
One of the possibilities is the eventual cooling of the hot circumgalactic medium. 
If strong CR heating were sufficient to cause the evaporation of cold inflows, such filaments could only start to be re-instated once the CR emission for the galaxy is diminished.
 Presumably this would arise some time after the end of star-formation, once any outflow activity and substantial CR production had abated.
 
Prior to the end of star-formation, the large amount of advected CRs, ultraviolet (UV) radiation from the young stars and/or X-rays from the stars and their remnants 
  in a starburst galaxy~\citep[see][]{Hashimoto2018Nat} 
  could heat and ionise the circumgalactic medium and the IGM.  
Thus a hot, ionised bubble would be carved out around the galaxy. 
Without a supply of cold gas, star-formation in the galaxy would be quenched~\citep{Peng2015Natur}. 
 \citet{Hashimoto2018Nat} considered a uniform IGM and UV escape fraction of 20\% 
  and estimated that a \cite{Stromgren1939ApJ} sphere up to a radius 0.4 Mpc from MACS1149-JD1  
  would be created by the stellar UV radiation alone. 
If the observed later star-formation episode 
  were re-ignited by the inflow of cold, neutral IGM gas from outside this Str\"{o}mgren sphere,
  a free-fall velocity of around $4,000~\text{km}~\text{s}^{-1}$ (i.e. about 1\% of the speed of light)   
  would be required in order to reach the galaxy in the required time (100 Myr).  
This speed is excessive, given that an inflow velocity of only a few hundred ${\rm km}~{\rm s}^{-1}$ was obtained 
   from the measurement of a blueshift in the Lyman-$\alpha$ line compared to the O\;III 88-$\mu$m rest-frame 
   \citep[see also][for further details and applications of this technique to detect inflows]{Dijkstra2006ApJ, Verhamme2015A&A}. 

However, if the gas within the \cite{Stromgren1939ApJ} sphere around MACS1149-JD1 
  could be cooled sufficiently, 
  its inflow would fuel the star-formation process after the quiescent period. 
The thermal free-free cooling timescale of a hot ionised gas is roughly given by 
\begin{equation} 
  \tau_{\rm cool} \approx 100 \;\! \left(\frac{n_{\rm e}}{10^{-2} \;\! {\rm cm}^{-3}} \right)^{-1} \left(\frac{T_{\rm e}}{10^5 \;\! {\rm K}} \right)^{1/2} {\rm Myr} \ ,   
\end{equation} 
  where $n_{\rm e}$ is the electron number density and $T_{\rm e}$ is the electron temperature. 
Taking $n_{\rm e} = \langle n_{\rm H} \rangle \approx 10^{-2} \;\! {\rm cm}^{-3}$ (where $n_{\rm H}$ is the number density of the medium)
  the value inferred from the \cite{Stromgren1939ApJ} sphere estimation in \citet{Hashimoto2018Nat} 
  and adopting an electron temperature $T_{\rm e} =10^5 \;\! {\rm K}$ 
  gives a cooling time equal to the duration of the quiescent period inferred for MACS1149-JD1.  
Although, in reality, $T_{\rm e}$ could be higher than $10^5 \;\! {\rm K}$ in a gas photo-ionised by UV photons and/or keV X-rays, 
  the cooling times would be shortened if the electron density increases. 
Ionised dense clumps and filaments embedded within the ionised circumgalactic medium or IGM could be cooled more quickly, and  
  these could provide the gas reservoir required to refuel subsequent star-formation. 
While the Str\"{o}mgren sphere prescription would hold in the low-density regions between filaments 
   with the H\;\!II region still extending to Mpc scales, 
   the extent of the H\;\!II region would be much reduced in the direction of the clumps and filaments. 
For an ionisation distance in the direction of the filaments of around 10\% of the Str\"{o}mgren radius 
 of MACS1149-JD1, 
 an inflow velocity of around 400 km s$^{-1}$, consistent with the velocity inferred from the offset of the Lyman-$\alpha$ line, 
  would be sufficient to account for the re-ignition of star-formation activity within 100 Myr. 

\section{Conclusions}
\label{sec:conclusions}

The heating power derived from CRs undergoing hadronic interactions has the potential to exceed that due to radiative heating processes from stellar and diffuse X-ray emission by one or two orders of magnitude, with CR heating reaching powers of around $10^{-25}~\text{erg}~\text{cm}^{-3}~\text{s}^{-1}$ (when adopting a SN-rate $\mathcal{R}_{\rm SN} = 0.1~\text{yr}^{-1}$). In this work, we have shown that this remains the case, even when a strong galactic outflow develops. A galactic outflow is able to remove around 10\% of the CRs from the ISM overall in a model protogalaxy, although this advection of CRs is predominantly focussed within the cone of the outflow itself. This points towards a `Two-Zone' picture: one region which is predominantly advective, where the CRs are transported by the bulk motion of the outflow wind in which they are entrained, and another in which CR transport is predominantly diffusive where CRs are contained and deposit much of their energy into their host galaxy's ISM.

The enhanced heating power of CRs contained by a protogalactic magnetic field can have a range of important consequences for the future evolution of the host galaxy and its neighbours. This effect can particularly have implications for subsequent star formation and the initial mass function of stars in the host after CR containment, as well as an enhanced X-ray emission due to the radiative inverse Compton losses of secondary particles produced in CR interaction showers~\citep[see also][]{Schober2015MNRAS}. 
The degree to which these effects arise and their relative importance can only be understood with more detailed modelling. While the advective transport of CRs modelled in the present paper does reduce some of the effects discussed above, we calculate that the heating rates and CR containment expected in the presence of strong galactic outflow activity are not vastly changed, and are only substantially reduced in the parts of the galaxy directly affected by the outflow wind. This means that the phenomenological picture painted above is largely unchanged.

Further to this, the level of CR heating observed and the fraction of CRs which may be transported by advection to heat the surroundings implies that the effects on the IGM in the vicinity of the host galaxy are non-negligible. This opens up new questions about the impacts of CR heating and ionisation on, e.g. pre-heating for cosmic reionisation~\citep{Sazonov2015MNRAS, Leite2017MNRAS} or on amplifying intra-cluster and/or intergalactic magnetic fields~\citep[e.g.][]{Miniati2011ApJ, Beck2013MNRAS_B, Lacki2015MNRAS}, and the ability for such processes to be maintained. The influence of the advected CRs may be even more important than this. If around 10\% of CRs are indeed able to escape from highly star-forming galaxies in the presence of outflows, the balance of hydrostatic equilibrium may be changed in the host galaxy's surroundings due to the additional CR pressure this would introduce. This may be particularly important in larger scale groups of galaxies and clusters containing one or more actively star-forming protogalaxies if the additional CR pressure provided is non-negligible and maintained for a sufficient length of time.

\section*{Acknowledgements}

ERO is supported by a UK Science and Technology Facilities Council PhD Studentship. 
XJ's visit to UCL-MSSL was supported by a UCL MAPS Dean's Summer Research Scholarship and a MSSL Summer Studentship. 
SC's visit to UCL-MSSL was supported by a MSSL Summer Studentship. 
We thank Prof Chung-Ming Ko (NCU) for discussions regarding transport of cosmic rays, 
    Prof Mat Page (UCL-MSSL) for SN rates in local starburst systems 
    and Prof Daisuke Kawata (UCL-MSSL) for implications of cosmic-ray heating in galaxies. 
    We thank the anonymous referee for their constructive and informative comments which led to refinements and improvements of our model and discussion.
This research has made use of NASA's Astrophysics Data Systems.




\bibliographystyle{mnras}
\bibliography{references} 



\appendix

\section{Numerical Scheme for the Transport Equation}
\label{sec:appendixa}

In \S~\ref{subsec:transport} we invoke a numerical scheme to solve the differential equations~\ref{eq:advection_eq} and~\ref{eq:diffusion_equation2}. Here we outline the numerical scheme used in each case in more detail.

\subsection{Advection Regime}
\label{subsec:advection}

In this case, we aim to solve
\begin{equation}
\frac{\partial Z}{\partial h} = \frac{1}{v(h)}\left\{ Z \frac{\partial b(E, h)}{\partial E} + b(E, h) \frac{\partial Z}{\partial E} -  {\rm c} ~Z ~\hat{\sigma}_{\rm p\pi}(E)~n_{\rm p}(h)\right\} \ ,
\end{equation}
where $b(E, h) = -{{\rm d}E}/{{\rm d}t}$ is known analytically from the total cooling processes (see \S~\ref{sec:cr_spectrum}) but, apart from adiabatic cooling, is negligible in the case of CR protons, and where $\partial b(E, h) / \partial E$ immediately follows from this analytically. We retain these terms in the following treatment so as to describe a general scheme applicable to other particles where these cooling terms may not be negligible.
 
The equation is discretised according to a numerical grid over 170 points in energy $E_j$ distributed linearly between $E_0 = 1~{\rm GeV}$ and $E_{\rm max} = 10^6~{\rm GeV}$ and 10,000 points in position $h_i$ distributed linearly between $0.1~\text{kpc}$ and $100~\text{kpc}$, indexed by the notation $j$ and $i$ respectively. This gives the first-order difference equation
\begin{align}
Z_{i, j+1} = Z_{i,j} + \frac{\Delta h}{v(h_j)} \Bigg\{ Z_{i,j} \frac{\partial b_{i,j}}{\partial E} & + b_{i,j} \frac{Z_{i+1, j}-Z_{i-1, j}}{E_{i+1}-E_{i-1}} \nonumber \\
 & -  {\rm c} ~Z_{i,j} ~\hat{\sigma}_{\rm p\pi}(E_i)~n_{\rm p}(h_j) \Bigg\} \ .
\label{eq:difference_advection_eq}
\end{align}
Each calculated point $Z_{i,j}$ requires the adjacent points in the previous $h$-step over three energies. This allows the gradient $\partial Z /\partial E$ to be estimated, and the gradient at the central point may then be propagated forwards. In the `edge-cases' we find it is sufficient to simply estimate the gradient from two of the three available points; the energy bins are of sufficient resolution that any small inaccuracies from taking fewer points to calculate the gradient at the edges of the grid are rapidly suppressed. 
Our tests with higher order difference schemes did not yield noticeably different results.

We use a 4th order Runge-Kutta (RK4) scheme~\citep{Press2007book} with 5th order error estimation to arrive at the numerical solution at each grid point, subject to the boundary conditions discussed in the main part of the text.


\subsection{Diffusion Regime}
\label{subsec:diffusion}

Here, we aim to solve the second order differential equation
\begin{align}
\frac{\partial^2 Z}{\partial h^2} = -\frac{1}{D(E)} & \Bigg\{ Z \frac{\partial b(E, h)}{\partial E} + b(E, h) \frac{\partial Z}{\partial E} \nonumber \\
& -  {\rm c} ~Z ~\hat{\sigma}_{\rm p\pi}(E)~n_{\rm p}(h)\Bigg\} + \frac{2}{h} \frac{\partial Z}{\partial h} - \frac{2 Z}{h^2} \ ,
\end{align}
which can be done in a similar way to the advection equation above (again, subject to the earlier boundary conditions) and splitting into an `inner' and `outer' scheme. We may use the RK4 method to find the numerical solution ${\rm d}Z/{\rm d}h$ in the `inner' scheme according to the difference equation
\begin{align}
\frac{{\rm d}Z}{{\rm d}h}\bigg\rvert_{i, j+1} = & \frac{{\rm d}Z}{{\rm d}h}\bigg\rvert_{i, j} + \frac{\Delta h}{D_{i,j} \left(1-\frac{2}{h}\right)} \Bigg\{ Z_{i,j} \frac{\partial b_{i,j}}{\partial E} + b_{i,j} \frac{Z_{i+1, j}-Z_{i-1, j}}{E_{i+1}-E_{i-1}} \nonumber \\
 & -  {\rm c} ~Z_{i,j} ~\hat{\sigma}_{\rm p\pi}(E_i)~n_{\rm p}(h_j) \Bigg\}- \frac{2 Z_{i,j}}{h}\left(\frac{1}{h-2}\right)\ ,
\label{eq:difference_diffusion_eq1}
\end{align}
and a further `outer' step is required to arrive at the numerical result $Z$, with difference equation
\begin{equation}
 Z_{i,j+1} = Z_{i,j} + \frac{{\rm d}Z}{{\rm d}h}\bigg\rvert_{i, j} \Delta h \ .
 \label{eq:difference_diffusion_eq2}
 \end{equation}
Again, tests with higher order difference schemes indicated the approach adopted here was adequate. We note that additional Neumann boundary conditions are needed for the inner numerical scheme, i.e. step 1 in equation~\ref{eq:difference_diffusion_eq1}, as detailed in the main text -- see \S~\ref{subsec:transport}.





\bsp	
\label{lastpage}
\end{document}